\documentclass[a4paper,12pt]{article}
\usepackage{natbib,amsmath,amsthm,amssymb,amsfonts}
\makeatletter

\@addtoreset{equation}{section}
\makeatother

\def\ep{\varepsilon}

\def\R{\mathbb R}
\def\S{\mathbb S}
\def\N{\mathbb N}
\def\pa{\partial}
\def\b{\backslash}
\def\A{\mathbf A}

\begin{document}

\noindent{\bf\Large On inverse problems in electromagnetic field}

\noindent{\bf\Large in classical mechanics at fixed energy}
\vskip 7mm

\noindent Alexandre Jollivet

\vskip 1cm

\noindent{\bf Abstract.} In this paper, we consider inverse scattering and inverse boundary value problems at sufficiently large and 
fixed energy for the 
multidimensional relativistic and nonrelativistic  Newton equations in a static external electromagnetic field  $(V,B)$, $V\in C^2,$ $B\in C^1$
in classical mechanics.  Developing the approach going back to Gerver-Nadirashvili 1983's work on an inverse problem of mechanics,
we obtain, in particular, theorems of uniqueness.
\vskip 1cm

\section{Introduction}

\noindent 1.1 {\it Relativistic Newton equation.}
Consider the Newton-Einstein equation in a static electromagnetic field in an open subset $\Omega$ of $\R^n,$
$n\ge 2,$
\begin{equation}
\label{1.1}\begin{array}{l}
\dot p=-\nabla V(x)+{1\over c} B(x)\dot x,\\
p={\dot x \over \sqrt{1-{|\dot x|^2 \over c^2}}},
\end{array}
\end{equation}
where $x=x(t)$ is a $C^1$
function with values in $\Omega,$ $\dot p={dp\over dt},$ $\dot x={dx\over dt},$ and 
$V\in C^2(\bar \Omega,\R)$ (i.e. there exists $\tilde{V}\in C^2(\R^n,\R)$ such that $\tilde{V}$ restricted to $\bar \Omega$ is equal to V),
$B\in {\cal F}_{mag}(\bar \Omega)$ where ${\cal F}_{mag}(\bar \Omega)$ is the family of magnetic fields on $\bar \Omega,$ i.e. 
${\cal F}_{mag}(\bar \Omega)=\{B'\in C^1(\bar \Omega, A_n(\R)) \ |\ B'=(B'_{i,k}),\ 
{\pa\over \pa x_i}B'_{k,l}(x)+{\pa\over \pa x_l}B'_{i,k}(x)+{\pa\over \pa x_k}B'_{l,i}(x)=0,\ x\in \bar D, \ i,k,l=1\ldots n\}$ 
and $A_n(\R)$ denotes the space of $n\times n$ real
antisymmetric matrices.

By $\|V\|_{C^2,\Omega}$ we denote the supremum of the set 
$\{|\pa_x^j V(x)|\ |\ x\in \Omega,\ $ $j=(j_1,..,j_n)\in (\N\cup \{0\})^n,\sum_{i=1}^n j_i\le 2\}$ and by $\|B\|_{C^1,\Omega}$ we denote the supremum of the set 
$\{|\pa_x^j B_{i,k}(x)|\ |\ x\in \Omega,\ $ $i,k=1\ldots n,$ $j=(j_1,..,j_n)\in (\N\cup \{0\})^n,\sum_{i=1}^n j_i\le 1\}$.

The equation (1.1) is an equation for $x=x(t)$ and is the equation of motion in $\R^n$ of a relativistic particle of mass
$m=1$ and charge $e=1$ in an external electromagnetic field described by  $V$ and $B$
(see [E] and, for example, Section 17 of [LL]). In this equation $x$ is the position of the particle, $p$ is its impulse, $t$ is
the time and $c$ is the speed of light.

For the equation (1.1) the energy
\begin{equation}
\label{1.2}E=c^2\sqrt{1+{|p(t)|^2 \over c^2}}+V(x(t))
\end{equation}
is an integral of motion.
We denote by $B_c$ the euclidean open ball whose radius is c and whose centre is 0.

In this paper we consider the equation (1.1) in two situations.
We study equation (1.1) when
\begin{equation}
\label{1.3a}\begin{array}{l}
\Omega=D \textrm{ where } D \textrm{ is a bounded strictly convex in the strong sense}\\
\textrm{open domain of }\R^n, n\ge 2,\textrm{with }C^2 \textrm{ boundary}.
\end{array}
\end{equation}

And we study equation (1.1) when 
\begin{equation}
\label{1.3b}\begin{array}{l}
\Omega=\R^n \textrm{ and }|\pa_x^{j_1}V(x)|\le \beta_{|j_1|}(1+|x|)^{-\alpha-|j_1|},\ x\in \R^n,\ n\ge 2,\\
|\pa_x^{j_2}B_{i,k}(x)|\le \beta_{|j_2|+1}(1+|x|)^{-\alpha-1-|j_2|},\ x\in \R^n,
\end{array}
\end{equation}
for $|j_1|\le 2,|j_2|\le 1,$ $i,k=1\ldots n$ and some $\alpha>1$ (here $j$ is the multiindex $j\in (\N\cup \{0\})^n,$ $|j|=\sum_{i=1}^nj_i$ and $\beta_{|j|}$ are positive real
constants).

For the equation (1.1) under condition \eqref{1.3a}, we consider boundary data. For
equation (1.1) under condition \eqref{1.3b}, we consider scattering data.

\vskip 4mm

\noindent 1.2 {\it Boundary data.}
For the equation (1.1) under condition \eqref{1.3a}, one can prove that at
sufficiently large energy $E$ (i.e. $E>E(\|V\|_{C^2,D},\|B\|_{C^1,D},D)$), the solutions $x$ at energy
$E$ have the following properties (see Properties 2.1 and 2.2 in Section 2 and see Section 6):
\begin{equation}
\label{1.4}\begin{array}{l}
\textrm{for each solution }x(t) \textrm{ there are }t_1,t_2\in \R, t_1<t_2, \textrm{ such that }\\
x\in C^3([t_1,t_2],\R^n),x(t_1), x(t_2)\in \pa D, x(t)\in D
\textrm{ for }t\in ]t_1,t_2[,\\
x(s_1)\not=x(s_2) {\rm \ for\ } s_1,s_2\in [t_1,t_2], s_1\not= s_2;
\end{array}
\end{equation}
\begin{equation}
\label{1.5}\begin{array}{l}
\textrm{for any two distinct points }q_0,q\in \bar D, \textrm{ there is one and only one solution}\\
x(t)=x(t,E,q_0,q)\textrm{ such that }x(0)=q_0, x(s)=q \textrm{ for some }s>0.
\end{array}
\end{equation}
Let $(q_0,q)$ be two distinct points of $\pa D$. By $s_{V,B}(E,q_0,q)$ we denote the time at which $x(t,E,q_0,q)$ reaches $q$ from $q_0$.
By $k_{0,V,B}(E,q_0,q)$ we denote the velocity vector $\dot x(0,E,q_0,q).$
By $k_{V,B}(E,q_0,q)$ we denote the velocity vector $\dot x(s_{V,B}(E,q_0,q),E,q_0,q).$ We consider $k_{0,V,B}(E,q_0,q),$ $k_{V,B}(E,q_0,q),$ $q_0,q\in\pa D, q_0\not=q,$ as the boundary value data. 

\vskip 2mm

{\bf Remark 1.1.} For 
$q_0,q\in\pa D, q_0\not=q,$
the trajectory of $x(t,E,q_0,q)$ and the trajectory of $x(t,E,q,q_0)$ are distinct, in general. 

\vskip 2mm
Note that in the present paper we always assume that the aforementioned real constant $E(\|V\|_{C^2,D},\|B\|_{C^1,D},D)$, considered 
as function of  

\noindent $\|V\|_{C^2,D}$ and $\|B\|_{C^1,D}$, satisfies
\begin{equation}
E(\lambda_1,\lambda_2,D)\le E(\lambda_1',\lambda_2',D) \textrm{ if }\lambda_1\le  \lambda_1'\textrm{ and } \lambda_2\le  \lambda_2',
\label{1.5*}
\end{equation}
for $\lambda_1,\lambda_2,\lambda_1',\lambda_2'\in [0,+\infty[$.
\vskip 4mm

\noindent 1.3 {\it Scattering data.}
For the equation (1.1) under condition \eqref{1.3b}, the following is valid (see [Y]): for any 
$(v_-,x_-)\in B_c\times\R^n,\ v_-\neq 0,$
the equation (1.1)  has a unique solution $x\in C^2(\R,\R^n)$ such that
\begin{equation}
{x(t)=v_-t+x_-+y_-(t),}\label{1.6}
\end{equation}
where $\dot y_-(t)\to 0,\ y_-(t)\to 0,\ {\rm as}\ t\to -\infty;$  in addition for almost any 
$(v_-,x_-)\in B_c\times \R^n,\ v_-\neq 0,$
\begin{equation}
{x(t)=v_+t+x_++y_+(t),}\label{1.7}
\end{equation}
where $v_+\neq 0,$ $|v_+|<c,$ $v_+=a(v_-,x_-),$ $x_+=b(v_-,x_-),$ $\dot y_+(t)\to 0,$ $y_+(t)\to 0,$ as $t \to +\infty$.

For an energy $E>c^2,$ the map 
$S_E: \S_E\times\R^n \to \S_E\times\R^n $ (where 
$\S_E=\{v\in B_c\ |\ |v|=c\sqrt{1-\left({c^2\over E}\right)^2} \}$)
given by the formulas
\begin{equation}
{v_+=a(v_-,x_-),\ x_+=b(v_-,x_-)},\label{1.8}
\end{equation}
is called the scattering map at fixed energy $E$ for the equation (1.1) under condition \eqref{1.3b}. 
By ${\cal D}(S_E)$ we denote the domain of definition of $S_E.$  The data
$a(v_-,x_-),$ $b(v_-,x_-)$ for $(v_-,x_-)\in {\cal D}(S_E)$ are called the scattering data at fixed energy $E$ for the equation (1.1) 
under condition \eqref{1.3b}.

\vskip 4mm

\noindent 1.4 {\it Inverse scattering and boundary value problems.}
In the present paper, we consider the following inverse boundary value problem at fixed energy for the equation (1.1) under condition 
\eqref{1.3a}:
\begin{eqnarray*}
{\bf Problem\ }1: 
\textrm{ given }k_{V,B}(E,q_0,q),\ k_{0,V,B}(E,q_0,q) \textrm{ for all } q_0,q\in\pa D,&&\\
q_0\not=q,\textrm{ at fixed
 sufficiently large energy }E,
\textrm{ find } V \textrm{ and }B.&&
\end{eqnarray*}
The main results of the present work include the following theorem of uniqueness for Problem 1.
\vskip 2mm
{\bf Theorem 1.1.} {\it At fixed} $E>E(\|V\|_{C^2,D},\|B\|_{C^1,D},D)$, {\it the boundary data} $k_{V,B}(E,q_0,q)$, 
$(q_0,q)\in \pa D\times \pa D,$ $q_0\not=q$,
{\it uniquely determine $V,B$. 

At fixed} $E>E(\|V\|_{C^2,D},\|B\|_{C^1,D},D)$, {\it the boundary data} $k_{0,V,B}(E,q_0,q)$, 
$(q_0,q)\in \pa D\times \pa D,$ $q_0\not=q$,
{\it uniquely determine $V,B$. 
}
\vskip 2mm

Theorem 1.1 follows from Theorem 3.1 given in Section 3.

In the present paper, we also consider the following inverse scattering problem at fixed energy for the equation (1.1) under condition 
\eqref{1.3b}:
$$
{\bf Problem\ }2: {\rm\ given\ }S_E {\rm\ at\ fixed\ energy\ }E, {\rm\ find\ }V {\rm\ and\ }B.
$$
The main results of the present work include the following theorem of uniqueness for Problem 2.
\vskip 2mm 

{\bf Theorem 1.2.}
{\it Let $\lambda\in \R^+$ and let $D$ be a bounded strictly convex in the strong sense 
open domain of $\R^n$ with $C^2$ boundary.  Let $V_1, V_2\in C^2_0(\R^n,\R),$
$B_1, B_2\in C^1_0(\R^n,A_n(\R))\cap{\cal F}_{mag}(\R^n),$ 
$\max(\|V_1\|_{C^2,D},\|V_2\|_{C^2,D},\|B_1\|_{C^1,D} ,\|B_2$ $\|_{C^1,D})\le \lambda,$ and ${\rm supp}(V_1)\cup {\rm supp}
(V_2)$ $\cup{\rm supp}(B_1)\cup{\rm supp}(B_2)\subseteq D.$ Let $S^\mu_E$ be the scattering map at fixed energy $E$ subordinate to 
$(V_\mu,B_\mu)$ for $\mu=1,2.$ 
Then there exists a nonnegative real
constant $E(\lambda,D)$ such that for any
$E>E(\lambda,D),$ $(V_1,B_1)\equiv (V_2,B_2)$ if and only if $S^1_E\equiv S^2_E.$
}

\vskip 2mm 

Theorem 1.2 follows from Theorem 1.1, \eqref{1.5*} and Proposition 2.1 of Section 2.

\vskip 2mm

{\bf Remark 1.2.} Note that for $V\in C^2_0(\R^n,\R)$, if $E< c^2+\sup\{V(x)\ |\ x\in \R^n\}$ then $S_E$ does not determine uniquely $V.$
\vskip 2mm
 
{\bf Remark 1.3.} Theorems 1.1 and 1.2 give uniqueness results. In this paper we do not prove and do not obtain stability results for 
Problem 1 and for Problem 2.

\vskip 4mm

\noindent 1.5 {\it Historical remarks.}
An inverse boundary value problem at fixed energy and at high energies was studied in [GN] for the multidimensional nonrelativistic Newton equation
(without magnetic field) in a bounded open
strictly convex domain. In [GN] results of uniqueness and stability for the inverse boundary value problem at fixed energy are derived from results for the
problem of determining an isotropic Riemannian metric from its hodograph (for this geometrical problem, see [MR], [B] and [BG]). 

Novikov [N2] studied inverse scattering for nonrelativistic multidimensional Newton equation (without magnetic field). Novikov [N2] gave, in particular, a connection between the inverse scattering
problem at fixed energy and Gerver-Nadirashvili's inverse boundary value problem at fixed energy.
Developing the approach of [GN] and [N2], the author [J3] studied an inverse boundary problem and inverse scattering problem for the multidimensional
relativistic Newton equation (without magnetic field) at fixed energy. In [J3] results of uniqueness and stability are obtained.

Inverse scattering at high energies for the relativistic multidimensional Newton equation was studied by the author (see [J1], [J2]).

As regards analogs of Theorems 1.1, 1.2 and Proposition 2.1 for the case $B\equiv0$ for nonrelativistic quantum mechanics see [N1], 
[NSU], [N3] and further references therein. As regards an analog of Theorem 1.2 for the case $B\equiv0$  
for relativistic quantum mechanics see [I]. As regards analogs of Theorems 1.1, 1.2 for the case $B\not\equiv 0$
for nonrelativistic quantum mechanics, see [ER], [NaSuU] and further references
given therein.  

As regards results given in the
literature on inverse scattering in quantum mechanics at high energy limit see references given in [J2].

\vskip 4mm

\noindent 1.6 {\it Structure of the paper.}
The paper is organized as follows. In Section 2, we give some properties of boundary data and scattering data
and we connect the inverse scattering problem at fixed energy to the inverse boundary value problem at fixed energy.
In Section 3, we give, actually, a proof of Theorem 1.1 (based on Theorem 3.1 formulated in Section 3). 
Section 4 is devoted to the proof of Lemma 3.1 and Theorem 3.1 formulated in Section 3.
Section 5 is devoted to the proof of Lemma 2.1 and Proposition 3.1 formulated in Section 2 and in Section 3.
Section 6 is devoted to the proof of Properties (2.1) and (2.2).
In Section 7, we give results similar to Theorems 1.1, 1.2 for the nonrelativistic case.
\vskip 4mm

\noindent{\bf Acknowledgement.} This work was fulfilled in the framework of Ph. D. thesis research under the direction of R.G. Novikov.

\section{Scattering data and boundary value data}

\noindent 2.1 {\it Properties of the boundary value data.} 
Let $D$ be a bounded strictly convex open subset of $\R^n$, 
$n\ge 2$, with $C^2$ boundary.

At fixed sufficiently large energy $E$ (i.e. $E>E(\|V\|_{C^2,D},\|B\|_{C^1,D},D)$ $\ge c^2+\sup_{x\in \bar D}V(x)$) solutions $x(t)$
of the equation (1.1) under condition \eqref{1.3a} have the following properties (see Section 6):

\begin{equation}
\label{2.1}
\begin{array}{l}
{\rm for\ each\ solution\ }x(t)\ {\rm there\ are\ }t_1,t_2\in \R, t_1<t_2, {\rm\ such\ that\ }\\
x\in C^3([t_1,t_2],\R^n),x(t_1), x(t_2)\in \pa D, x(t)\in D
{\rm \ for\ }t\in ]t_1,t_2[,\\
x(s_1)\not=x(s_2) {\rm \ for\ } s_1,s_2\in [t_1,t_2], s_1\not= s_2,\dot x(t_1)\circ N(x(t_1))<0\\
{\rm and\ }\dot x(t_2)\circ N(x(t_2))>0,{\rm\ where\ }N(x(t_i)) {\rm\ is\ the\ unit\ outward}\\
{\rm normal\ vector\ of \ }\pa D{\rm\ at\ }x(t_i) {\rm\ for\ }i=1,2;
\end{array}
\end{equation}
\begin{equation}
\label{2.2}
\begin{array}{l}
{\rm for\ any\ two\ points\ }q_0,q\in \bar D, q\not=q_0, {\rm\ there\ is\ one\ and\ only\ one\ solution}\\
x(t)=x(t,E,q_0,q){\rm\ such\ that\ }x(0)=q_0, x(s)=q {\rm\ for\ some\ }s>0;\\
\dot x(0,E,q_0,q)\in C^1((\bar D\times\bar D)\backslash \bar G,\R^n),{\rm\ where\ } \bar G {\rm\ is\ the\ diagonal\ in\ }
\bar D\times\bar D;
\end{array}
\end{equation}
where $\circ$ denotes the usual scalar product on $\R^n$
(and where by ``$\dot x(0,E,q_0,q)$ $\in C^1((\bar D\times\bar D)\backslash \bar G,\R^n)$" we mean that $\dot x(0,E,q_0,q)$ is the 
restriction to $(\bar D\times\bar D)\backslash \bar G$ of a function which belongs to 
$C^1((\R^n\times \R^n)\backslash\Delta)$ where $\Delta$ is the diagonal of $\R^n\times \R^n$).

\vskip 2mm
{\bf Remark 2.1.} If $B\in C^1(\bar D,A_n(\R))$ and $B\not\in {\cal F}_{mag}(\bar D)$ (where $A_n(\R)$ denotes the space of $n\times n$ real
antisymmetric matrices), then at fixed energy $E>E(\|V\|_{C^2,D},\|B\|_{C^1,D},D)$ 
solutions $x(t)$ of equation (1.1) under condition \eqref{1.3a} also have properties (2.1), (2.2) (see Section 6).

\vskip 2mm

We remind that the aforementioned real constant $E(\|V\|_{C^2,D},\|B\|_{C^1,D},$ 

\noindent $D)$, considered  
as function of  $\|V\|_{C^2,D}$ and $\|B\|_{C^1,D}$, satisfies \eqref{1.5*}. 
In addition, real constant $E(\|V\|_{C^2,D},\|B\|_{C^1,D},D)$ has the
following property: for any $C^2$ continuation $\tilde{V}$ of $V$ on $\R^n,$ and for any $\tilde{B}\in C^1(\R^n,A_n(\R))$ 
such that $\tilde{B}\equiv B$ on $\bar D$,  one has 
\begin{equation}
E(\|\tilde{V}\|_{C^2,D_{x_0,\ep}},\|\tilde{B}\|_{C^1,D_{x_0,\ep}},D_{x_0,\ep})\to E(\|V\|_{C^2,D},\|B\|_{C^1,D},D), \textrm{ as }\ep\to 0,
\label{1.6*}
\end{equation}
where  $D_{x_0,\ep}=\{x_0+(1+\ep)(x-x_0)\ |\ x\in D\}$ for any $x_0\in D$ and $\ep >0$ (note that $D_{x_0,\ep}$ is a bounded, open, strictly convex (in
the strong sense) domain of $\R^n$ with $C^2$ boundary).

Let $E> E(\|V\|_{C^2,D},\|B\|_{C^1,D},D)$. Consider the solution $x(t,E,q_0,q)$ from (2.2) for $q_0,q\in \bar D,$ $q_0\not=q.$ 
We define vectors $k_{V,B}(E,$ $q_0,q)$ and $k_{0,V,B}(E,q_0,q)$ by 
\begin{eqnarray*}
k_{V,B}(E,q_0,q)&=&\dot x(s_{V,B}(E,q_0,q),E,q_0,q),\\
k_{0,V,B}(E,q_0,q)&=&\dot x(0,E,q_0,q),
\end{eqnarray*}
where we define
$s=s_{V,B}(E,q_0,q)$ as the root of the
equation 
$$
x(s,E,q_0,q)=q,\ \ s>0.
$$ 
For $q_0=q\in \bar D$, we put $s_{V,B}(E,q_0,q)=0.$

Note that 
\begin{equation}
\label{2.3}
\begin{array}{l}
|k_{0,V,B}(E,q_0,q)|=c\sqrt{1-\left({E-V(q_0)\over c^2}\right)^{-2}},\\
|k_{V,B}(E,q_0,q)|=c\sqrt{1-\left({E-V(q)\over c^2}\right)^{-2}},
\end{array}
\end{equation}
for $(q,q_0)\in (\bar D\times \bar D)\backslash \bar G.$

Using Properties (2.1) and (2.2), we obtain
\vskip 2mm
{\bf Lemma 2.1.} {\it At fixed $E> E(\|V\|_{C^2,D},\|B\|_{C^1,D},D),$ we have that 

\noindent $s_{V,B}(E,q_0,q)\in C(\bar D\times \bar D,\R),$
$s_{V,B}(E,q_0,q)\in C^1((\bar D\times \bar D)\backslash \bar G,\R)$ and

\noindent $k_{V,B}(E,q_0,q)\in C^1((\bar D\times \bar D)\backslash \bar G,\R^n)$.
}
\vskip 2mm

We consider $k_{V,B}(E,q_0,q),$ $k_{0,V,B}(E,q_0,q),$ $q_0,q\in \pa D,$ $q_0\not=q$ as the boundary value data.
\vskip 2mm
{\bf Remark 2.2.} Note that if $x(t)$ is solution of \eqref{1.1} under condition \eqref{1.3a}, then $x(-t)$ is solution of \eqref{1.1}
with $B$ replaced by $-B\in{\cal F}_{mag}(\bar D)$ under condition \eqref{1.3a}.
Hence the following equalities are valid: at fixed $E> E(\|V\|_{C^2,D},\|B\|_{C^1,D},D),$
\begin{eqnarray}
k_{0,V,B}(E,q_0,q)&=&-k_{V,-B}(E,q,q_0),\label{P1}\\
k_{V,B}(E,q_0,q)&=&-k_{0,V,-B}(E,q,q_0),\label{P2}\\
s_{V,B}(E,q_0,q)&=&s_{V,-B}(E,q,q_0),\label{P3}
\end{eqnarray}
for $q_0,q\in \bar D,$ $q_0\not=q.$

\vskip 4mm
\noindent 2.2 {\it Properties of the scattering operator.}
For equation (1.1) under condition \eqref{1.3b}, the following is valid (see [Y]): for any 
$(v_-,x_-)\in B_c\times\R^n,\ v_-\neq 0,$
the equation (1.1) under condition \eqref{1.3b} has a unique solution $x\in C^2(\R,\R^n)$ such that
\begin{equation}
{x(t)=v_-t+x_-+y_-(t),}\label{2.4}
\end{equation}
where $\dot y_-(t)\to 0,\ y_-(t)\to 0,\ {\rm as}\ t\to -\infty;$  in addition for almost any 
$(v_-,x_-)\in B_c\times \R^n,\ v_-\neq 0,$
\begin{equation}
{x(t)=v_+t+x_++y_+(t),}\label{2.5}
\end{equation}
where $v_+\neq 0,\ |v_+|<c,\ v_+=a(v_-,x_-),\ x_+=b(v_-,x_-),\ \dot y_+(t)\to 0,\ \ y_+(t)$ $\to 0,{\rm\ as\ }t \to +\infty$.

The map $S: B_c\times\R^n \to B_c\times\R^n $ given by the formulas
\begin{equation}
{v_+=a(v_-,x_-),\ x_+=b(v_-,x_-)}\label{2.6}
\end{equation}
is called the scattering map for the equation (1.1) under condition \eqref{1.3b}. The functions $a(v_-,x_-),$ 
$b(v_-,x_-)$ are called the scattering data for the equation (1.1) under condition \eqref{1.3b}.

By ${\cal D}(S)$ we denote the domain of definition of $S$; by ${\cal R}(S)$ we denote the range of $S$ (by definition,
if $(v_-,x_-)\in {\cal D}(S)$, then $v_-\neq 0$ and $a(v_-,x_-)\neq 0$). 

The map $S$  has the following simple properties (see [Y]):
${\cal D}(S)$  is an open set of $B_c \times \R^n$ and ${\rm Mes}((B_c \times \R^n) \b {\cal D}(S))=0$ for the Lebesgue
measure on $B_c \times \R^n$  induced by the Lebesgue measure on $\R^n\times\R^n$;
the map $S:{\cal D}(S)\to {\cal R}(S)$ is continuous and preserves the element of volume; for any $(v,x)\in {\cal D}(S)$,
$a(v,x)^2=v^2.$

For $E>c^2,$ the map $S$ restricted to 
$$
\Sigma_E=\{(v_-,x_-)\in B_c\times \R^n\ |\ |v_-|=c\sqrt{1-\left({c^2\over E}\right)^2}\}
$$
is the scattering operator at fixed energy $E$ and is denoted by $S_E$.

We will use the fact that the map $S$ is uniquely determined by its restriction to ${\cal M}(S)={\cal
D}(S)\cap {\cal M},$ where
$$
{\cal M}=\{(v_-,x_-)\in B_c \times \R^n|v_-\neq 0, v_-x_-=0\}.
$$
This observation is based on the fact that if $x(t)$ satisfies (1.1), then $x(t+t_0)$
also satisfies (1.1) for any $t_0\in\R$.
In particular, the map $S$ at fixed energy $E$ is uniquely determined by its restriction to ${\cal M}_E(S)={\cal D}(S)\cap {\cal M}_E,$
where ${\cal M}_E=\Sigma_E\cap {\cal M}.$

\vskip 4mm

\noindent 2.3 {\it Relation between scattering data and boundary value data.}
Assume that 
\begin{equation}
\label{2.7}
V\in C^2_0(\bar D,\R),\ B\in C^1_0(\bar D,A_n(\R)), \ B\in {\cal F}_{mag}(\bar D).
\end{equation}
We consider equation (1.1) under condition \eqref{1.3a} and equation (1.1) under condition \eqref{1.3b}. We shall connect  
the boundary value data $k_{V,B}(E,q_0,q),$ $k_{0,V,B}(E,q_0,q)$ for $E>E(\|V\|_{C^2,D},\|B\|_{C^1,D},D)$ and 
$(q,q_0)\in (\pa D\times \pa D)\backslash \pa G,$ to the scattering data $a,b.$

\vskip 2mm

{\bf Proposition 2.1.} 
{\it Let $E>E(\|V\|_{C^2,D},\|B\|_{C^1,D},D).$ Under condition} \eqref{2.7}, {\it the following statement is valid:
$s_{V,B}(E,q_0,q),$ $k_{V,B}(E,q_0,q),$ $k_{0,V,B}(E,$ $q_0,q)$ given for all $(q,q_0)\in (\pa D\times \pa D)\backslash \pa G,$ are determined uniquely by 
the scattering data 
$a(v_-,x_-)$, $b(v_-,x_-)$ given for all $(v_-,x_-)\in {\cal M}_E(S).$
The converse statement holds:  $s_{V,B}(E,q_0,q),$ $k_{V,B}(E,q_0,q),$ $k_{0,V,B}(E,q_0,q)$ given for all $(q,q_0)\in (\pa D\times \pa D)\backslash 
\pa G,$ determine uniquely the scattering data 
$a(v_-,x_-)$, $b(v_-,x_-)$ for all $(v_-,x_-)\in {\cal M}_E(S).$
}

\begin{proof}[Proof of Proposition {\rm 2.1}]
First of all we introduce functions $\chi,$ $\tau_-$ and $\tau_+$ dependent on $D.$

For $(v,x)\in \R^n\backslash \{0\}\times\R^n$, $\chi(v,x)$ denotes the nonnegative number of points contained in the intersection of
$\pa D$ with the straight line parametrized by $\R\to \R^n, t\mapsto tv+x.$ As $D$ is a strictly convex open subset of $\R^n$, 
$\chi(v,x)\le 2$ for all $v,x\in \R^n,$ $v\not=0.$

Let $(v,x)\in \R^n\backslash \{0\}\times\R^n$. Assume that $\chi(v,x)\ge 1$. The real $\tau_-(v,x)$ denotes the smallest real number 
$t$ such that $\tau_-(v,x)v+x\in \pa D$, and the real $\tau_+(v,x)$ denotes the greatest real number $t$ such
that $\tau_+(v,x)v+x\in \pa D$ (if $\chi(v,x)=1$ then $\tau_-(v,x)=\tau_+(v,x)$).

\noindent {\it Direct statement.} Let $(q_0,q)\in (\pa D\times \pa D)\backslash \pa G.$
Under conditions \eqref{2.7} and from (2.1) and (2.2), it follows that there exists an unique $(v_-,x_-)\in {\cal M}_E(S)$ such that
\begin{equation*}
\begin{array}{l}
\chi(v_-,x_-)=2,\\
q_0=x_-+\tau_-(v_-,x_-)v_-,\\
q=b(v_-,x_-)+\tau_+(a(v_-,x_-), b(v_-,x_-))a(v_-,x_-).
\end{array}
\end{equation*}
In addition, $s_{V,B}(E,q_0,q)=\tau_+(a(v_-,x_-),b(v_-,x_-))-\tau_-(v_-,x_-)$ and $k_{V,B}(E,$ $q_0,q)=a(v_-,x_-)$ and $k_{0,V,B}(E,q_0,q)=v_-.$

\noindent {\it Converse statement.} Let $(v_-,x_-)\in {\cal M}_E(S).$ 
Under conditions \eqref{2.7}, if $\chi(v_-,$ $x_-)$ $\le 1$ then $(a(v_-,x_-),b(v_-,x_-))=(v_-,x_-).$

Assume that $\chi(v_-,x_-)=2.$ Let 
$$
q_0=x_-+\tau_-(v_-,x_-)v_-.
$$
From (2.1) and (2.2) it follows that there is one and only one solution  of the equation
\begin{equation}
k_{0,V,B}(E,q_0,q)=v_-,\ q\in \pa D,\ q\not=q_0.\label{2.8}
\end{equation}
We denote by $q(v_-,x_-)$ the unique solution of \eqref{2.8}.
Hence we obtain 
\begin{equation*}
\begin{array}{ll}
a(v_-,x_-)=&k_{V,B}(E,q_0,q(v_-,x_-)),\\
b(v_-,x_-)=&q(v_-,x_-)-k_{V,B}(E,q_0,q(v_-,x_-))(s_{V,B}(E,q_0,q(v_-,x_-))\\
&+\tau_-(v_-,x_-)).
\end{array}
\end{equation*}
Proposition 2.1 is proved.
\end{proof}

\vskip 2mm

For a more complete discussion about connection between scattering data and boundary value data, see [N2] considering the nonrelativistic
Newton equation (without magnetic field).

\vskip 1cm

\section{Inverse boundary value problem}

\noindent In this Section, Problem 1 of Introduction is studied.
\vskip 4mm
\noindent 3.1 {\it Notations.}
For $x\in \bar D,$ and for $E>V(x)+c^2,$  we define
$$r_{V,E}(x)=c\sqrt{\left({E-V(x)\over c^2}\right)^2-1}.$$

At $E>E(\|V\|_{C^2,D},\|B\|_{C^1,D},D)$ for $q_0,q\in \bar D\times \bar D,$ $q_0\not=q,$ we define the vectors $\bar k_{V,B}(E,q_0,q)$ 
and $\bar k_{0,V,B}(E,q_0,q)$ by
\begin{equation}
\begin{array}{l}
\bar k_{V,B}(E,q_0,q)
={k_{V,B}(E,q_0,q)\over \sqrt{1-{|k_{V,B}(E,q_0,q)|^2\over c^2}}},\\
\bar k_{0,V,B}(E,q_0,q)
={k_{0,V,B}(E,q_0,q)\over \sqrt{1-{|k_{0,V,B}(E,q_0,q)|^2\over c^2}}}.
\end{array}\label{3.01}
\end{equation}
and $\bar k_{V,B}(E,q_0,q)=(\bar k^1_{V,B}(E,q_0,q),\ldots, \bar k^n_{V,B}(E,q_0,q)),$ 
$\bar k_{0,V,B}(E,q_0,q)=($ $\bar k^1_{0,V,B}(E,q_0,q),\ldots, \bar k^n_{0,V,B}(E,q_0,q))$.
Note that from \eqref{2.3}, it follows that
\begin{equation}
\begin{array}{l}
|\bar k_{V,B}(E,q_0,q)|
=r_{V,E}(q),\\
|\bar k_{0,V,B}(E,q_0,q)|
=r_{V,E}(q_0).
\end{array}\label{3.03}
\end{equation}

For $B\in {\cal F}_{mag}(\bar D),$ let ${\cal F}_{pot}(D, B)$ be the set of $C^1$ magnetic potentials for the magnetic field $B$, i.e. 
\begin{equation}
\begin{array}{l}
{\cal F}_{pot}(D, B):=\{\A\in C^1(\bar D,\R^n)\ |\ 
B_{i,k}(x)={\pa \over \pa x_i}\A_k(x)-{\pa \over \pa x_k}\A_i(x),\ x\in \bar D,\\
i,k=1\ldots n\}.
\end{array}\label{3.02}
\end{equation}
(The set ${\cal F}_{pot}(D, B)$ is not empty: take, for example, $\A(x)=-\int_0^1sB(x_0+s(x-x_0))(x-x_0)ds,$ for $x\in \bar D$ and some
fixed point $x_0$ of $\bar D$.)

\vskip 4mm

\noindent 3.2 {\it Hamiltonian mechanics.} 
Let $\A\in {\cal F}_{pot}(D, B)$. 
The equation (1.1) in $D$ is the Euler-Lagrange equation for the Lagrangian $L$ defined by
$L(\dot x,x)=-c^2\sqrt{1-{\dot x^2\over
c^2}}+c^{-1}\A(x)\circ\dot x-V(x),$ $\dot x\in B_c$ and $x\in D,$ where $\circ$ denotes the usual scalar product on $\R^n.$
The Hamiltonian $H$ associated to the Lagrangian $L$ by Legendre's transform (with
respect to $\dot x$) is 
$H(P,x)=c^2\left(1+c^{-2}|P-\right.$ 

\noindent $\left.c^{-1}\A(x)|^2\right)^{1/2}$ $+V(x)$
where $P\in \R^n$ and $x\in D.$
Then equation (1.1) in $D$ is equivalent to the Hamilton's equation
\begin{equation}
\begin{array}{l}
\dot x={\pa H\over \pa P}(P,x),\\
\dot P=-{\pa H\over \pa x}(P,x),
\end{array}
\label{3.1}
\end{equation} 
for $P\in \R^n,$ $x\in D.$

For a solution $x(t)$ of equation (1.1) in $D,$ we define the impulse vector 
$$
P(t)={\dot x(t)\over \sqrt{1-{|\dot x(t)|^2\over c^2}}}+c^{-1}\A(x(t)).
$$
Further for $q_0,q\in \bar D,$ $q_0\not= q,$  and $t\in[0,s(E,q_0,q)],$ we consider 
\begin{equation}
P(t,E,q_0,q)={\dot x(t,E,q_0,q)\over \sqrt{1- {|\dot x(t,E,q_0,q)|^2\over c^2}}}
+c^{-1}\A(x(t,E,q_0,q)),
\end{equation}
where $x(.,E,q_0,q)$ is the solution given by \eqref{2.2}.
From Maupertuis's principle (see [A]), it follows that if $x(t),$ $t\in[t_1,t_2],$ is a solution of $\eqref{1.1}$ in $D$ with energy
$E$, then $x(t)$ is a critical point of the functional ${\cal A}(y)=\int_{t_1}^{t_2}\left[r_{V,E}(y(t))|\dot y(t)|+c^{-1}
\A(y(t))\circ\dot y(t)
\right] dt$ defined on the set of the functions $y\in C^1([t_1,t_2], D),$ with boundary conditions $y(t_1)=x(t_1)$ and $y(t_2)=x(t_2).$
Note that for $q_0,q\in D,$ $q_0\not=q,$ functional ${\cal A}$ taken along the trajectory of the solution $x(.,E,q_0,q)$ 
given by \eqref{2.2} is equal  to the reduced action ${\cal S}_{0_{V,\A,E}}(q_0,q)$ from $q_0$ to $q$ at fixed energy $E$  for
\eqref{3.1},
where
\begin{equation}
{\cal S}_{0_{V,\A,E}}(q_0,q)=\left\lbrace
\begin{array}{ll}
0,& \textrm{ if } q_0=q,\\
\int_0^{s(E,q_0,q)}P(s,E,q_0,q)\circ\dot x(s,E,q_0,q)ds,&\textrm{ if } q_0\not=q,
\end{array}
\right.
\end{equation}
for $q_0,q\in \bar D$.
\vskip 4mm
\noindent 3.3 {\it Properties of ${\cal S}_{0_{V,\A,E}}$ at fixed and sufficiently large energy $E$.}
The following Propositions 3.1, 3.2 give properties of ${\cal S}_{0_{V,\A,E}}$ at fixed and sufficiently large energy $E$.
\vskip 2mm
{\bf Proposition 3.1.}
{\it 
Let 
$E>E(\|V\|_{C^2,D}, \|B\|_{C^1,D},D).$ The following 

\noindent statements are valid:
\begin{eqnarray}
&&{\cal S}_{0_{V,\A,E}}\in C(\bar D\times \bar D,\R),\label{3.5}\\
&&{\cal S}_{0_{V,\A,E}}\in C^2((\bar D\times \bar D)\backslash \bar G,\R),\label{3.6}\\
&&{\pa {\cal S}_{0_{V,\A,E}}\over \pa x_i}(\zeta,x)=\bar k^i_{V,B}(E,\zeta,x)+c^{-1}\A_i(x),\label{3.7}\\
&&{\pa {\cal S}_{0_{V,\A,E}}\over \pa \zeta_i}(\zeta,x)=-\bar k^i_{0,V,B}(E,\zeta,x)-c^{-1}\A_i(\zeta),\label{3.8}\\
&&{\pa^2 {\cal S}_{0_{V,\A,E}}\over \pa \zeta_i\pa x_j}(\zeta,x)=-{\pa \bar k^i_{0,V,B}\over \pa x_j}(E,\zeta,x)=
{\pa \bar k^j_{V,B}\over \pa \zeta_i}(E,\zeta,x)
,\label{3.9}
\end{eqnarray}
for $(\zeta,x)\in (\bar D\times \bar D)\backslash \bar G,$ $\zeta=(\zeta_1,..,\zeta_n),$ $x=(x_1,..,x_n),$ and $i,j=1\ldots n.$
In addition, 
\begin{eqnarray}
&&\max(|{\pa {\cal S}_{0_{V,\A,E}}\over \pa x_i}(\zeta,x)|,|{\pa {\cal S}_{0_{V,\A,E}}\over \pa \zeta_i}(\zeta,x)|)\le M_1,\label{3.10}\\
&&|{\pa^2 {\cal S}_{0_{V,\A,E}}\over \pa \zeta_i\pa x_j}(\zeta,x)|\le {M_2\over |\zeta -x|},\label{3.11}
\end{eqnarray}
for $(\zeta,x)\in (\bar D\times \bar D)\backslash \bar G,$ $\zeta=(\zeta_1,..,\zeta_n),$ $x=(x_1,..,x_n),$ and $i,j=1\ldots n,$ 
and where $M_1$ and $M_2$ depend on $V,$ $B$ and $D.$
}

\vskip 2mm
Proposition 3.1 is proved in Section 5.

Equalities \eqref{3.7} and \eqref{3.8} are known formulas of classical Hamiltonian mechanics (see Section 46 and further Sections
of [A]). 
\vskip 2mm
{\bf Proposition 3.2.} 
{\it 
Let 
$E>E(\|V\|_{C^2,D}, \|B\|_{C^1,D},D).$
The map $\nu_{V,B,E}:\pa D\times D\to \S^{n-1},$  defined by
\begin{equation}
\nu_{V,B,E}(\zeta,x)=-{k_{V,B}(E,\zeta,x)\over|k_{V,B}(E,\zeta,x)|}, \ {\it for }\ (\zeta,x)\in \pa D\times D, \label{3.12}
\end{equation}
has the following properties:
\begin{equation}
\begin{array}{l}
\nu_{V,B,E}\in C^1(\pa D\times D,\S^{n-1}),\\
{\it the\ map\  }\nu_{V,B,E,x}:\pa D\to \S^{n-1},\ \zeta\mapsto \nu_{V,B,E}(\zeta,x),{\it \ is\ a}\\
C^1\ {\it orientation\ preserving\ diffeomorphism\ from\ }\pa D \ {\it onto}\ \S^{n-1}
\end{array}
\label{3.13}
\end{equation}
for $x\in D$ (where we choose the canonical orientation of $\S^{n-1}$ and the orientation of $\pa D$ given by the canonical orientation of $\R^n$ and 
the unit outward normal vector).
}

\vskip 2mm
Proposition 3.2 follows from \eqref{1.1}, \eqref{1.2} and properties \eqref{2.1}, \eqref{2.2}.
\vskip 2mm
{\bf Remark 3.1.} Taking account of \eqref{3.7} and \eqref{3.8}, we obtain the following formulas: 
at $E>E(\|V\|_{C^2,D}, \|B\|_{C^1,D},D),$ for any $x,\ \zeta \in \bar D,$ $x\not=\zeta,$
\begin{eqnarray}
B_{i,j}(x)&=&-c\left({\pa \bar k^j_{V,B}\over \pa x_i}(E,\zeta,x)-{\pa \bar k^i_{V,B}\over \pa x_j}(E,\zeta,x)\right),\label{3.14}\\
B_{i,j}(x)&=&-c\left({\pa \bar k^j_{0,V,B}\over \pa x_i}(E,x,\zeta)-{\pa \bar k^i_{0,V,B}\over \pa x_j}(E,x,\zeta)\right).\label{3.15}
\end{eqnarray}

\noindent 3.4  {\it Results of uniqueness.}
We denote by $\omega_{0,V,B}$ the $n-1$ differential form on $\pa D\times D$ obtained in the following manner:

- for $x\in D$, let $\omega_{V,B,x}$ be the pull-back of $\omega_0$ by $\nu_{V,B,E,x}$ where $\omega_0$ denotes the canonical 
orientation form on $\S^{n-1}$ (i.e. $\omega_0(\zeta)(v_1,..,v_{n-1})={\rm det}(\zeta,v_1,..,v_{n-1}),$ for $\zeta\in \S^{n-1}$ and 
$v_1,..,v_{n-1}\in T_\zeta\S^{n-1}$),

- for $(\zeta,x)\in \pa D\times D$ and for any $v_1,..,v_{n-1}\in T_{(\zeta,x)}(\pa D\times D),$  
$$\omega_{0,V,B}(\zeta,x)(v_1,..,v_{n-1})=\omega_{V,B,x}(\zeta)(\sigma'_{(\zeta,x)}(v_1),..,\sigma'_{(\zeta,x)}(v_{n-1})),$$
where $\sigma :\pa D\times D\to \pa D,$
$(\zeta',x')\mapsto \zeta',$ and $\sigma'_{(\zeta,x)}$ denotes the derivative (linear part) of $\sigma$ at $(\zeta,x)$.

From smoothness of $\nu_{V,B,E},$ $\sigma$ and $\omega_0$, it follows that $\omega_{0,V,B}$ is a continuous $n-1$ form on $\pa D\times D.$

Now let $\lambda\in \R^+$ and $V_1, V_2\in C^2(\bar D,\R),$ $B_1, B_2\in {\cal F}_{mag}(\bar D),$ such that 
$\max(\|V_1\|_{C^2,D},\|V_2\|_{C^2,D},\|B_1\|_{C^1,D},\|B_2\|_{C^1,D})\le \lambda$. For $\mu=1,2,$ let $\A_\mu\in {\cal
F}_{pot}(D, B_\mu).$ 

Let $E>E(\lambda,\lambda,D)$ where $E(\lambda,\lambda,D)$ is defined in \eqref{1.5*}.
Consider $\beta^1,$ $\beta^2$ the differential one forms defined on $(\pa D\times \bar D)\backslash{\bar G}$ by
\begin{equation}
\label{3.16}\beta^\mu(\zeta,x)=\sum_{j=1}^n\bar k^j_{V_\mu,B_\mu}(E,\zeta,x)dx_j,
\end{equation}
for $(\zeta,x)\in (\pa D\times \bar D)\backslash{\bar G},$ $x=(x_1,\ldots,x_n)$ and $\mu=1,2.$

Consider the differential forms $\Phi_0$ on 
$(\pa D\times \bar D)\backslash{\bar G}$ and $\Phi_1$ on $(\pa D\times \bar D)\backslash{\bar G}$ defined by
\begin{eqnarray}
\Phi_0(\zeta,x)&=&-(-1)^{n(n+1)\over 2}(\beta^2-\beta^1)(\zeta,x)\wedge d_{\zeta}({\cal S}_{0_{V_2,\A_2,E}}-{\cal S}_{0_{V_1,\A_1,E}})(\zeta,x)\nonumber\\
&&\wedge \sum_{p+q=n-2}(dd_\zeta {\cal S}_{0_{V_1,\A_1,E}}(\zeta,x))^p\wedge(dd_\zeta {\cal S}_{0_{V_2,\A_2,E}}(\zeta,x))^q, \label{3.17}
\end{eqnarray}
for $(\zeta,x)\in (\pa D\times \bar D)\backslash{\bar G},$ where $d=d_\zeta+d_x,$
\begin{eqnarray}
\Phi_1(\zeta,x)&=&-(-1)^{n(n-1)\over 2}\left[\beta^1(\zeta,x)\wedge (dd_\zeta {\cal S}_{0_{V_1,\A_1,E}}(\zeta,x))^{n-1}\right.
\nonumber\\
&&+\beta^2(\zeta,x)\wedge (dd_\zeta {\cal S}_{0_{V_2,\A_2,E}}(\zeta,x))^{n-1}
-\beta^1(\zeta,x)\label{3.18}\\
&&\left.\wedge (dd_\zeta {\cal S}_{0_{V_2,\A_2,E}}(\zeta,x))^{n-1}-\beta^2(\zeta,x)\wedge (dd_\zeta {\cal S}_{0_{V_1,\A_1,E}}(\zeta,x))^{n-1}
\right],\nonumber
\end{eqnarray}
for $(\zeta,x)\in (\pa D\times \bar D)\backslash{\bar G},$ where $d=d_\zeta+d_x.$

Consider the $C^2$ map 
$incl:(\pa D\times \pa D)\backslash{\pa G} \to (\pa D\times \bar D)\backslash{\bar G},$ $(\zeta,x)\mapsto (\zeta,x).$

From \eqref{3.6}, \eqref{3.10} and \eqref{3.11}, it follows that $\Phi_0$ is continuous on  $(\pa D\times \bar D)\backslash{\bar G}$ and  $incl^*(\Phi_0)$ 
is integrable on $\pa D\times \pa D$
and $\Phi_1$ is continuous on 
$(\pa D\times \bar D)\backslash{\bar G}$ and integrable on $\pa D\times \bar{D}$ (where $incl^*(\Phi_0)$ is the pull-back of the
differential form $\Phi_0$ by the inclusion map $incl$).

\vskip 2mm

{\bf Lemma 3.1.} {\it Let $\lambda\in \R^+$ and $E>E(\lambda,\lambda,D).$  
Let $V_1,V_2\in C^2(\bar D,\R),$ $B_1, B_2\in {\cal F}_{mag}(\bar D)$ such that 
$\max(\|V_1\|_{C^2,D},\|V_2\|_{C^2,D},\|B_1\|_{C^1,D},\|B_2\|_{C^1,D})\le \lambda$.
For $\mu=1,2$, let $\A_\mu\in {\cal F}_{pot}(D,B_\mu) $.}
{\it The following equalities are valid:
\begin{equation}
\int_{\pa D\times \pa D} incl^*(\Phi_0)=\int_{\pa D\times \bar{D}}\Phi_1;\label{3.19}
\end{equation}
\begin{eqnarray}
{1\over (n-1)!}\Phi_1(\zeta,x)&=&\left(r_{V_1,E}(x)^n\omega_{0,V_1,B_1}(\zeta,x)
+r_{V_2,E}(x)^n\omega_{0,V_2,B_2}(\zeta,x)\right.\nonumber \\
&&-{\bar k_{V_1,B_1}}(E,\zeta,x)\circ{\bar k_{V_2,B_2}}(E,\zeta,x)\label{3.20}\\
&&\left.\times\left(r_{V_1,E}(x)^{n-2}\omega_{0,V_1,B_1}(\zeta,x)
+r_{V_2,E}(x)^{n-2}\right.\right.\nonumber\\
&&\left.\left.\times\omega_{0,V_2,B_2}(\zeta,x)\right)
\right)\wedge dx_1\wedge..\wedge dx_n,\nonumber
\end{eqnarray}
for $(\zeta,x)\in \pa D\times D.$
}

\vskip 2mm

Equality \eqref{3.19} follows from regularization and Stokes' formula.
Proof of Lemma 3.1 is given in Section 4.

Taking account of Lemma 3.1, Proposition 3.2 and Remark 3.1, we obtain the following Theorem of uniqueness. 

\vskip 2mm

{\bf Theorem 3.1.} {\it Let $\lambda\in \R^+$ and $E>E(\lambda,\lambda,D).$
Let $V_1,V_2\in C^2(\bar D,\R),$ $B_1, B_2\in {\cal F}_{mag}(\bar D)$
such that $\max(\|V_1\|_{C^2,D},\|V_2\|_{C^2,D},\|B_1\|_{C^1,D},\|B_2\|_{C^1,D})\le \lambda
$. For $\mu=1,2$, let $\A_\mu\in {\cal F}_{pot}(D,B_\mu) $.
}
{\it The following estimate is valid:
\begin{eqnarray}
&&\int_D\left(r_{V_1,E}(x)-r_{V_2,E}(x)\right)\left(r_{V_1,E}(x)^{n-1}-r_{V_2,E}(x)^{n-1}\right)dx\le\nonumber\\
&&{\Gamma({n\over 2})\over 2\pi^{n\over 2}(n-1)!}
\int_{\pa D\times \pa D}incl^*(\Phi_0).\label{3.21}
\end{eqnarray}
In addition, the following statements are valid:

if $k_{V_1,B_1}(E,\zeta,x)=k_{V_2,B_2}(E,\zeta,x)$ for $\zeta,x\in \pa D,$ $\zeta\not=x,$ then $V_1\equiv V_2$ and $B_1\equiv B_2$ on
$\bar D$;
if $k_{0,V_1,B_1}(E,\zeta,x)=k_{0,V_2,B_2}(E,\zeta,x)$ for $\zeta,x\in \pa D,$ $\zeta\not=x,$ then $V_1\equiv V_2$ and $B_1\equiv B_2$ on
$\bar D$.
}
\vskip 2mm
Proof of Theorem 3.1 is given in Section 4.

If $B_1\equiv 0,$ $B_2\equiv 0$ and $V_1,$ $V_2,$ and $D$ are smoother than $C^2$, then Lemma 3.1 and Theorem 3.1 follow from results
of  [B] and [GN].

\section{Proof of Theorem 3.1 and Lemma 3.1}
Using Lemma 2.1, \eqref{2.3}, Propositions 3.1, 3.2 and Lemma 3.1, we first prove Theorem 3.1.
\begin{proof}[Proof of Theorem 3.1]
From \eqref{3.20} and Proposition 3.2 and definition of  $\omega_{0,V_\mu,B_\mu}$, $\mu=1,2,$ it follows that
\begin{eqnarray}
&&\label{46b}{1\over (n-1)!}\int_{\pa D\times \bar{D}}\Phi_1=\\
&&\int_D r_{V_1,E}(x)^n\int_{\S^{n-1}}\left(1+{w\circ\bar k_{V_2,B_2}
(E,\zeta_{1,x}(w),x)\over r_{V_1,E}
(x)}\right)d\sigma(w)dx\nonumber\\
&&+\int_D r_{V_2,E}(x)^n\int_{\S^{n-1}}\left(1+{\bar k_{V_1,B_1}(E,\zeta_{2,x}(w),x)\circ w\over r_{V_2,E}(x)}
\right)d\sigma(w)dx,\nonumber
\end{eqnarray}
where $d\sigma$ is the canonical measure on $\S^{n-1}$, and where $\circ$ denotes the usual scalar product on $\R^n$, and where for $x\in D$ and $w\in \S^{n-1}$  and $\mu=1,2,$  $\zeta_{\mu,x}(w)$ 
denotes the unique point $\zeta$ of $\pa D$ such that
$w=\nu_{V_\mu,B_\mu,E,x}(\zeta).$
Hence using Cauchy-Bunyakovski-Schwarz inequality and \eqref{3.03} and the equality $\int_{\S^{n-1}}d\sigma(w)=
{2\pi^{n\over 2}\over \Gamma({n\over 2})},$ 
we obtain
\begin{equation}
\label{46}{1\over (n-1)!}\int_{\pa D\times \bar{D}}\!\!\!\!\!\!\Phi_1\ge{2\pi^{n\over 2}\over \Gamma({n\over 2})}\int_D\!\!\!(r_{V_1,E}(x)-r_{V_2,E}(x))
(r_{V_1,E}(x)^{n-1}-r_{V_2,E}(x)^{n-1})
dx.
\end{equation}
Estimate \eqref{46} and equality \eqref{3.19} prove \eqref{3.21}.

Now assume that $k_{V_1,B_1}(E,\zeta,x)=k_{V_2,B_2}(E,\zeta,x)$ for $\zeta,x\in \pa D$, $\zeta\not=x.$ Then from \eqref{3.01} and 
\eqref{3.16}, it follows that 
the one form $incl^*(\beta^2-\beta^1)(\zeta,x)$ is null for any $\zeta,x\in \pa D,$ $\zeta\not=x.$ 
Hence from \eqref{3.17}, it follows that the $2n-2$ form $incl^*(\Phi_0)(\zeta,x)\textrm{ is null for any }\zeta,x\in \pa D,\
\zeta\not=x.$
Thus using \eqref{3.21}, we obtain
$\int_D(r_{V_1,E}(x)-r_{V_2,E}(x)) (r_{V_1,E}(x)^{n-1}-r_{V_2,E}(x)^{n-1})dx\le 0,$
and as $n\ge 2,$ this latter inequality implies that 
\begin{equation}
\label{46a}r_{V_1,E}\equiv r_{V_2,E}\textrm{ on } \bar D.
\end{equation}
Thus  $V_1\equiv V_2.$

Using \eqref{46a} and the equality $|\bar k_{V_i,B_i}(E,\zeta,x)|=r_{V_i,E}(x)$ for $i=1,2,$ $x\in D$ and $\zeta\in \pa D,$
and using \eqref{46b}, we obtain that 
\begin{eqnarray*}
{1\over (n-1)!}\int_{\pa D\times \bar{D}}\Phi_1&=&{1\over 2}\sum_{i=1}^2\int_D r_{V_i,E}(x)^{n-2}\int_{\S^{n-1}}\left|\bar
k_{V_1,B_1}(E,\zeta_{i,x}(w),x)\right.\\
&&\left.-\bar k_{V_2,B_2}(E,\zeta_{i,x}(w),x)\right|^2d\sigma(w)dx.
\end{eqnarray*} 
As $\int_{\pa D\times \bar{D}}\Phi_1=0$ (due to \eqref{3.19}),   we obtain that for any $x\in D,$ and any $w\in \S^{n-1}$, 
$\bar k_{V_1,B_1}(E,\zeta_{1,x}(w),x)=\bar k_{V_2,B_2}(E,\zeta_{1,x}(w),x).$ At fixed $x\in D$, we know that $\zeta_{1,x}$ is onto $\pa D.$
Hence the following equality is valid 
\begin{equation}
\label{46d}\bar k_{V_1,B_1}(E,\zeta,x)=\bar k_{V_2,B_2}(E,\zeta,x),\ \zeta\in \pa D,\ x\in D.
\end{equation}
From \eqref{46d} and \eqref{3.14}, it follows that $B_1\equiv B_2$ on $D$.

Now assume that $k_{0,V_1,B_1}(E,\zeta,x)=k_{0,V_2,B_2}(E,\zeta,x)$ for $\zeta,x\in \pa D$, $\zeta\not=x.$ Then 
using \eqref{P1} and replacing $B_i$ by $-B_i,$ $i=1,2,$ in the proof, we obtain $(V_1,B_1)\equiv (V_2,B_2).$
\end{proof}

Using Lemma 2.1, \eqref{2.3}, Propositions 3.1, 3.2, we prove Lemma 3.1.

\begin{proof}[Proof of Lemma 3.1]
We first prove \eqref{3.20}.
Let $U$ be an open subset of $\R^{n-1}$ and $\phi:U\to \pa D$ such that $\phi$ is a $C^2$ parametrization of $\pa D$.
Let $\phi_0:U\times D\to  \pa D \times D,$ $(\zeta,x)\mapsto(\phi(\zeta),x).$ We work in coordinates given by $(U\times D,\phi_0)$. 
Let $\mu,\mu'=1,2.$ 
On one hand from definition of $\omega_{0,V_\mu,B_\mu},$ definition of 
$\nu_{V_\mu,B_\mu, E,x}$ and \eqref{3.03},
we obtain
\begin{eqnarray}
&&\label{47}\omega_{0,V_\mu,B_\mu}(\zeta,x)\wedge dx_1\wedge \ldots\wedge dx_n=(-1)^nr_{V_\mu,E}(x)^{-n}\\
&&\times{\rm det}\left(
\bar k_{V_\mu,B_\mu}(E,\zeta,x),{\pa\bar k_{V_\mu,B_\mu}\over \pa \zeta_1}(E,\zeta,x)
,\ldots,{\pa\bar k_{V_\mu,B_\mu}\over \pa \zeta_{n-1}}(E,\zeta,x)
\right)\nonumber\\
&&d\zeta_1\wedge\ldots\wedge d\zeta_{n-1}\wedge dx_1\wedge \ldots\wedge dx_n\nonumber
\end{eqnarray}
for $\zeta\in U$ and $x\in D$ and $\mu=1,2.$

On the other hand straightforward calculations give ($dd_\zeta {\cal S}_{0_{V_\mu,\A_\mu,E}}(\zeta,x)=\sum_{m_1=1..n\atop m_2=1..n-1}
{\pa^2 {\cal S}_{0_{V_\mu,\A_\mu,E}}\over \pa x_{m_1}\pa \zeta_{m_2}}(\zeta,x)dx_{m_1}\wedge
d\zeta_{m_2}$)
\begin{eqnarray}
&&\label{47c}\beta^{\mu'}(\zeta,x)\wedge (dd_\zeta  {\cal S}_{0_{V_\mu,\A_\mu,E}}(\zeta,x))^{n-1}=
(-1)^{{(n-1)(n-2)\over 2}}(n-1)!\\
&&\times
{\rm det}\left(
\bar k_{V_{\mu'},B_{\mu'}}(E,\zeta,x),{\pa^2 {\cal S}_{0_{V_\mu,\A_\mu,E}}\over\pa \zeta_1\pa x}(\zeta,x),\ldots,
{\pa^2 {\cal S}_{0_{V_\mu,\A_\mu,E}}\over\pa \zeta_{n-1}\pa x}(\zeta,x)
\right)\nonumber\\
&&d\zeta_1\wedge\ldots\wedge d\zeta_{n-1}\wedge dx_1\wedge \ldots \wedge dx_n,\nonumber
\end{eqnarray}
for $\zeta\in U$, $x\in D$.
Note that due to \eqref{3.03} and Proposition 3.2, $\bar k_{V_\mu,B_\mu}(E,\zeta,x)$ is orthogonal to 
${\pa \over \pa \zeta_m} \bar k_{V_\mu,B_\mu}(E,\zeta,x)$, $m=1\ldots n-1,$ and that $(\bar k_{V_\mu,B_\mu}(E,\zeta,x),$
${\pa \over \pa \zeta_1} \bar k_{V_\mu,B_\mu}(E,\zeta,x),$ $\ldots,$ ${\pa \over \pa \zeta_{n-1}} \bar k_{V_\mu,B_\mu}(E,\zeta,x))$ 
is a basis of $\R^n.$ 
Hence from \eqref{47}, \eqref{47c} and \eqref{3.7}, we obtain 
\begin{eqnarray}
\label{47d}\beta^{\mu'}(\zeta,x)\wedge (dd_\zeta {\cal S}_{0_{V_\mu,\A_\mu,E}}(\zeta,x))^{n-1}=-(-1)^{n(n-1)\over 2}(n-1)!r_{V_\mu,E}(x)^n
&&\\
\times {\bar k_{V_{\mu'},B_{\mu'}}(E,\zeta,x)\circ\bar k_{V_\mu,B_\mu}(E,\zeta,x)\over r_{V_\mu,E}(x)^2}
\omega_{0,V_\mu,B_\mu}(\zeta,x)\wedge dx_1\wedge \ldots\wedge dx_n,\nonumber&&
\end{eqnarray}
for $\zeta\in U$, $x\in D$.
Definition \eqref{3.18} and equality \eqref{47d} proves \eqref{3.20}.

We sketch the  proof of \eqref{3.19}.
Let $\ep\in ]0,1[$ and $x_0\in D.$
We consider
\begin{equation}
\label{48}D_\ep=\{x_0+\ep (x-x_0)\ |\ x\in D\}.
\end{equation}
As $D$ is a strictly convex (in the strong sense) open domain of $\R^n$, with $C^2$ boundary, it follows that 
$D_\ep$ is also a  strictly convex (in the strong sense) open domain of $\R^n$, with $C^2$ boundary, and in addition, as $0<\ep<1$,
\begin{eqnarray}
\label{49a}&&\bar D_\ep\subseteq D,\\
\label{49b}&&dist(\pa D,\pa D_\ep)=(1-\ep)dist(\pa D,x_0).
\end{eqnarray}
where $dist(\pa D,\pa D_\ep)=\inf\{|x-y|\ |\ x\in D_\ep, \ y\in \pa D\}$ and $dist(\pa D,x_0)= \inf\{|y-x_0|\ |\ y\in \pa D\}>0.$

For ${\cal M}$ and ${\cal N}$ two finite-dimensional oriented $C^2$ manifold (with or without boundary), we consider on ${\cal M}
\times {\cal N}$ the
differential product structure induced by  the already given differential structures of 
${\cal M}$ and ${\cal N}$ and we consider the orientation of ${\cal M}\times {\cal N}$ given by
the already fixed orientation of ${\cal M}$ and of ${\cal N}$.
The orientation of $\pa D_\ep$ is given by the unit outward normal vector and $\pa D\times \bar D_\ep$ is a $C^2$ manifold with boundary
$\pa D\times \pa D_\ep$ (which is a 
$C^2$ manifold without boundary).
Let $incl_\ep\in C^2(\pa D\times \pa D_\ep,\pa D\times \bar D_\ep)$ be defined by $incl_\ep(\zeta,x)=(\zeta,x),$ 
$(\zeta,x)\in \pa D\times \pa D_\ep.$
Here we omit the details of the proof of the following statement: $\int_{\pa D\times \bar D_\ep}\Phi_1\to$ 
$\int_{\pa D\times \bar D}\Phi_1$ and $\int_{\pa D\times \pa D_\ep}incl_\ep^*(\Phi_0)\to$ 
$\int_{\pa D\times \pa D}incl^*(\Phi_0)$ as $\ep\to 1^-.$ These statements follow from \eqref{3.10}, \eqref{3.11} and \eqref{3.7}.
We shall prove that
\begin{equation}
\label{50}\int_{\pa D\times \bar D_\ep}\Phi_1=\int_{\pa D\times \pa D_\ep}incl_\ep^*(\Phi_0).
\end{equation}

For $\mu=1,2,$ let ${\cal S}_{0_\mu}\in C^2((\R ^n\times\R^n)\backslash\{(x,x)\ |\ x\in \R^n\},\R)$ such that 
${\cal S}_{0_\mu}(\zeta,x)={\cal S}_{0_{V_\mu,\A_\mu,E}}(\zeta,x),\ (\zeta,x)\in (\bar D\times \bar D)\backslash \bar G$
(from \eqref{3.6}, it follows that such a function ${\cal S}_{0_\mu}$ exists).
Let $\delta_\ep =dist(\pa D,\pa D_\ep).$ Let $W_{1,\delta_\ep}$ be the open subset $\pa D+B(0,{\delta_\ep\over
2})=\{x+y\ |\ x\in \pa D,\ y\in \R^n,\ |y|<{\delta_\ep\over 2}\}$ and let $W_{2,\delta_\ep}$ be the open subset $D_\ep+
B(0,{\delta_\ep\over 2})=\{x+y\ |\ x\in
D_\ep,\ y\in \R^n,\ |y|<{\delta_\ep\over 2}\}$. Note that $W_{1,\delta_\ep}$ is an open neighborhood of $\pa D$ which does not 
intersect $W_{2,\delta_\ep}$
which is an open neighborhood of $\bar D_\ep.$ Hence ${\cal S}_{0_\mu}\in C^2(W_{1,\delta_\ep}\times W_{2,\delta_\ep},\R)$ and 
there exists a sequence of functions $({\cal S}_{0_{\mu,m}})$ such that 
\begin{eqnarray}
\label{49d}&&{\cal S}_{0_{\mu,m}}\in C^3(W_{1,\delta_\ep}\times W_{2,\delta_\ep},\R),\\
\label{49e}&&\sup_{(x,y)\in \pa D\times \bar D_\ep\atop {\alpha=(\alpha_1, \alpha_2)\in \N^2\atop|\alpha|=
\alpha_1+\alpha_2\le 2 }}|{\pa^{|\alpha|} ({\cal S}_{0_{\mu,m}}-{\cal S}_{0_\mu})\over \pa x_{\alpha_1}\pa y_{\alpha_2}}(x,y)|\to 0,\textrm{ as
}m\to+\infty.
\end{eqnarray} 

Fix $m\in \N.$
Let $\mu=1,2.$ We define the differential one-form, $\beta^\mu_m$ on $(\pa D\times \bar D_\ep)$ by
\begin{eqnarray}
\label{50a}\beta^\mu_m(\zeta,x)&=&d_x{\cal S}_{0_{\mu,m}}(\zeta,x)-{1\over c}\sum_{j=1}^n\A^j_\mu(x)dx_j,
\end{eqnarray}
for $(\zeta,x)\in \pa D\times \bar D_\ep$ and $x=(x_1,\ldots,x_n)$ and $\A_\mu(x)=(\A_\mu^1(x),\ldots,\A_\mu^n(x))$ 
and where $d=d_\zeta+d_x$ is the De Rham differential operator on $\pa D\times
\bar D_\ep.$

We define the continuous differential $2n-2$ form $\Phi_{0,m}$ on $\pa D\times \bar D_\ep$ by
\begin{eqnarray}
\Phi_{0,m}(\zeta,x)&=&-(-1)^{n(n+1)\over 2}(\beta^2_m-\beta^1_m)(\zeta,x)\wedge 
d_\zeta({\cal S}_{0_{2,m}}-{\cal S}_{0_{1,m}})(\zeta,x)\nonumber\\
\label{50b}&&\wedge \sum_{p+q=n-2}(dd_\zeta {\cal S}_{0_{1,m}}(\zeta,x))^p\wedge(dd_\zeta {\cal S}_{0_{2,m}}(\zeta,x))^q,
\end{eqnarray}
for $\zeta\in \pa D,$ $x\in \bar D_\ep,$  where $d=d_\zeta+d_x$ is the De Rham differential operator on $\pa D\times
\bar D_\ep.$
We define the continuous differential $2n-1$ form $\Phi_{1,m}$ on $\pa D\times \bar D_\ep$ by
\begin{eqnarray}
\Phi_{1,m}(\zeta,x)&=&-(-1)^{n(n-1)\over 2}\left[\beta^1_m(\zeta,x)\wedge (dd_\zeta {\cal S}_{0_{1,m}}(\zeta,x))^{n-1}\right.
+\beta^2_m(\zeta,x)\nonumber \\
&&\wedge(dd_\zeta {\cal S}_{0_{2,m}}(\zeta,x))^{n-1}
-\beta^1_m(\zeta,x)\wedge (dd_\zeta {\cal S}_{0_{2,m}}(\zeta,x))^{n-1}\nonumber\\
\label{50c}&&\left.-\beta^2_m(\zeta,x)\wedge (dd_\zeta {\cal S}_{0_{1,m}}(\zeta,x))^{n-1}
\right],
\end{eqnarray}
for $(\zeta,x)\in \pa D\times \bar D_\ep.$

From \eqref{50a}-\eqref{50c}, \eqref{3.7}, \eqref{49e}, it follows that
\begin{eqnarray}
\label{50d}\int_{\pa D\times \bar D_\ep}\Phi_{1,m}&\to&\int_{\pa D\times \bar D_\ep}\Phi_1,\  \textrm{ as }m\to +\infty,\\
\label{50e}\int_{\pa D\times \pa D_\ep} incl_\ep^*(\Phi_{0,m})&\to&\int_{\pa D\times \pa D_\ep} incl_\ep^*(\Phi_0),
\  \textrm{ as }m\to +\infty.
\end{eqnarray}
If we prove that $\int_{\pa D\times \bar D_\ep}\Phi_{1,m}=\int_{\pa D\times \pa D_\ep} incl_\ep^*(\Phi_{0,m}),$
then formula
\eqref{50} will follow from \eqref{50d} and \eqref{50e}.

From \eqref{49d}, it follows that
\begin{eqnarray}
\label{51a}&&dd_\zeta {\cal S}_{0_{\mu,m}} \textrm{ is a } C^1\textrm{ form on }\pa D\times\bar D_\ep,\\
\label{51b}&&d(dd_\zeta {\cal S}_{0_{\mu,m}})=0,
\end{eqnarray}
where $d$ is the De Rham differential operator on $\pa D\times \bar D_\ep$.
From \eqref{50a}, it follows that 
\begin{equation}
\label{51c}d\beta^\mu_m(\zeta,x)=-dd_\zeta {\cal S}_{0_{\mu,m}}(\zeta,x)-{1\over c}\sum_{1\le j_1< j_2\le n}B_{j_1,j_2}^\mu(x)
dx_{j_1}\wedge dx_{j_2},
\end{equation}
for $(\zeta,x)\in \pa D\times \bar D_\ep$ and $x=(x_1,\ldots,x_n)$ ($B_{j_1,j_2}^\mu(x)$ denotes the elements of $B_\mu(x)$).
From \eqref{51a}-\eqref{51c}, it follows that $\Phi_{0,m}$ is $C^1$ on $\pa D\times \bar D_\ep$ and that 
\begin{eqnarray}
&&\label{51d}d\Phi_{0,m}(\zeta,x)=(-1)^{n-1}\Phi_{1,m}(\zeta,x)+\omega(\zeta,x),\\
&&\omega(\zeta,x)=(-1)^{{n(n+1)\over 2}}{1\over c}\sum_{1\le j_1< j_2\le n}(B_{j_1,j_2}^2(x)-B_{j_1,j_2}^1(x))dx_{j_1}\wedge dx_{j_2}\nonumber\\
&&\wedge d_\zeta({\cal S}_{0_{2,m}}-{\cal S}_{0_{1,m}})(\zeta,x)
\wedge \sum_{p+q=n-2}(dd_\zeta {\cal S}_{0_{1,m}}(\zeta,x))^p\wedge(dd_\zeta {\cal S}_{0_{2,m}}(\zeta,x))^q,\nonumber
\end{eqnarray}
for $(\zeta,x)\in \pa D\times \bar D_\ep$ and $x=(x_1,\ldots,x_n)$.
Note that as $B_\mu,$ $\mu=1,2,$ is continuously differentiable on $\bar D,$ then the $2$-form defined on $\pa D\times \bar D_\ep$ by 
$\sum_{1\le j_1< j_2\le n}(B_{j_1,j_2}^2(x)-B_{j_1,j_2}^1(x))dx_{j_1}\wedge dx_{j_2},$ $(\zeta,x)\in \pa D\times \bar D_\ep$ is
also $C^1.$ 
Note that ($\bar D_\ep$ is a $n$-dimensional $C^2$ manifold and \eqref{51b})
\begin{equation}
\label{51f}\omega(\zeta,x)=d\tilde{\omega}(\zeta,x),
\end{equation}
where
\begin{eqnarray}
&&\tilde{\omega}(\zeta,x)=
(-1)^{{n(n+1)\over 2}}{({\cal S}_{0_{2,m}}-{\cal S}_{0_{1,m}})(\zeta,x)\over c}\sum_{1\le j_1< j_2\le n}(B_{j_1,j_2}^2(x)-B_{j_1,j_2}^1(x))
\nonumber\\
&&\label{51g}dx_{j_1}\wedge dx_{j_2}\wedge \sum_{p+q=n-2}(dd_\zeta {\cal S}_{0_{1,m}}(\zeta,x))^p\wedge(dd_\zeta {\cal S}_{0_{2,m}}(\zeta,x))^q,
\end{eqnarray}
for $(\zeta,x)\in \pa D\times \bar D_\ep$ and $x=(x_1,\ldots,x_n)$.
Since $\pa D_\ep$ is a $(n-1)$-dimensional $C^2$ manifold, it follows that
\begin{equation}
\label{51h}incl_\ep^*\tilde{\omega}(\zeta,x)=0
\end{equation}
for $(\zeta,x)\in \pa D\times \pa D_\ep.$
Using \eqref{51d}, \eqref{51f}, \eqref{51h}, we obtain by Stokes' formula the equality 
$\int_{\pa D\times \bar D_\ep}\Phi_{1,m}=\int_{\pa D\times \pa D_\ep} incl_\ep^*(\Phi_{0,m}).$
\end{proof}

\section{Proof of Lemma 2.1 and Proposition 3.1}
\noindent  5.1 {\it Continuation of $(V,B)$ and notations.}
Let $\tilde{V}\in C^2(\R^n,\R)$ be such that $\tilde{V}\equiv V$ on $\bar D$ and 
$\|\tilde{V}\|_{C^2,\R^n}<\infty$. Let $\tilde{B}\in C^1(\R^n,A_n(\R))$ (where $A_n(\R)$ denotes the space of real antisymmetric matrices) such
that   $\tilde{B}\equiv B$ on $\bar D$  and 
$\|\tilde{B}\|_{C^1,\R^n}<\infty$.
Let $\psi$ be the flow for the differential system
\begin{eqnarray}
\label{4}\dot x={p\over \sqrt{1+{p^2\over c^2}}},\\
\dot p=-\nabla \tilde{V}(x)+{1\over c}\tilde{B}(x){p\over \sqrt{1+{p^2\over c^2}}},\nonumber 
\end{eqnarray}
for $x\in \R^n$ and $p\in \R^n$ (it means that a solution  of \eqref{4}, $(x(t),p(t))$, $t\in]t_-,t_+[,$ which 
passes through $(x_0,p_0)\in \R^n\times\R^n$ at time $t=0$, is written as $(x(t),p(t))=\psi(t,x_0,p_0)$ for $t\in]t_-,t_+[$). 
For equation \eqref{4}, the energy 
\begin{equation}
\label{4a}E=c^2\sqrt{1+|p(t)|^2\over c^2}+\tilde{V}(x(t))
\end{equation}
is an integral of motion.

Under the conditions on $\tilde{V}$ and $\tilde{B}$, $\psi$ is defined on $\R\times\R^n\times\R^n$ and $\psi\in C^1(\R\times \R^n\times
\R^n,\R^n\times \R^n)$, and a solution $x(t),$ $t\in ]t_-,t_+[,$ of \eqref{1.1} which starts at $q_0\in D$ at time $0$ with velocity $v$
is written as $x(t)=\psi_1(t,x,{v\over\sqrt{1-{v^2\over c^2}}}),$ $t\in ]t_-,t_+[$ 
(we write $\psi=(\psi_1,\psi_2)$ where $\psi_i=(\psi_i^1,\ldots,\psi_i^n)\in C^1(\R\times \R^n\times\R^n,\R^n),$ $i=1,2$).

For $v\in \R^n$ and $x\in \R^n,$ we define the vector $F(x,v)$ of $\R^n$ by
\begin{equation}
\label{5.0}F(x,v)=-\nabla\tilde{V}(x)+{1\over c} \tilde{B}(x)v.
\end{equation}
For $x\in \R^n$ and $E>c^2+\tilde V(x),$ we denote by $r_{\tilde{V},E}(x)$ the positive real number
\begin{equation}
\label{5.02}
r_{\tilde{V},E}(x)=c\sqrt{\left({E-\tilde V(x)\over c^2}\right)^2-1},
\end{equation}
and we denote by $\S^{n-1}_{x,E}$ the following sphere of $\R^n$ of center $0$ 
\begin{equation}
\label{5.19}\S^{n-1}_{x,E}=\{p \in \R^n| |p|=r_{\tilde{V},E}(x)\}.
\end{equation}
\vskip 4mm
\noindent 5.2 {\it Growth estimates for a function $g$.}
Consider the function $g: \R^n \to B_c$ defined by
\begin{equation}
\label{B}g(x)={x\over \sqrt{1+{|x|^2\over c^2}}}
\end{equation}
where $x\in \R^n.$ This function was considered for example in [J1].

We remind that $g$ has the following simple properties:
\begin{eqnarray}
\label{5.16}|\nabla g_i(x)|^2&\le& {1\over{1+{|x|^2\over c^2}}},\\
\label{5.17}|g(x)-g(y)|&\le& \sqrt{n} \sup\limits_{\ep\in [0,1]}{1\over\sqrt{1+{|\ep x+(1-\ep) y|^2\over c^2}}}|x-y|,\\
\label{5.18}|\nabla g_i(x)-\nabla g_i(y)|&\le& {3 \sqrt{n}\over c} \sup\limits_{\ep\in [0,1]}{1\over1+{|\ep x+(1-\ep) y|^2\over c^2}} |x-y|,
\end{eqnarray}
for  $x,\ y\in \R^n,\ i=1\ldots n,$ and $g=(g_1,\ldots,g_n).$ The function $g$ is an infinitely smooth diffeomorphism from $\R^n$ onto 
$B_c$, and its inverse is given by 
$g^{-1}(x)={x\over \sqrt{1-{|x|^2\over c^2}}},$
for $x\in B_c.$

\vskip 4mm

\noindent 5.3 {\it Proof of Lemma 2.1}.
For $q_0,q\in \bar D,$ $q_0\not=q$, let $t_{+,q_0,q}=\sup\{t>0\  |\ \psi_1(t,q_0,\bar k_{0,V,B}(E,q_0,q))$$\in D\}.$
From Properties (2.1) and (2.2), it follows that $k_{0,V,B}(E,.,.)$ is continuous on $(\bar D\times \bar D)\backslash \bar G$ and
for $q_0,q\in \bar D,$ $q_0\not=q$, and any $s_1,s_2\in[0,t_{+,q_0,q}[,$
$s_1\not=s_2,$ $\psi_1(s_1,q_0,\bar k_{0,V,B}(E,q_0,q))\not=\psi_1(s_2,q_0,\bar k_{0,V,B}(E,$ $q_0,q))$ and 
$N(q_+)\circ{\pa \over \pa t}\psi_1(t,q_0,$ $\bar k_{0,V,B}(E,q_0,q))_{|t=t_{+,q_0,q}}$ is positive, where
$q_+=\psi_1(t_{+,q_0,q},q_0,\bar k_{0,V,B}(E,q_0,q))$ (and where $\circ$ denotes the usual scalar product on $\R^n$).
Using also continuity of $\psi_1,$ one obtains that $s_{V,B}(E,q_0,q)$ is continuous on $(\bar D\times \bar D)\backslash \bar G.$
Then we obtain that $s_{V,B}(E,q_0,$ $q)\in C^1((\bar D\times \bar D)\backslash \bar G,\R)$ by applying implicit function theorem on 
maps  $m_i: \R\times ((\R^n\times \R^n)\backslash \Delta),$ $(t,x,y)\to y_i-
\psi_1^i(t,x,\tilde k_{0,V,B}(E,x,y)),$ $i=1\ldots n,$ where $\Delta=\{(x,x)\ |\ x\in \R^n\}$ and $\tilde k_{0,V,B}(E,.,.)$ is a
$C^1$ continuation of
$\bar k_{0,V,B}(E,.,.)$ on $(\R^n\times \R^n)\backslash \Delta$ (such a continuation exists thanks to (2.2), and note that for any $q_0,q\in
\bar D,$ $q\not=q_0,$ $\bar k_{V,B}(E,q_0,q)\not=0$).
Note that $k_{V,B}(E,q_0,q)=g(\psi_2(s_{V,B}(E,q_0,q),q_0,\bar k_{0,V,B}(E,$ $q_0,q))),$ $q_0,q\in \bar D,$ $q_0\not=q.$
It remains to prove that  $s_{V,B}(E,q_0,q)$ is
continuous on $G=\{(q',q')\ |\ q'\in \bar D\}$. 
Let $q_0=q\in D.$
Let $(q_{0,m})$ and $(q_m)$ be two sequences of points of $\bar D$ such that $q_{0,m}\not=q_0$ for all $m$ and $q_{0,m}$ goes to $q_0$
and $q_m$ goes to $q=q_0$ as $m\to +\infty$.
Let $R=\lim \sup_{m\to +\infty}s_{V,B}(E,q_{0,m},q_m)\in[0,+\infty]$. We shall prove that $R=0.$
Assume that $R>0.$
Note that by conservation of energy $|\bar k_{0,V,B}(E,q_{0,m},q_m)|\le c\sqrt{\left({E+\|V\|_{\infty}\over c^2}\right)^2-1}$. Using definition of $R$ and compactness 
of the closed ball of $\R^n$ whose radius is 
$c\sqrt{\left({E+\|V\|_{\infty}\over c^2}\right)^2-1}$ and whose centre is $0$, we obtain that there exist 
subsequences of $q_{0,m}$ and $q_m$ (respectively still denoted by $q_{0,m}$ and $q_m$) such that
\begin{eqnarray}
&&\label{12a}\lim_{m\to +\infty}s_{V,B}(E,q_{0,m},q_m)=R,\\
&&\label{12b}\bar k_{0,V,B}(E,q_{0,m},q_m) \textrm{ converges to some } k\in \R^n.
\end{eqnarray}
Using conservation of energy, we obtain that 
\begin{equation}
\label{12c}|k|=c\sqrt{\left({E-V(q_0)\over c^2}\right)^2-1}.
\end{equation}

Using \eqref{12b} and \eqref{12a} and continuity of $\psi_1,$ we obtain that
\begin{equation}
\label{13}\psi_1(t,q_0,k)=\lim_{m\to+\infty} \psi_1(t,q_{0,m},\bar k_{0,V,B}(E,q_{0,m},q_m)), \textrm{ for all }t\in[0,R[.
\end{equation}
For all $m$ and $t\in[0,s_{V,B}(E,q_{0,m},q_m)[,$ $\psi_1(t,q_{0,m},\bar k_{0,V,B}(E,q_{0,m},q_m))\in \bar D$. Hence using \eqref{13}, 
we obtain that
\begin{equation}
\label{14}\psi_1(t,q_0,k)\in \bar D,\ t\in[0,R[.
\end{equation}

In addition,
\begin{equation}
\label{18}\psi_1(0,q_0,k)=q_0\in D.
\end{equation}
Then $R\not=+\infty$ (otherwise this would contradict \eqref{2.1}, in particular the fact that the solution of \eqref{1.1}
under condition \eqref{1.3a} with energy
$E$, which starts at time $0$ at $q_0=\psi_1(0,q_0,k)$, reaches the boundary $\pa D$ at a time $t_+>0$ and satisfies the estimate
${\pa \psi_1\over \pa t}(t_+,q_0,k)\circ N(\psi_1(t_+,q_0,k))>0$) .

Using continuity of $\psi_1$ and $\lim_{m\to+\infty} q_{0,m}=q_0,$ $\lim_{m\to+\infty} q_m=q_0,$ $\lim_{m\to+\infty}s_{V,B}(E,q_{0,m},q_m)=R
$, $\lim_{m\to+\infty}\bar k_{0,V,B}(E,q_{0,m},q_m)=k$ and the definition of $s_{V,B}(E,q_{0,m},q_m)$, we obtain that 
\begin{eqnarray}
\psi_1(R,q_0,k)&=&\lim_{m\to+\infty}\psi_1(s_{V,B}(E,q_{0,m},q_m),q_0,\bar k_{0,V,B}(E,q_{0,m},q_m))\nonumber\\
\label{20}&=&\lim_{m\to+\infty}q_m=q_0.
\end{eqnarray}
Properties \eqref{20}, \eqref{18}, \eqref{14} and \eqref{2.1} imply $R=0$, which contradicts the assumption $R>0.$
Finally we proved that $s_{V,B}(E,.,.)\in C((\bar D\times \bar D)\backslash \pa G, \R).$

Let $x_0\in D.$ From \eqref{1.6*}, it follows that for sufficiently small positive $\ep$, 
$E$ is greater than $E(\|\tilde{V}\|_{C^2,D_{x_0,\ep}},\|\tilde{B}\|_{C^1,D_{x_0,\ep}},D_{x_0,\ep})$ where 
$D_{x_0,\ep}=\{x_0+(1+\ep)(x-x_0)\ |\ x\in D\}$. 
Hence  
one obtains that solutions of energy $E$ for equation \eqref{4} in $D_{x_0,\ep}$ also have properties \eqref{2.1} and \eqref{2.2}; and 
replacing $V$, $B$, and $D$ by $\tilde{V},$ $\tilde{B}$ and $D_{x_0,\ep}$ above in the proof, one obtains that $s_{\tilde{V},\tilde{B}}(E,.,.)$ is 
continuous on $(\bar D_{x_0,\ep}\times \bar D_{x_0,\ep})\backslash\{(q,q)\ |\ q\in \pa D_{x_0,\ep}\}$ 
($s_{\tilde{V},\tilde{B}}(E,q_0',q'),$ $(q_0',q')\in \bar D_{x_0,\ep}\times \bar D_{x_0,\ep}$, are defined as 
$s_{V,B}(E,q_0,q),$ $(q_0,q)\in \bar D\times \bar D$, are defined
in Subsection 2.1).
Now, using also $\bar D\subseteq D_{x_0,\ep}$ and the equality $ s_{V,B}(E,q_0,q)=s_{\tilde{V},\tilde{B}}(E,q_0,q)$ for $q_0,$ $q\in \bar D$, one obtains
$s_{V,B}(E,.,.)\in C(\bar D\times \bar D, \R)$ (the equality $s_{V,B}(E,q_0,q)=s_{\tilde{V},\tilde{B}}(E,q_0,q)$ for $q_0,$ $q\in \bar D$, follows from
the fact that if $(x(t),p(t))$ is solution of \eqref{4} in $D$, then $(x(t),p(t))$ is also solution of \eqref{4} in $D_{x_0,\ep}$).

Lemma 2.1 is proved.
\hfill$\Box$

\vskip 4mm

\noindent 5.4 {\it Proof of Proposition 3.1}.
From Lemma 2.1, $\psi \in C^1(\R\times \R^n\times \R^n,\R^n\times\R^n),$ $ \A\in C^1(\bar D,\R^n),$ it follows that
$ {\cal S}_{0_{V,\A,E}}\in C(\bar D\times \bar D,\R)$ and ${\cal S}_{0_{V,\A,E}}\in 
C^1((\bar D\times \bar D)\backslash \bar G,\R).$
Equalities \eqref{3.7} and \eqref{3.8} are known equalities (see Section 46 and further Sections of [A]).
Statements \eqref{3.6}, \eqref{3.9}, \eqref{3.10} follow from  \eqref{3.7} and \eqref{3.8}.
We shall prove \eqref{3.11}. We omit indices $_{V,B}$ for $s_{V,B},$ $\bar k_{0,V,B}$ and $\bar k_{V,B}$ where
$\bar k_0,$ $\bar k$ are defined by \eqref{3.01}.
Using the equality $y-x=\int_0^{s(E,x,y)}{\pa \psi_1\over \pa t}(t,x,$ $\bar k_0(E,x,y))dt$ and estimate 
$|{\pa \psi_1\over \pa t}(t,x,\bar k_0(E,x,y))|\le c$, we
obtain
\begin{equation}
\label{5.00}|y-x|\le cs(E,x,y), \textrm{ for all }x,y\in \bar D,\ y\not=x.
\end{equation}
Derivating equality $\psi_1(s(E,x,y),x,\bar k_0(E,x,y))=y$ with respect to $y_i,$ we obtain that 
\begin{eqnarray}
\label{5.000}e_i&=&({\pa \psi_1^j\over \pa \bar k}(s(E,x,y),x,\bar k_0(E,x,y))\circ{\pa \bar 
k_0\over \pa y_i}(E,x,y))_{j=1..n}
\end{eqnarray}
\begin{eqnarray}
&&+{\pa s\over \pa y_i}(E,x,y){\pa \psi_1\over \pa s}(s(E,x,y),x,\bar k_0(E,x,y)),\nonumber
\end{eqnarray}
for any $x,y\in \bar D,$ $x\not=y$ and where $(e_1,\ldots, e_n)$ is the canonical basis of $\R^n$
(and where $\circ$ denotes the usual scalar product on $\R^n$).
For $t\in \R,$  $x\in \bar D,$ and $k\in \R^n$ and $j=1..n$ the following equality is valid:
$\psi_1^j(t,x,k)=x_j+tg_j(k)+$$\int_0^t\left[g_j(k+\int_0^s F(\psi_1(\sigma,x,k),g(\psi_2(\sigma,x,k)))
d\sigma)-g_j(k)\right]ds.$
Hence we obtain that for $t\in \R,$  $x\in \bar D,$ $k=(k_1,\ldots,k_n)\in \R^n,$ and  $j=1..n,$ 
\begin{eqnarray}
&&{\pa \psi_1^j\over \pa k_l}(t,x,k)=t{\pa g_j\over \pa k_l}(k)+\int_0^t\left[{\pa g_j\over\pa k_l}
(k+\int_0^sF(\psi_1(\sigma,x,k),g(\psi_2(\sigma,x,k)))\right.\nonumber\\
&&\left.d\sigma)-{\pa g_j\over\pa
k_l}(k)\right]ds+\int_0^t\nabla g_j(k+\int_0^s F(\psi_1(\sigma,x,k),g(\psi_2(\sigma,x,k)))d\sigma)\circ\nonumber\\
\label{5.1}&&\int_0^s{\pa\over\pa k_l}F(\psi_1,g(\psi_2))_{|(\sigma,x,k)}d\sigma
ds,
\end{eqnarray}
for any $l=1..n.$
Define
\begin{eqnarray*}
&&R=\sup_{(x',y')\in \bar D}s(E,x',y'),\\
&&M_3=\sup_{t\in[0,R],x'\in \bar D, l=1\ldots n\atop |k|\le c\sqrt{\left({\sup_{x'\in\bar D}E-V(x')\over c^2}\right)^2-1}}
|{\pa\over\pa k_l}F(\psi_1,g(\psi_2))_{|(t,x',k)}|,\\
&&M_4=\max(M_3,\sqrt{n}\|V\|_{C^2,D}+n\|B\|_{C^1,D}).
\end{eqnarray*}
Then using \eqref{5.1} and growth properties of $g$, we obtain that
\begin{eqnarray}
&&|{\pa \psi_1^j\over \pa k_l}(s(E,x,y),x,\bar k_0(E,x,y))-s(E,x,y){\pa g_j\over \pa k_l}(\bar k_0(E,x,y))|\nonumber\\
\label{5.2}&&\le M_4s(E,x,y)^2(1+{3\sqrt{n}\over c}),
\end{eqnarray}
for $x,y\in \bar D,$ $x\not=y$ and $j,l=1..n$. where $\bar k_0$ is defined by \eqref{3.01}.

Let $x,y\in \bar D,$ $x\not=y$.
Using the identity 
\begin{eqnarray*}
&&\bar k(E,x,y)=\bar k_0(E,x,y)+\int_0^{s(E,x,y)}\left(-\nabla V(\psi_1(s,x,\bar k_0(E,x,y)))\right.\\
&&
+\left.{1\over c}B(\psi_1(s,x,\bar k_0(E,x,y)))g(\psi_2(s,x,\bar k_0(E,x,y)))\right)ds,
\end{eqnarray*}
we obtain the following estimate
\begin{equation}
\label{5.3}|\bar k(E,x,y)-\bar k_0(E,x,y)|\le M_5s(E,x,y),
\end{equation}
where $M_5=\sqrt{n}\|V\|_{C^2,D}+n\|B\|_{C^1,D}.$
Using \eqref{3.03}, we obtain that 
\begin{equation}
\label{5.4}\bar k_0(E,x,y)\circ{\pa \bar k_0\over \pa y_i}(E,x,y)={1\over 2}{\pa |\bar k_0|^2 \over \pa y_i}(E,x,y)=0,
\end{equation}
for $i=1..n$.
From \eqref{3.03}, \eqref{5.3} and \eqref{5.4}, it follows that
\begin{eqnarray}
|{\bar k(E,x,y)\over |\bar k(E,x,y)|}\circ{\pa \bar k_0\over \pa y_i}(E,x,y)|
\le{1\over |\bar k(E,x,y)|}|(\bar k(E,x,y)-\bar k_0(E,\nonumber&&\\
x,y))\circ{\pa \bar k_0\over \pa y_i}(E,x,y)|
+{1\over |\bar k(E,x,y)|}|\bar k_0(E,x,y)\circ{\pa \bar k_0\over \pa y_i}(E,x,y)|&&\nonumber\\
\label{5.5}\le c^{-1}\left(\left({\inf_{x'\in \bar D} E-V(x')\over c^2}\right)^2-1\right)^{-{1\over 2}}M_5s(E,x,y)|{\pa \bar k_0\over
\pa y_i}(E,x,y)|,&&
\end{eqnarray}
for $i=1..n.$

Let $(v^1,\ldots,v^{n-1})$ be an orthonormal family of $\R^n$ such that $({\bar k(E,x,y)\over |\bar k(E,x,y)|},
v^1,$ $\ldots,v^{n-1})$ is an orthonormal basis of $\R^n.$
Note that using definition of $g$ and \eqref{5.4}, we obtain 
\begin{eqnarray}
\label{5.6}&&(\nabla g_j(\bar k_0(E,x,y))\circ{\pa \bar k_0\over \pa y_i}(E,x,y))_{j=1..n}
=\\
&&\left(1+{|\bar k_0(E,x,y)|^2\over c^2}\right)^{-1/2}{\pa \bar k_0\over \pa y_i}(E,x,y), \ i=1..n.\nonumber
\end{eqnarray}
Hence using \eqref{5.6}, \eqref{5.2} and \eqref{5.000} (and $k(E,x,y)\circ v^h=0),$ we obtain
\begin{eqnarray}
&&s(E,x,y)|{\pa \bar k_0\over \pa y_i}(E,x,y)\circ v^h|=
\sqrt{1+{|\bar k_0(E,x,y)|^2\over c^2}}
s(E,x,y)\nonumber\\
&&\times|( \nabla g_j(\bar k_0(E,x,y))\circ{\pa \bar k_0\over \pa y_i}(E,x,y))_{j=1..n}\circ v^h|\nonumber\\
&\le&\sqrt{1+{|\bar k_0(E,x,y)|^2\over c^2}}\left[nM_4s(E,x,y)^2(1+{3\over c})|{\pa \bar k_0\over \pa y_i}(E,x,y)|\right.\nonumber
\end{eqnarray}
\begin{eqnarray}
&&+\left.|({\pa \psi_1^j\over \pa k}(s(E,x,y),x,\bar k_0(E,x,y))\circ{\pa \bar k_0\over \pa y_i}(E,x,y))_{j=1..n}\circ v^h|
\right]\nonumber\\
&=&\sqrt{1+{|\bar k_0(E,x,y)|^2\over c^2}}\left[
nM_4s(E,x,y)^2(1+{3\over c})|{\pa \bar k_0\over \pa y_i}(E,x,y)|+|v^h_i|\right]\nonumber\\
\label{5.7}&\le&{\sup_{x'\in \bar D}E-V(x')\over c^2}(nM_4(1+{3\over c})s(E,x,y)^2|{\pa \bar k_0\over \pa y_i}(E,x,y)|+1),
\end{eqnarray}
for $i=1\ldots n$.
Let
$M_6=c^{-1}\left(\left(c^{-2}\inf_{x'\in \bar D} (E-V(x'))\right)^2-1\right)^{-{1\over 2}}M_5+
c^{-2}\times$

\noindent $\sup_{x'\in \bar D}(E-V(x'))(nM_4(1+{3\over c})+1).$
From \eqref{5.5} and \eqref{5.7}, it follows that
\begin{equation}
\label{5.8}s(E,x,y)|{\pa \bar k_0\over \pa y_i}(E,x,y)|\le \sqrt{n}M_6(1+s(E,x,y)(s(E,x,y)|{\pa \bar k_0\over \pa y_i}(E,x,y)|)),
\end{equation}
for $i=1\ldots n.$

Using uniform continuity of $s(E,.,.)$ on $\bar D\times \bar D,$ we obtain that there exists some $\eta >0$ such that if $x,y\in \bar
D,$ $|x-y|<\eta$, then $\sqrt{n}M_6s(E,x,y)\le {1\over 2}.$
Then, using \eqref{5.8}, we obtain that $s(E,x,y)|{\pa \bar k_0\over \pa y_i}(E,x,y)|\le 2\sqrt{n}M_6,$
for  $x,y\in \bar
D,$ $|x-y|<\eta$ and $i=1..n$.
Now using the continuous differentiability of $\bar k_0(E,.,.)$ on $(\bar D\times \bar D)\backslash \bar G,$ we obtain that 
$|{\pa \bar k_0\over \pa y_i}(E,x,y)|\le {M'_i
\over s(E,x,y)}$
for $x,y\in \bar
D,$ $x\not=y$ and where $M'_i=\max(2M_6\sqrt{n},R\sup_{x',y'\in \bar D, |x'-y'|\ge\eta}|{\pa \bar k_0\over \pa y_i}(E,x,y)|).$
Putting $M_2=\sup_{i=1..n}cM'_i $
and using \eqref{5.00} and \eqref{3.9}, we obtain \eqref{3.11}.
\hfill $\Box$


\section{Proof of Properties (2.1) and (2.2)}
In this Section we first consider solutions $x(t)$ of \eqref{4} in an open bounded subset $\Omega$ of $\R^n$ (see Subsection 6.1) and we give
properties of these solutions at fixed and sufficiently large energy (see Subsections 6.3 and 6.4) ($\Omega$ should be thought as an
open neighborhood of $D$). Using these properties we prove Properties 2.1 and 2.2 (see Subsection 6.5).
Subsections 6.6, 6.7, 6.8, 6.9 are devoted to the proof of Propositions 6.1, 6.2, 6.3, 6.4 formulated in Subsection 6.4.

We keep notations of Subsections 5.1, 5.2.

\vskip 4mm
\noindent 6.1 {\it Additional notations.}
Let $\Omega$ be a bounded open subset of $\R^n$ with frontier $\pa \Omega$.
We define a positive number $\delta(\Omega)$ by 
\begin{equation}
\label{6.4b}\delta(\Omega)=\sup_{x\in \Omega}|x|.
\end{equation}
We consider the following equation in $\Omega:$ 
\begin{equation}
\label{6.01}
\dot p = F(x,\dot x),\ p={\dot x \over \sqrt{1-{|\dot x|^2 \over c^2}}},\ x\in \Omega,\ p\in
\R^n.
\end{equation}
where the force $F(x,\dot x)= -\nabla \tilde{V}(x)+{1\over c}\tilde{B}(x)\dot x$ is defined by \eqref{5.0}.
For the equation \eqref{6.01}, the energy 
$E=c^2\sqrt{1+|p(t)|^2\over c^2}+\tilde{V}(x(t))$
is an integral of motion.

Note that if $x(t),$ $t\in ]t_-,t_+[,$ is a solution of \eqref{6.01} which starts at $x_0\in \Omega$ 
at time $0$ with velocity $v$ then $x(t)=\psi_1(t,x_0,g^{-1}(v)),$ $t\in]t_-,t_+[$, where the function $g$ is defined by \eqref{B}
and where $\psi=(\psi_1,\psi_2)$ is the flow of the differential system
\eqref{4}. We obtain, in particular, $x(t)\to  
\psi_1(t_\pm,x_0,g^{-1}(v)),$ as $t\to t_\pm$, and 
$\dot x(t)\to g(\psi_2(t_\pm,x_0,g^{-1}(v))),$ as $t\to t_\pm$.

We denote by $\Lambda$ the open subset of $\R\times \Omega\times \R^n$ where the flow of the differential system \eqref{6.01} is
defined, i.e. 
$$
\Lambda=\{(t,x,p)\in \R\times \Omega\times \R^n\ |\ \forall s\in[0,t]\ \psi_1(s,x,p)\in \Omega\},
$$
where $\psi=(\psi_1,\psi_2)$ is the flow of the differential system
\eqref{4}.

For $E>c^2+\sup_{x\in \Omega}\tilde V(x),$ we denote by  ${\cal V}_E$
the following smooth $2n-1$-dimensional submanifold of $\R^{2n}$
\begin{equation}
\label{56.1}{\cal V}_E=\{(x,p)\in \Omega\times \R^n\ |\ |p|=r_{\tilde V,E}(x)\},
\end{equation}
where $r_{\tilde V,E}(x)$ is defined by \eqref{5.02}.

For $E>c^2+\sup_{x\in \Omega}\tilde V(x),$ we also consider the map $\varphi_E\in C^1(\Lambda\cap\left(]0,+\infty[\times {\cal V}_E\right)
,
\Omega\times \Omega),$ defined by
\begin{equation}
\label{C}\varphi_E(t,x,p)= (x,\psi_1(t,x,p)),\ \textrm{ for }(t,x,p)\in \Lambda\cap\left(]0,+\infty[\times {\cal V}_E\right).
\end{equation}

\vskip 4mm
\noindent 6.2 {\it Estimates for the force $F$.}
We define the nonnegative real number $\beta(\tilde{V},\tilde{B},$ $\Omega)$ by
\begin{equation}
\label{18a}\beta(\tilde{V},\tilde{B},\Omega)=\max\left(\sup_{{x\in \Omega\atop\alpha\in (\N\cup\{0\})^n}\atop|\alpha|\le 2}|\pa_x^\alpha\tilde{V}(x)|,
\sup_{{x\in \Omega\atop\alpha'\in (\N\cup\{0\})^n}\atop|\alpha'|\le 1}|\pa_x^{\alpha'}\tilde B_{i,j}(x)|\right).
\end{equation}
The following estimates are valid:
\begin{equation}
\label{18c}|F(x,v)|\le n\beta(\tilde{V},\tilde{B},\Omega)({1\over c}|v|+1),
\end{equation}
\vskip -2mm
\begin{equation}
\label{18d}|F(x,v)-F(x',v')|\le n\beta(\tilde{V},\tilde{B},\Omega)\left[|x-x'|\left(1+{|v'|\over c}\right)+{1\over c }|v'-v|\right],
\end{equation}
for $x,x'\in \Omega$ and $v,v'\in \R^n.$
\vskip 4mm

\noindent 6.3 {\it Some constants.} 
For $x\in \Omega$ and $E>c^2+\sup_{x'\in \Omega}V(x')$, we define the following real constants 
\begin{eqnarray}
\label{6.6ia}&&C_1=2c^2+\sup_{x'\in \Omega}\left(\tilde V(x')+8|x'| \left(|\nabla \tilde V(x')|+ \sum_{i,j=1\ldots n}|\tilde{B}_{i,j}(x')|\right)
\right),\\
\label{6.6b}&&C_2=c^2\sqrt{1+{800 n^2\beta(\tilde{V},\tilde{B},\Omega)^2\delta(\Omega)^2\over c^4}}+\sup_{x'\in
\Omega}\tilde V(x'),\\
&&\label{6.47aa}C_3=C_4\left(1+
5\delta(\Omega)C_5 
e^{5 \delta(\Omega)C_5}\right)\\
&&\label{6.47a}C_4=
\left(1+{10\sqrt{2}n^{3/2}\delta(\Omega)\beta(\tilde{V},\tilde{B},\Omega)\over  E-\tilde V(x)}
e^{{10\sqrt{2}\delta(\Omega)n^{3/2}\beta(\tilde{V},\tilde{B},\Omega)\over E-\tilde
V(x)}}\right)
\\
&&\times{10n^{3/2}\delta(\Omega)^2\beta(\tilde{V},\tilde{B},\Omega)\over  E-\tilde V(x)}\left(5+
{600n^{3/2}\delta(\Omega)\beta(\tilde{V},\tilde{B},\Omega)\over  E-\tilde V(x)}
+24n^{1/2}\right)\nonumber\\
&&+{20\sqrt{2}n^2\delta(\Omega)\beta(\tilde{V},\tilde{B},\Omega)\over E-\tilde V(x)}
e^{{10\sqrt{2}\delta(\Omega)n^{3/2}\beta(\tilde{V},\tilde{B},\Omega)\over E-\tilde V(x)}}
+{40n^{3/2}\delta(\Omega)\beta(\tilde{V},\tilde{B},\Omega)\over c^2\sqrt{\left({E-\tilde V(x)\over c^2}\right)^2-1}},\nonumber
\end{eqnarray}
\begin{eqnarray}
&&\label{6.47b}C_5=\left(1+{10\sqrt{2}n^{3/2}\delta(\Omega)\beta(\tilde{V},\tilde{B},\Omega)\over E-\tilde V(x)}
e^{{10\sqrt{2}\delta(\Omega)n^{3/2}\beta(\tilde{V},\tilde{B},\Omega)\over E-\tilde
V(x)}}\right)\\
&&\times{20n^{3/2}\beta(\tilde{V},\tilde{B},\Omega)\delta(\Omega)\over E-\tilde V(x)}
\left(1+{120n^{3/2}\beta(\tilde{V},\tilde{B},\Omega)\delta(\Omega)\over E-\tilde V(x)}\right),\nonumber\\
\label{56.41a}&&C_6=\inf_{x'\in \Omega}\left(\sqrt{1-\left({c^2\over E-\tilde V(x')}\right)^2}
-{20n^2\beta(\tilde{V},\tilde{B},\Omega)\delta(\Omega)\over E-\tilde
V(x')}\right),\\
&&\label{56.41b}C_7=\inf_{x'\in \Omega}\left(c\sqrt{1-\left({c^2\over E-\tilde V(x')}\right)^2}-{5c(n+1)^{1/2}
n^2\delta(\Omega)\over E-\tilde
V(x')}\right.\\
&&\times\beta(\tilde{V},\tilde{B},\Omega)
e^{{10n^{3/2}\beta(\tilde{V},\tilde{B},\Omega)\delta(\Omega)\over E-\tilde V(x')}
\left(1+2cn^{1/2}e^{10n^{3/2}\beta(\tilde{V},\tilde{B},\Omega)\delta(\Omega)\over E-\tilde V(x')}\right)}\nonumber\\
&&\times\left.{cr_{\tilde V,E}(x')\over E-\tilde V(x')}\left[12\sqrt{n}+1
+10\sqrt{n}\delta(\Omega)\right]\right).\nonumber
\end{eqnarray}

Now assume that $\Omega$ is a bounded strictly convex (in the strong sense) open domain of $\R^n$ with $C^2$ boundary. 
Let $\chi_\Omega$ be a $C^2$ defining function for $\Omega$, i.e. $\Omega=\chi_\Omega^{-1}(]-\infty,0[)$ and $\pa \Omega=\chi_\Omega^{-1}(\{0\})$ and for all 
$x\in \pa \Omega$ $\nabla\chi_\Omega(x)\not=0$
and the Hessian matrix $Hess \chi_\Omega(x)$ of $\chi_\Omega$ at $x$ is a symmetric positive definite matrix.
For $E> c^2+\sup_{x\in \Omega}\tilde{V}(x),$ we define the real constant $C_8(E,\tilde V,\tilde B,$ $\Omega)$ by
\begin{equation}
\label{56.41c}C_8=C_{10}(\Omega)\left(1-\left({c^2\over E-\sup_{y\in \Omega}\tilde V(y)}\right)^2\right)-
{4nC_9(\Omega)\beta(\tilde{V},\tilde{B},\Omega)\over E-\sup_{y\in \Omega}\tilde V(y)},
\end{equation} 
where $C_9(\Omega)$ and $C_{10}(\Omega)$ are the two positive real numbers defined by
\begin{eqnarray}
\label{56.31a}C_9(\Omega)&=&\sup_{x\in\pa \Omega}|\nabla  \chi_\Omega(x)|,\\
\label{56.31b}C_{10}(\Omega)&=&\inf_{x\in \pa \Omega\atop v\in  \S^{n-1}}|Hess\chi_\Omega(x)(v,v)|.
\end{eqnarray}

Note that  from \eqref{6.47aa}-\eqref{56.41c}, it follows that 
\begin{equation}
\label{56.46}
\begin{array}{l}
\sup_{x\in \Omega}C_3(E,x,\tilde{V},\tilde{B},\Omega)\to 0, \textrm{ as }E\to +\infty,\\
C_6(E,\tilde V,\tilde B,\Omega)\to 1>0,\textrm{ as }E\to +\infty,\\
C_7(E,\tilde V,\tilde B,\Omega)\to c>0,\textrm{ as }E\to +\infty,\\
C_8(E,\tilde V,\tilde B,\Omega)\to C_{10}(\Omega)>0,\textrm{ as }E\to +\infty.\end{array}
\end{equation}

When $\Omega$ is strictly convex in the strong sense with $C^2$ boundary, then one can relate 
an upper bound for the real constant
$E(\|\tilde{V}\|_{C^2,\Omega},\|\tilde{B}\|_{C^1,\Omega},\Omega)$ (mentioned in Subsection 2.1)
with constants $C_1,$ $C_2,$ $\sup_{x\in \Omega}C_3,$ $C_6,$ $C_7$ and $C_8$ 
(see Subsections 6.4 and 6.5).  

\vskip 2mm 
{\bf Remark 6.1.} We remind that 
$\tilde{V}$ is a $C^1$ continuation of $V$ on $\R^n$ and that 
$\tilde{B}\in C^1(\R^n,A_n(\R))$ is such that $\tilde{B}\equiv B$ on $\bar D.$ 
Note that from \eqref{6.6ia}-\eqref{56.41c} it follows that
$C_1(\tilde V,\tilde B,D)$, $C_2(\tilde V,\tilde B,D)$, $\sup_{x\in D}C_3(E,x,\tilde{V},\tilde{B},D)$, $C_6(E,\tilde V,\tilde B,D),$ $C_7(E,\tilde V,\tilde B,D)$ and 
$C_8(E,\tilde V,\tilde B,D)$ depend only on $(V,B)$ and $D$.

\vskip 4mm

\noindent 6.4 {\it Properties  of the first component of the flow of \eqref{6.01} at fixed and sufficiently large energy $E$.}
The following Proposition 6.1 gives an upper bound for living time for solutions of \eqref{6.01} with energy $E$ when $E$ is 
sufficiently large.

\vskip 2mm
{\bf Proposition 6.1.}
{\it Let 
\begin{equation}
\label{6.6}E\ge C_1(\tilde V,\tilde B,\Omega),
\end{equation}
where $C_1$ is defined by \eqref{6.6ia}.
Let $x: ]t_-,t_+[\to \Omega$ be a solution of \eqref{6.01} with energy $E$, where $t_\pm\in \R\cup\{\pm \infty\}$. 
Then the following statement holds}:
$t_-,t_+$ {\it are finite and they satisfy the following estimate} 
\begin{equation}
\label{6.6a}|t_+-t_-|\le {5\delta(\Omega)\over c},
\end{equation}
{\it where $\delta(\Omega)$ is defined by \eqref{6.4b}.}

\vskip 2mm

A proof of Proposition 6.1 is given in Subsection 6.6.

\vskip 2mm

For $E\ge C_1(\tilde V,\tilde B,\Omega)$ ($C_1$ is defined by \eqref{6.6ia})
and for $(x,p)\in{\cal V}_E$, we define the real numbers  $t_{+,x,p}$ and $t_{-,x,p}$ by
\begin{eqnarray}
t_{+,x,p}&=&\sup\{t>0\ |\ (t,x,p)\in \Lambda\},\label{6.t+}\\
t_{-,x,p}&=&\inf\{t<0\ |\ (t,x,p)\in \Lambda\}.\label{6.t-}
\end{eqnarray}

The following Proposition 6.2 gives, in particular, a one-to-one property of the map $\varphi_E$ defined by \eqref{C}.
 
\vskip 2mm
{\bf Proposition 6.2.}
{\it Let 
\begin{equation}
\label{6.21}
E\ge \max(C_1(\tilde V,\tilde B,\Omega),C_2(\tilde V,\tilde B,\Omega)),
\end{equation}
where constants $C_1$ and $C_2$ are defined by \eqref{6.6ia} and \eqref{6.6b}.
Let $x\in \Omega$ and let $p_1,p_2\in \S^{n-1}_{x,E}$ (defined by \eqref{5.19}).
Then the following estimate is valid:
\begin{equation}
\label{6.22}
\left||\psi_1(t_1,x,p_1)-\psi_1(t_2,x,p_2)|-|t_1v_1-t_2v_2|\right|\le C_3|t_1v_1-t_2v_2|,
\end{equation}
for $(t_1,t_2)\in [0,t_{+,x,p_1}[\times[0,t_{+,x,p_2}[$, where $v_i={p_i\over \sqrt{1-{p_i^2\over c^2}}},$ $i=1,2,$
and where $C_3=C_3(E,x,\tilde{V},\tilde{B},\Omega)$ is defined by  \eqref{6.47aa}.
}

\vskip 2mm
A proof of Proposition 6.2 is given in Subsection 6.7.
We remind that 
\begin{equation}
\label{D}\sup_{x\in \Omega}C_3(E,x,\tilde{V},\tilde{B},\Omega)\to 0, \textrm{ as }E\to +\infty\textrm{ (see }\eqref{56.46}\textrm{)}.
\end{equation}
Taking account of \eqref{D} and the equality $\psi_1(0,x,p)=x$ for any $(x,p)\in{\cal V}_E$ and taking account of 
Proposition 6.2, we obtain that at fixed and sufficiently large energy $E$ the map $\varphi_E$ defined by \eqref{C} is one-to-one and
its range is included in 
$(\Omega\times \Omega)\backslash\{(x,x)\ |\ x\in \Omega\}$.

\vskip 2mm
The following Proposition 6.3 is proved in  Subsection 6.8.
\vskip 2mm

{\bf Proposition 6.3.}
{\it Assume that
\begin{equation}
\label{56.2}
\begin{array}{l}
E\ge C_1(\tilde V,\tilde B,\Omega),\\
E\ge c^2\sqrt{1+{400 n^2\beta(\tilde{V},\tilde{B},\Omega)^2\delta(\Omega)^2\over c^4}}+\sup_{x\in \Omega}\tilde{V}(x),\\
\min(C_6(E,\tilde{V},\tilde{B},\Omega),C_7(E,\tilde{V},\tilde{B},\Omega))>0,
\end{array}
\end{equation}
where $C_1,$ $C_6$ and $C_7$ are defined by \eqref{6.6ia}, \eqref{56.41a} and \eqref{56.41b}. 
Then the map $\varphi_E$ defined by \eqref{C} is a local 
$C^1$ diffeomorphism at any point $(t,x,p)\in \Lambda\cap\left(]0,+\infty[\times {\cal V}_E\right).$ 
}
\vskip 2mm

Now assume that $\Omega$ is a bounded strictly convex (in the strong sense) open domain of $\R^n$ with $C^2$ boundary. 
Let $\chi_\Omega$ be a $C^2$ defining function for $\Omega$. For $E> c^2+\sup_{x\in \Omega}\tilde{V}(x),$ 
real constant $C_8(E,\tilde V,\tilde B,\Omega)$ is defined by \eqref{56.41c} with respect to 
$\chi_\Omega.$
 
The following Proposition 6.4 gives a surjectivity property of the map $\varphi_E$ defined by \eqref{C} at fixed and sufficiently large
energy $E.$
\vskip 2mm

{\bf Proposition 6.4.} {\it
Let 
\begin{equation}
\label{56.41}
\begin{array}{l}
E\ge C_1(\tilde V,\tilde B,\Omega),\\
E\ge c^2\sqrt{1+{400 n^2\beta(\tilde{V},\tilde{B},\Omega)^2\delta(\Omega)^2\over c^4}}+\sup_{x\in \Omega}\tilde V(x),\\
\min(C_6(E,\tilde V,\tilde B,\Omega),C_7(E,\tilde V,\tilde B,\Omega),C_8(E,\tilde V,\tilde B,\Omega))>0.
\end{array}
\end{equation}
where $C_1,$ $C_6,$ $C_7$ and $C_8$ are defined by \eqref{6.6ia}, \eqref{56.41a}, \eqref{56.41b} and \eqref{56.41c}.

Then $(\Omega\times \Omega)\backslash \{(x,x) \ | \ x\in \Omega\}$ is included in the range of the map
$\varphi_E$ defined by \eqref{C}.
}

\vskip 2mm
A proof of Proposition 6.4 is given in Subsection 6.9.

Taking account of Propositions 6.2, 6.3, 6.4, we obtain, in particular, that
at fixed and sufficiently large
energy $E$
the map $\varphi_E$ defined by \eqref{C} is a $C^1$ diffeomorphism from $\Lambda\cap\left(]0,+\infty[\times {\cal V}_E\right)$ onto
$(\Omega\times \Omega)\backslash \{(x,x) \ | \ x\in \Omega\}$.
 
Now we are ready to prove Properties 2.1 and 2.2.

\vskip 4mm

\noindent 6.5 {\it Final part of the proof of Properties (2.1) and (2.2).}
Let $\chi_D$ be a $C^2$ defining function for $D$, i.e. $\chi_D\in C^2(\R^n,\R)$ and $D=\chi_D^{-1}(]-\infty,0[)$, and $\pa
D=\chi_D^{-1}(\{0\})$, and for all $x\in \pa D$ $\nabla \chi_D(x)\not=0$ and the Hessian matrix $Hess\chi_D(x)$ of $\chi_D$ at $x$ is a symmetric definite positive matrix.

Let $x_0\in D.$
For $\ep >0,$ we define the open neighborhood $\Omega_\ep$ of $\bar D$ by
$\Omega_\ep =\{x_0+(1+\ep)(x'-x_0)\ |\ x'\in D\}.$
Then $\Omega_\ep$ is also a bounded strictly convex in the strong sense open domain of $\R^n$ with $C^2$ boundary and the map
$\chi_{\Omega_\ep}\in C^2(\R^n,\R)$ defined by
$\chi_{\Omega_\ep}(x)=\chi_D(x_0+{x-x_0\over 1+\ep}),\ x\in \R^n,$   
is a $C^2$ defining function for $\Omega_\ep.$ 
In addition, note that

\begin{equation}
\label{56.49}
\begin{array}{l}
x\in \pa \Omega_\ep\Leftrightarrow x_0+{x-x_0\over 1+\ep}\in \pa D,\\
\nabla \chi_{\Omega_\ep}(x)=(1+\ep)^{-1}\nabla \chi_D(x_0+{x-x_0\over 1+\ep}),\ x \in \R^n,\\
Hess \chi_{\Omega_\ep}(x)=(1+\ep)^{-2}Hess \chi_D(x_0+{x-x_0\over 1+\ep}),\ x \in \R^n,\\
\sup_{x\in \Omega_\ep}\left(\inf\{|x-y|\ |\ y\in \bar D\}\right)=\ep \sup\{|x-x_0|\ |\ x\in D\}\underset{\ep\to 0^+}{\longrightarrow} 0.
\end{array}
\end{equation}
Note also that $\Omega_{\ep_2}\subseteq \Omega_{\ep_1}$ if $0<\ep_2<\ep_1.$

Let $E>c^2+\sup_{x\in D}V(x).$
Assume that
\begin{equation}
\label{56.51}
\begin{array}{l}
E> \max(C_1(\tilde V,\tilde B,D),C_2(\tilde V,\tilde B,D) ),\\
\sup_{x\in \Omega}C_3(E,x,\tilde{V},\tilde{B},D)<1,\\
\min(C_6(E,\tilde V,\tilde B,D),C_7(E,\tilde V,\tilde B,D),C_8(E,\tilde V,\tilde B,D))>0,
\end{array}
\end{equation}
where $C_1,$ $C_2,$ $C_3,$ $C_6$, $C_7$ and $C_8$ are defined by \eqref{6.6ia}, \eqref{6.6b}, \eqref{6.47aa}, \eqref{56.41a}, \eqref{56.41b}
 and \eqref{56.41c}
(taking account of \eqref{56.46}, we obtain that if $E$ is sufficiently large, then \eqref{56.51} is satisfied).

Let $\ep>0.$ We denote by  $\Lambda_\ep$ the open subset of $\R\times \Omega_\ep\times \R^n$ 
defined by
$\Lambda_\ep=\{(t,x,p)\in \R\times \Omega\times \R^n\ |\ \forall s\in[0,t]\ \psi_1(s,x,p)\in \Omega_\ep\},$ 
and we denote by ${\cal V}_{E,\ep}$ the following smooth $2n-1$-dimensional submanifold of $\R^{2n}$
${\cal V}_{E,\ep}=\{(x,p)\in \Omega_\ep\times \R^n\ |\ |p|=r_{\tilde V,E}(x)\}.$
From \eqref{56.49} and continuity of $\pa^\alpha_x\tilde V$ and $\pa^{\alpha'}_x\tilde B$ for $\alpha,\alpha'\in (\N\cup\{0\})^n,$
$|\alpha|\le 2,$ $|\alpha'|\le 1,$ 
and from \eqref{6.6ia}-\eqref{56.41c}, it follows that
\begin{eqnarray*}
&&C_i(\tilde V,\tilde B,\Omega_\ep)\to C_i(\tilde V,\tilde B,D),\textrm{ as }\ep\to 0^+,\textrm{ for }i=1,2,\\
&&\sup_{x\in \Omega_\ep}C_3(E,x,\tilde{V},\tilde{B},\Omega_\ep)\to \sup_{x\in D}C_3(E,x,\tilde{V},\tilde{B},D),\textrm{ as }\ep\to 0^+,
\\
&&C_i(E,\tilde V,\tilde B,\Omega_\ep)\to C_i(E,\tilde V,\tilde B,D),\textrm{ as }\ep\to 0^+,\textrm{ for }i=6,7,8.
\end{eqnarray*}
Taking also account of \eqref{56.51} and Propositions 6.2, 6.3, 6.4, we obtain that there exists $\ep_0>0$ such that 
\begin{equation}
\label{56.53}
\begin{array}{l}
\varphi^\ep_E:\Lambda_\ep\cap\left(]0,+\infty[\times {\cal V}_{E,\ep}\right)\to \Omega_\ep\times \Omega_\ep, (t,x,p)\mapsto (x,\psi_1(t,x,p)),
\textrm{ is a }\\
C^1 \textrm{ diffeomorphism from }\Lambda_\ep\cap\left(]0,+\infty[\times {\cal V}_{E,\ep}\right) \textrm{ onto }
(\Omega_\ep\times \Omega_\ep)
\backslash\{(x,x)\ |\\
 x\in  \Omega_\ep\}\textrm{ for any }\ep\in]0,\ep_0[.
\end{array}
\end{equation}

Let $q_0,q\in \bar D,$ $q_0\not=q.$ 
Let $\ep_1\in]0,\ep_0[.$ 
From \eqref{56.53}, it follows that there exists an unique $p_{\ep_1}\in \S^{n-1}_{q_0,E}$ and an unique positive
real number $t_{\ep_1}$ such that $q=\psi_1(t_{\ep_1},q_0,p_{\ep_1})$ and $(t_{\ep_1},q_0,p_{\ep_1})\in \Lambda_{\ep_1}.$
Consider the function $m\in C^2(\R,\R),$ defined by 
$m(t)=\chi_D(\psi_1(t,q_0,p_{\ep_1})),$ $t\in \R.$ Derivating twice $m$, we obtain 
\begin{eqnarray}
\label{A}\ddot m(t)=Hess\chi_D(\psi_1(t,q_0,p_{\ep_1}))(g(\psi_2(t,q_0,p_{\ep_1})),g(\psi_2(t,q_0,p_{\ep_1})))&&\\
+\left(1+{|\psi_2(t,q_0,p_{\ep_1})|^2\over c^2}\right)^{-1/2}\nabla \chi_D(\psi_1(t,q_0,p_{\ep_1}))\circ
F\left(\psi_1(t,q_0,p_{\ep_1}),\right.\nonumber&&\\
\left.g(\psi_2(t,q_0,p_{\ep_1}))\right)-{\psi_2(t,q_0,p_{\ep_1})\circ F\left(\psi_1(t,q_0,p_{\ep_1}),g(\psi_2(t,q_0,p_{\ep_1}))\right)
\over c^2\left(1+{|\psi_2(t,q_0,p_{\ep_1})|^2\over c^2}\right)^{3/2}}
\nonumber&&\\
\times\nabla \chi_D(\psi_1(t,q_0,p_{\ep_1}))\circ \psi_2(t,q_0,p_{\ep_1}),\nonumber&&
\end{eqnarray}
for $t\in \R$,
where $g$ is the function defined by \eqref{B} and $\circ$ denotes the usual scalar product on $\R^n$ (we used \eqref{4}). 
In addition, note that using the fact that $\chi_D$ is a $C^2$ defining function of $D$, we obtain that for $t\in \R$
\begin{eqnarray*}
&&\psi_1(t,q_0,p_{\ep_1})\in D\Leftrightarrow m(t)<0,\\
&&\psi_1(t,q_0,p_{\ep_1})\in \pa D\Leftrightarrow m(t)=0.
\end{eqnarray*}

Assume that there exists some $s\in]0,t_{\ep_1}[$ such that $\psi_1(s,q_0,p_{\ep_1})\not\in D$ (i.e. $m(s)\ge 0$). Let
$s_0=\sup\{s'\in[0,s]\ |\ \psi_1(s',x,p_{\ep_1})\in \bar D\}.$ Hence
\begin{eqnarray}
\label{A1}&&\psi_1(s_0,q_0,p_{\ep_1})\in \pa D, \textrm{ (i.e. }m(s_0)=0\textrm{)}, \\
\label{A2}&&m(t)\le 0,\ \textrm{for }t\in[0,s_0]. 
\end{eqnarray}
From \eqref{A1}, \eqref{A}, \eqref{4a}, and the estimates \eqref{18c}, $|g(\psi_2(t,q_0,p_{\ep_1}))|<c,$   
and definition \eqref{56.41c}, it follows that
$\ddot m(s_0)\ge c^2C_8(E,\tilde V,$ $\tilde B,D)>0$  (we used \eqref{56.51}).
From \eqref{A2} and from the estimate $\ddot m(s_0)>0$ and Taylor expansion of $m$ at $s_0$
($m(t)=\dot m(s_0)(t-s_0)+{1\over 2}\ddot m(s_0)(t-s_0)^2+o((t-s_0)^2),\ t\in \R$) it follows that
$\dot m(s_0)>0$.
Using also the equality $m(s_0)=0,$ we obtain that there exists $\ep'>0$ such that $s_0+\ep'<t_{\ep_1}$ and $m(s_0+\ep')>0$ 
which implies that
$\psi_1(s_0+\ep',q_0,p_{\ep_1})\not\in \bar D.$  Then, due to $\sup_{z\in \Omega_\ep}\inf\{|z-z'|\ |\ z'\in \bar D\}\to 0$ as
$\ep\to 0$, there exists $\ep_2\in]0,\ep_1[$ such that $\psi_1(s_0+\ep',q_0,p_{\ep_1})\not\in \Omega_{\ep_2}$ 
and using also \eqref{56.53}, we obtain that there exists 
$(p_{\ep_2},t_{\ep_2})\in \S^{n-1}_{q_0,E}\times ]0,+\infty[$ such that $(p_{\ep_2},t_{\ep_2})\not=(p_{\ep_1},t_{\ep_1})$ and 
$(t_{\ep_2},q_0,p_{\ep_2})\in \Lambda_{\ep_2}$ and 
$q=\psi_1(t_{\ep_2},q_0,p_{\ep_2}),$ which contradicts unicity of $(p_{\ep_1},t_{\ep_1}).$

We finally proved that 
\begin{equation}
\label{56.60}\psi_1(s,q_0,p_{\ep_1})\in D\textrm{ for all }s\in ]0,t_{\ep_1}[.
\end{equation}
Now consider
\begin{eqnarray}
\label{56.61}t_2&=&\sup\{t\in ]0,+\infty[\ |\ \psi_1(s,q_0,p_{\ep_1})\in D \textrm{ for all }s\in]0,t]\},\\
\label{56.62}t_1&=&\inf\{t\in \R\ |\ \psi_1(s,q_0,p_{\ep_1})\in D \textrm{ for all }s\in[t,t_{\ep_1}[\}
\end{eqnarray}
(using Proposition 6.1 and \eqref{56.51}, we obtain that $t_2$ and $t_1$ are real numbers that satisfy $t_2-t_1\le{5\delta(D)\over c}$).
Then for $i=1,2,$ from \eqref{A}, \eqref{4a}, \eqref{56.41c} and \eqref{56.51}, it follows that
$\ddot m(t_i)\ge c^2C_8(E,\tilde V,\tilde B,D)>0$ .
For all $t\in]t_1,t_2[,$ $\psi_1(t,q_0,p)\in D.$ Hence 
$m(t)<0,\ t\in]t_1,t_2[.$ This latter  estimate and the estimate  $\ddot m(t_i)>0$ and
Taylor expansion of $m$ at $t_i$ ($m(t)=\dot m(t_i)(t-t_i)+{1\over 2}\ddot m(t_i)(t-t_i)^2+o((t-t_i)^2),\ t\in \R$) for $i=1,2,$
imply
that $\dot m(t_2)>0$ and $\dot m(t_1)<0,$ i.e. 
\begin{equation}
\label{56.66}
\begin{array}{l}
{\pa \psi_1\over \pa t}(t,q_0,p_{\ep_1})_{|t=t_1}\circ N(t_1)<0,\\
{\pa \psi_1\over \pa t}(t,q_0,p_{\ep_1})_{|t=t_2}\circ N(t_2)>0,
\end{array}
\end{equation}
where $N(t_i)=
{\nabla \chi(\psi_1(t,q_0,p_{\ep_1}))\over \left|\nabla \chi(\psi_1(t,q_0,p_{\ep_1}))\right|},$ $i=1,2.$

Statement \eqref{56.60} with \eqref{56.66} and \eqref{56.53} (with ``$\ep=\ep_1$") proves \eqref{2.1} and \eqref{2.2}.
\hfill $\Box$

\vskip 4mm

\noindent 6.6 {\it Proof of Proposition 6.1.}
We denote ${\dot x(t)\over \sqrt{1-{{\dot x(t)}^2\over c^2}}}$ by $p(t)$ for $t\in ]t_-,t_+[.$

Let $I(t)={1\over 2}|x(t)|^2,$ 
for $t\in ]t_-,t_+[.$
Derivating twice $I$ and using \eqref{6.01}, we obtain
\begin{eqnarray}
\label{6.8}\ddot I(t)&=&{p(t)^2\over{1+{p(t)^2\over c^2}}}+{1\over \sqrt{1+{p(t)^2\over c^2}}}F\left(x(t),{p(t)\over\sqrt{1+{p(t)^2\over
c^2}}}\right)\circ x(t)\\
&&-{p(t)\circ x(t)\over c^2(1+{p(t)^2\over c^2})^{3/2}}p(t)\circ F\left(x(t),{p(t)\over\sqrt{1+{p(t)^2\over c^2}}}\right)\nonumber
\end{eqnarray}
for $t\in ]t_-,t_+[$, where $\circ$ denotes the usual scalar product in $\R^n.$
From the estimate ${|p(t)|\over \sqrt{1+{p(t)^2\over c^2}}}<c$, $t\in]t_-,t_+[,$ and from
\eqref{6.8} and \eqref{4a}, it follows that $\ddot I(t)\ge c^2\left(1-{1\over ({E-\tilde{V}(x(t))\over c^2})^2} \right)
-2{|x(t)|\left(|\nabla \tilde V(x(t))|+\sum_{i,j=1\ldots n}|\tilde B_{i,j}(x(t))|\right)\over {E-\tilde{V}(x(t))\over c^2}},$
for $t\in ]t_-,t_+[,$ which with  \eqref{6.6} implies
\begin{equation}
\label{6.11}\ddot I(t)\ge {c^2\over 2},
\end{equation}
for $t\in ]t_-,t_+[.$

Let $t,s\in ]t_-,t_+[,$ $s\le t.$
From \eqref{6.11} and the equality 
$I(t)=I(s)+\dot I(s)(t-s)+\int_s^t\int_s^\tau \ddot I(\sigma)d\sigma d\tau,
$
it follows that 
\begin{equation}
\label{6.12a}I(t)-I(s)\ge \dot I(s)(t-s)+{c^2\over 4}(t-s)^2.
\end{equation}
Using \eqref{6.4b} and the estimate $|\dot x(s)|<c,$ we obtain
$\dot I(s)=x(s)\circ\dot x(s)\ge -|x(s)||\dot x(s)|\ge  -c\delta(\Omega).$
Using \eqref{6.4b}, we obtain $I(t)-I(s)={1\over 2}(|x(t)|^2-|x(s)|^2)\le \delta(\Omega)^2.$
From \eqref{6.12a} and the two latter inequalities, it follows that
$0\ge -\delta(\Omega)^2-c\delta(\Omega)(t-s)+{c^2\over 4}(t-s)^2,$
which implies that 
$t-s\le {\delta(\Omega)\over c}(2\sqrt{2}+2)<{5\delta(\Omega)\over c}$
(the roots of $-\delta(\Omega)^2-c\delta(\Omega)X+{c^2\over 4}X^2$ are 
$(\delta(\Omega)/c)(2\pm 2\sqrt{2})$).
As $t\to t_+$ and $s\to t_-$,  the latter inequality proves \eqref{6.6a}.
Proposition 6.1 is proved.\hfill $\Box$

\vskip 4mm

\noindent 6.7 {\it Proof of Proposition 6.2.}
Throughout this Subsection, we denote by $\gamma_{x,p_i}(t)$ the point of $\R^n$ defined by
$$
\gamma_{x,p_i}(t)=\psi_1(t,x,p_i),
$$
for any $t\in \R$ and $i=1,2$, where $\psi=(\psi_1,\psi_2)$ is the flow of the differential system \eqref{4}.

From \eqref{4}, it follows that
\begin{eqnarray}
\gamma_{x,p_i}(t)&=&x+tv_i+\int_0^{t_i}\left(g(p_i+\int_0^\sigma F(\gamma_{x,p_i}(\tau),\dot\gamma_{x,p_i}(\tau))d\tau)
\right.\nonumber\\
\label{6.25}&&\left.-g(p_i)\right)
d\sigma
\end{eqnarray}
for $t\in [0,t_{+,x,p_i}[$  and $i=1,2,$ where $t_{+,x,p_i}$ is defined by \eqref{6.t+} for $i=1,2.$

From \eqref{6.25}, it follows that 
\begin{equation}
\label{6.26}|t_1v_1-t_2v_2|-\Delta(t_1,t_2) \le|\gamma_{x,p_1}(t_1)-\gamma_{x,p_2}(t_2)|\le|t_1v_1-t_2v_2|+\Delta(t_1,t_2)
\end{equation}
where
\begin{eqnarray}
\label{6.26'}\Delta(t_1,t_2)=\left|\int_0^{t_1}\left(g(p_1+\int_0^\sigma F(\gamma_{x,p_1}(\tau),\dot\gamma_{x,p_1}(\tau)) d\tau)-g(p_1)\right)
d\sigma\right.&&\\
-\left.\int_0^{t_2}\left(g(p_2+\int_0^\sigma F(\gamma_{x,p_2}(\tau),\dot\gamma_{x,p_2}(\tau) )d\tau)-g(p_2)\right)d\sigma\right|,
\nonumber&&
\end{eqnarray}
for $t_1\in [0,t_{+,x,p_1}[$ and $t_2\in [0,t_{+,x,p_2}[$.
We shall look for an upper bound of $\Delta(t_1,t_2),$ $t_1\in [0,t_{+,x,p_1}[$ and  $t_2\in [0,t_{+,x,p_2}[,$ $t_2\le t_1.$  

{\it First case: $v_1\circ v_2\le 0.$} 
Using \eqref{5.17}, we obtain that
\begin{eqnarray}
\label{6.28}
\left|\int_0^{t_i}\left(g(p_i+\int_0^\sigma F(\gamma_{x,p_i}(\tau),\dot\gamma_{x,p_i}(\tau)) d\tau)-g(p_i)\right)d\sigma\right|&&\\
\le\sqrt{n}\int_0^{t_i}\sup_{\ep\in[0,1]}
\left(1+c^{-2}\left|p_i+\ep\int_0^\sigma F(\gamma_{x,p_i}(s),\dot\gamma_{x,p_i}(s))ds\right|^2\right)^{-1/2}\nonumber&&\\
\times\int_0^{\sigma}|F(\gamma_{x,p_i}(s),\dot\gamma_{x,p_i}(s))|dsd\sigma,\nonumber&&
\end{eqnarray}
for $i=1,2.$
From \eqref{18c} and \eqref{6.6a} and from the estimate $|g(\psi_2(s,x,p_i))|\le c$, it follows that
\begin{equation}
\label{6.26a}\int_0^\sigma|F(\gamma_{x,p_i}(s),\dot\gamma_{x,p_i}(s))|ds\le {10n\delta(\Omega)\beta(\tilde{V},\tilde{B},\Omega)\over c}.
\end{equation}
for $\sigma\in[0,t_i],$ $i=1,2.$
Using $p_i\in \S^{n-1}_{x,E}$ and \eqref{6.26a} and \eqref{6.21}, we obtain 
\begin{equation}
\label{6.26b}\left|p_i+\ep\int_0^\sigma\!\!\!\! F(\gamma_{x,p_i}(s),\dot\gamma_{x,p_i}(s))ds\right|
\ge {1\over 2}r_{\tilde V,E}(x),
\end{equation}
for $\ep\in [0,1]$ and $\sigma \in[0,t_i].$ From \eqref{6.26a}, \eqref{6.26b} and \eqref{6.28}, it follows that
\begin{eqnarray}  
\label{6.28a}&&\left|\int_0^{t_i}\left(g(p_i+\int_0^\sigma F(\gamma_{x,p_i}(s),\dot\gamma_{x,p_i}(s)) ds)-g(p_i)\right)d\sigma\right|\nonumber\\
&\le& t_i{20n^{3/2}c\delta(\Omega)\beta(\tilde{V},\tilde{B},\Omega)\over E-\tilde V(x)},
\end{eqnarray}
for $i=1,2.$
Using  $v_1\circ v_2\le 0$, we obtain that
$|v_i|s_i\le |s_1v_1-s_2v_2|$
for $i=1,2$ and for $s_1\ge 0,$ $s_2\ge0.$ 
Using this latter inequality and \eqref{6.26'} and \eqref{6.28a} and equality $|v_i|=c\sqrt{1-\left({E-\tilde V(x)\over c^2}\right)^{-2}},$ 
we obtain 
$$
\Delta(t_1,t_2)\le 
40c^{-1}n^{3/2}\delta(\Omega)\beta(\tilde{V},\tilde{B},\Omega)r_{\tilde V,E}(x)^{-1}|t_1v_1-t_2v_2|.
$$

{\it Second case: $v_1\circ v_2\ge 0.$} 

From  \eqref{6.26'}, it follows that
\begin{equation}
\label{6.29a}\Delta(t_1,t_2)\le \Delta_1(t_1,t_2)+\Delta_2(t_1,t_2),
\end{equation}
where
\begin{eqnarray}
\label{6.29b}\Delta_1(t_1,t_2)=\left|\int_{t_2}^{t_1}\!\!\!\left(g(p_1+\int_0^\sigma\!\!\!\! F(\gamma_{x,p_1}(\tau), \dot \gamma_{x,p_1}(\tau))
d\tau)-g(p_1)\right)d\sigma\right|&&\\
\label{6.29c}\Delta_2(t_1,t_2)=\left|\int_0^{t_2}\left[g(p_1+\int_0^\sigma F(\gamma_{x,p_1}(\tau), \dot \gamma_{x,p_1}(\tau)) d\tau)-g(p_1)
d\sigma\right.\right.&&\\
-\left.\left.\left(g(p_2+\int_0^\sigma F(\gamma_{x,p_2}(\tau), \dot \gamma_{x,p_2}(\tau))d\tau)-g(p_2)\right)\right]d\sigma\right|.\nonumber
&&
\end{eqnarray}

{\it An upper bound for $\Delta_1(t_1,t_2).$}
Using \eqref{6.29b} and \eqref{5.17}, we obtain that
\begin{eqnarray}
&&\label{6.30}\Delta_1(t_1,t_2)\le \sqrt{n}(t_1-t_2)\int_{t_2}^{t_1}|F(\gamma_{x,p_1}(s),\dot \gamma_{x,p_1}(s))|ds\\
&&\times\sup_{\ep\in[0,1]\atop\sigma\in [0,t_1]}
\left(1+c^{-2}\left|p_1+\ep\int_0^\sigma F(\gamma_{x,p_1}(s),\dot \gamma_{x,p_1}(s))ds\right|^2\right)^{-1/2}.\nonumber
\end{eqnarray}
In the same manner than in the first case ($v_1\circ v_2\le 0$), we obtain 
\begin{equation}
\label{6.30a}\Delta_1(t_1,t_2)\le (t_1-t_2){20n^{3/2}c\delta(\Omega)\beta(\tilde{V},\tilde{B},\Omega)\over E-\tilde V(x)}.
\end{equation}
Note that from $|v_1|=|v_2|$ and $t_i\ge 0, i=1,2$, it follows that $|v_1|(t_1-t_2)\le |t_1v_1-t_2v_2|$. Using this latter inequality with \eqref{6.30a}, 
we obtain 
\begin{equation}
\label{6.30b}\Delta_1(t_1,t_2)\le 
{20n^{3/2}\delta(\Omega)\beta(\tilde{V},\tilde{B},\Omega)\over c r_{\tilde V,E}(x)}|t_1v_1-t_2v_2|
\end{equation} 
(we use the equality $|v_1|=c\sqrt{1-\left({E-\tilde V(x)\over c^2}\right)^{-2}}$). 

{\it An upper bound for $\Delta_2(t_1,t_2).$}
Note that
$
g_j(p_i+\int_0^\sigma F(\gamma_{x,p_i}(\tau), \dot \gamma_{x,p_i}(\tau)) d\tau)$ $-g_j(p_i)
=\int_0^1\nabla g_j(p_i+\ep\int_0^\sigma F(\gamma_{x,p_i}(\tau), \dot \gamma_{x,p_i}(\tau)) d\tau)\circ$$
\int_0^\sigma F(\gamma_{x,p_i}(s),\dot \gamma_{x,p_i}(s)) ds$ $d\ep
$
for $i=1,2$ and $j=1..n,$ where $g=(g_1,\ldots,g_n).$
Hence
\begin{equation}
\label{6.31a}\Delta_2^j(t_1,t_2)
\le\int_0^{t_2}\Delta_{2,1,j}(\sigma)d\sigma+\int_0^{t_2}\Delta_{2,2,j}(\sigma)d\sigma,
\end{equation}
where $\Delta_2(t_1,t_2)=(\Delta_2^1(t_1,t_2),\ldots,\Delta_2^n(t_1,t_2))$ and 
\begin{eqnarray}
&&\label{6.31b}\Delta_{2,1,j}(\sigma)=\left|\int_0^1\left[\nabla g_j(p_1+\ep\int_0^\sigma F(\gamma_{x,p_1}(\tau), \dot \gamma_{x,p_1}(\tau)) d\tau)
-\nabla g_j\right.\right.\\
&&\left.\left.(p_2+\ep\int_0^\sigma F(\gamma_{x,p_2}(\tau), \dot \gamma_{x,p_2}(\tau))d\tau)\right]\circ
\int_0^\sigma F(\gamma_{x,p_1}(s),\dot \gamma_{x,p_1}(s)) ds d\ep\right|,\nonumber\\
&&\label{6.31c}\Delta_{2,2,j}(\sigma)=\left|\int_0^1\nabla g_j(p_2+\ep\int_0^\sigma F(\gamma_{x,p_2}(\tau), \dot
\gamma_{x,p_2}(\tau))d\tau)\circ\right.\\
&&\left.\int_0^\sigma\left[F(\gamma_{x,p_1}(s),\dot \gamma_{x,p_1}(s))-F(\gamma_{x,p_2}(s), \dot \gamma_{x,p_2}(s))\right]ds d\ep\right|\nonumber
\end{eqnarray}
for $\sigma \in [0,t_2]$ and $j=1\ldots n.$ 

We first look for an upper bound for $\Delta_{2,1,j}(\sigma)$.
Since $v_1\circ v_2\ge 0,$ we obtain $p_1\circ p_2\ge 0$. Using this latter inequality and the equality $p_1^2=p_2^2$, we obtain that 
\begin{eqnarray}
|\mu p_1+(1-\mu)p_2|=\sqrt{\mu^2p_1^2+(1-\mu)^2p_2^2+2\mu(1-\mu)p_1\circ p_2}\nonumber&&\\
\label{6.32}\ge\sqrt{\mu^2p_1^2+(1-\mu)^2p_2^2}\ge{1\over \sqrt{2}}|p_1|={r_{\tilde V,E}(x)\over \sqrt{2}},&&
\end{eqnarray}
for any $\mu\in [0,1].$

From \eqref{6.26a} and \eqref{6.32} and \eqref{6.21}, it follows that
\begin{eqnarray}
&&\left|\mu p_1+(1-\mu)p_2+\mu\epsilon\int_0^\sigma F(\gamma_{x,p_1}(\tau), \dot \gamma_{x,p_1}(\tau)) d\tau+(1-\mu)\epsilon
\right.\nonumber\\
&&\left. \times\int_0^\sigma F(\gamma_{x,p_2}(\tau),\dot \gamma_{x,p_2}(\tau))d\tau\right|\ge |\mu p_1+(1-\mu) p_2|-
{10n\delta(\Omega)\beta(\tilde{V},\tilde{B},\Omega)\over c}\nonumber\\
\label{6.33}&&\ge{1\over 2\sqrt{2}}r_{\tilde V,E}(x),
\end{eqnarray}
for $\mu,$ $\epsilon\in[0,1].$
From \eqref{4}, it follows that
$$\dot\gamma_{x,p_i}(\sigma)=g(p_i+\int_0^\sigma F(\gamma_{x,p_i}(\tau), \dot \gamma_{x,p_i}(\tau))d\tau)$$
for $\sigma\in[0,t_i],$ $i=1,2.$
Using this latter equality and \eqref{5.17}, we obtain 
\begin{eqnarray}
&&\label{6.35}|\dot\gamma_{x,p_1}(\sigma)-\dot\gamma_{x,p_2}(\sigma)|\le 
\\
&&\sqrt{n}\sup_{\mu\in [0,1]}\left(1+c^{-2}|\mu p_1+(1-\mu) p_2+
\mu\int_0^\sigma F(\gamma_{x,p_1}(\tau), \dot \gamma_{x,p_1}(\tau)) d\tau\right.\nonumber\\
&&+\left.(1-\mu)\int_0^\sigma F(\gamma_{x,p_2}(\tau), 
\dot \gamma_{x,p_2}(\tau))d\tau|^2\right)^{-1/2}\nonumber\\
&&\times|p_1-p_2+\int_0^\sigma 
\left(F(\gamma_{x,p_1}(\tau), \dot \gamma_{x,p_1}(\tau))-F(\gamma_{x,p_2}(\tau), \dot \gamma_{x,p_2}(\tau))\right)
d\tau|\nonumber
\end{eqnarray}
for $\sigma\in[0,t_2].$ 

Note that from \eqref{18d} and the inequality $|\dot \gamma_{x,p_2}(\tau)|\le c$, it follows that
\begin{eqnarray}
\label{6.36}
&&\left|F(\gamma_{x,p_1}(\tau), \dot \gamma_{x,p_1}(\tau))-F(\gamma_{x,p_2}(\tau), \dot \gamma_{x,p_2}(\tau))\right|
\le\\ 
&&n\beta(\tilde{V},\tilde{B},\Omega)(2|\gamma_{x,p_1}(\tau)-\gamma_{x,p_2}(\tau)|
+{1\over c }|\dot \gamma_{x,p_1}(\tau)-
\dot \gamma_{x,p_2}(\tau)|),\nonumber
\end{eqnarray}
for $\tau \in [0,t_2].$
 
Using \eqref{6.33}-\eqref{6.36}, we obtain
\begin{eqnarray}
\label{6.37}&&|\dot\gamma_{x,p_1}(\sigma)-\dot\gamma_{x,p_2}(\sigma)|\le  
{2^{3/2}\sqrt{n}c^2 \over E-\tilde V(x)}
\left[|p_1-p_2|+n\beta(\tilde{V},\tilde{B},\Omega)\right.\\
&&\left.\left(2\int_0^\sigma |\gamma_{x,p_1}(\tau)-\gamma_{x,p_2}(\tau)| d\tau+{1\over c}\int_0^\sigma |\dot \gamma_{x,p_1}(\tau)-
\dot \gamma_{x,p_2}(\tau)| d\tau\right)\right],\nonumber
\end{eqnarray}
for $\sigma\in[0,t_2].$ 

We shall use the following Gronwall's lemma.
\vskip 2mm
{\bf Gronwall's lemma.} {\it Let $a>0$ and let $\phi\in C([0,a], [0,+\infty[)$ be a continuous map  and let
$A,B\in [0,+\infty[$ be such that $\phi(t)\le A+B\int_0^t\phi(s)ds$ for all $t\in [0,a].$ Then 
$\phi(t)\le Ae^{Bt}$ for all $t\in[0,a].$   
}
\vskip 2mm

Taking account of \eqref{6.37}, Gronwall's lemma and \eqref{6.6a}, we obtain that
\begin{eqnarray}
\label{6.38}|\dot\gamma_{x,p_1}(\sigma)-\dot\gamma_{x,p_2}(\sigma)|&\le &  
{2^{3/2}\sqrt{n}c^2\over E-\tilde V(x)}
\left[|p_1-p_2|+2n\beta(\tilde{V},\tilde{B},\Omega)\right.\\
&&\left.\int_0^\sigma |\gamma_{x,p_1}(\tau)-\gamma_{x,p_2}(\tau)| d\tau\right]e^{{10\sqrt{2}\delta(\Omega)n^{3/2}\beta(\tilde{V},\tilde{B},\Omega)\over E-\tilde V(x)}},
\nonumber
\end{eqnarray}
for $\sigma \in[0,t_2].$ 

From \eqref{5.18} and \eqref{6.31b}, it follows that
\begin{eqnarray}
&&\label{6.39}\Delta_{2,1,j}(\sigma)\le\\
&&{3\sqrt{n}\over c}\int_0^1
\sup_{\mu\in[0,1]}\left(1+{1\over c^2}\left|\mu p_1+(1-\mu)p_2+\mu\epsilon\int_0^\sigma F(\gamma_{x,p_1}(\tau), 
\dot \gamma_{x,p_1}(\tau)) d\tau\right.\right.\nonumber
\end{eqnarray}
\begin{eqnarray}
&&+\left.\left.(1-\mu)\epsilon\int_0^\sigma F(\gamma_{x,p_2}(\tau), \dot \gamma_{x,p_2}(\tau))d\tau\right|^2\right)^{-1}\nonumber\\
&&\times|p_1-p_2+\ep\int_0^\sigma 
\left(F(\gamma_{x,p_1}(\tau), \dot \gamma_{x,p_1}(\tau))-F(\gamma_{x,p_2}(\tau), \dot \gamma_{x,p_2}(\tau))\right)
d\tau|\nonumber\\
&&\times\int_0^\sigma |F(\gamma_{x,p_1}(s),\dot \gamma_{x,p_1}(s))| dsd\ep,\nonumber
\end{eqnarray}
for $\sigma\in[0,t_2].$

From \eqref{18c} and the estimate $|\dot \gamma_{x,p_1}(s)|\le c$, it follows that
\begin{equation}
\label{6.39a}\int_0^\sigma|F(\gamma_{x,p_1}(s),\dot \gamma_{x,p_1}(s))|ds\le 2n\beta(\tilde{V},\tilde{B},\Omega)\sigma.
\end{equation}
for $\sigma\in[0,t_1].$

From \eqref{6.39}, \eqref{6.33}, \eqref{6.39a} and \eqref{6.36} and \eqref{6.38}, it follows that
\begin{eqnarray}
&&\label{6.40}\Delta_{2,1,j}(\sigma)\le
{48n^{3/2}c^3\beta(\tilde{V},\tilde{B},\Omega)\over (E-\tilde V(x))^2}
\left(|p_1-p_2|\sigma+ n\beta(\tilde{V},\tilde{B},\Omega)\sigma 
\right.\\
&&\left(2\int_0^\sigma |\gamma_{x,p_1}(\tau)-\gamma_{x,p_2}(\tau)|d\tau+{2^{3/2}\sqrt{n}c\over E-\tilde V(x)}
\left(|p_1-p_2|\sigma+2n\beta(\tilde{V},\tilde{B},\Omega)\right.\right.\nonumber\\
&&\left.\left.\left.\int_0^\sigma \int_0^s|\gamma_{x,p_1}(\tau)-\gamma_{x,p_2}(\tau)| d\tau
ds\right)\times e^{{10\sqrt{2}\delta(\Omega)n^{3/2}\beta(\tilde{V},\tilde{B},\Omega)\over E-\tilde V(x)}}\right)\right)\nonumber
\end{eqnarray}
for $\sigma\in[0,t_2].$

From  $|v_1|=|v_2|$ and $t_2\le t_1,$ it follows that
$|v_1-v_2|\sigma\le |v_1-v_2|t_2\le |t_1v_1-t_2v_2|,$
for $\sigma \in [0,t_2].$
Note that using these latter estimates and $p_i\in\S^{n-1}_{x,E},$ $i=1,2$, we obtain
\begin{equation}
\label{6.44}|p_1-p_2|\sigma\le {E-\tilde V(x)\over c^2}|t_1v_1-t_2v_2|,
\end{equation}
for  $\sigma\in[0,t_2].$

Using \eqref{6.44}, the estimate $\int_0^\sigma \int_0^s|\gamma_{x,p_1}(\tau)-\gamma_{x,p_2}(\tau)| d\tau
ds\le \sigma \int_0^\sigma|\gamma_{x,p_1}(\tau)$ $-\gamma_{x,p_2}(\tau)| 
d\tau $ and $\sigma\le{5\delta(\Omega)\over c}$ (due to \eqref{6.6a}) and \eqref{6.40}, we obtain
\begin{eqnarray}
&&\Delta_{2,1,j}(\sigma)\le\left(1+
{10\sqrt{2}n^{3/2}\delta(\Omega)\beta(\tilde{V},\tilde{B},\Omega)\over E-\tilde V(x)}
e^{{10\sqrt{2}\delta(\Omega)n^{3/2}\beta(\tilde{V},\tilde{B},\Omega)\over E-\tilde V(x)}}\right)\nonumber\\
&&\label{6.41a}\times\left(
{48n^{3/2}c\delta(\Omega)\beta(\tilde{V},\tilde{B},\Omega)\over E-\tilde V(x)}|t_1v_1-t_2v_2|\right.\\
&&+\left.{480n^{5/2}c^2\delta(\Omega)\beta(\tilde{V},\tilde{B},\Omega)^2\over (E-\tilde V(x))^2} 
\int_0^\sigma |\gamma_{x,p_1}(\tau)-\gamma_{x,p_2}(\tau)|d\tau
\right)\nonumber
\end{eqnarray}
for $\sigma\in[0,t_2].$

We look for an upper bound for $\Delta_{2,2,j}(\sigma),$ $\sigma\in[0,t_2],$ defined by \eqref{6.31c}.
Using \eqref{6.31c}, \eqref{5.16}, and \eqref{6.26b}, and  using \eqref{6.36}, \eqref{6.38} and \eqref{6.44}, we obtain 
\begin{eqnarray}
\Delta_{2,2,j}(\sigma)&\le&
\left(1+{10\sqrt{2}n^{3/2}\delta(\Omega)\beta(\tilde{V},\tilde{B},\Omega)\over E-\tilde V(x)}
e^{{10\sqrt{2}\delta(\Omega)n^{3/2}\beta(\tilde{V},\tilde{B},\Omega)\over E-\tilde
V(x)}}\right)\nonumber\\
\label{6.43}&&\times{4nc^2\beta(\tilde{V},\tilde{B},\Omega)\over E-\tilde V(x)}\int_0^\sigma|\gamma_{x,p_1}(\tau)-\gamma_{x,p_2}(\tau)|d\tau\\
&&+{2^{5/2}n^{3/2}c\beta(\tilde{V},\tilde{B},\Omega)\over E-\tilde V(x)}
e^{{10\sqrt{2}\delta(\Omega)n^{3/2}\beta(\tilde{V},\tilde{B},\Omega)\over E-\tilde V(x)}}|t_1v_1-t_2v_2|,\nonumber
\end{eqnarray}
for $\sigma\in[0,t_2].$

Note also that from \eqref{6.26}, it follows that
$\int_0^\sigma|\gamma_{x,p_1}(\tau)-\gamma_{x,p_2}(\tau)|d\tau
\le$$\int_0^\sigma\Delta$ $(\tau,\tau)d\tau+{\sigma^2\over 2}|v_1-v_2|,$
for $\sigma\in [0,t_2].$
Hence using also $\sigma\le{5\delta(\Omega)\over c}$  (due to \eqref{6.6a}) and $\sigma |v_1-v_2|\le |t_1v_1-t_2v_2|$, we obtain
\begin{equation}
\label{6.45}\int_0^\sigma|\gamma_{x,p_1}(\tau)-\gamma_{x,p_2}(\tau)|d\tau
\le\int_0^\sigma\Delta(\tau,\tau)d\tau+{5\delta(\Omega)\over 2c}|t_1v_1-t_2v_2|,
\end{equation}
for $\sigma\in [0,t_2].$
Note that $t_2\le {5\delta(\Omega)\over c}$ and note that from positiveness of $\Delta$ it follows that
$\int_0^{t_2}\int_0^\sigma\Delta(\tau,\tau)d\tau  d\sigma\le t_2 \int_0^{t_2}\Delta(\tau,\tau)d\tau.$ 
Hence using \eqref{6.45}, we obtain
\begin{equation}
\label{6.45a}\int_0^{t_2}\int_0^\sigma|\gamma_{x,p_1}(\tau)-\gamma_{x,p_2}(\tau)|d\tau d\sigma
\le{5\delta(\Omega)\over c}\int_0^{t_2}\Delta(\tau,\tau)d\tau+{25\delta(\Omega)^2\over 2c^2}|t_1v_1-t_2v_2|,
\end{equation}
for $\sigma\in [0,t_2].$

Combining \eqref{6.29a}, \eqref{6.30b}, \eqref{6.41a}, \eqref{6.43} and \eqref{6.45a}, we obtain
\begin{equation}
\label{6.46}\Delta(t_1,t_2)\le
C_4(E,x,\tilde{V},\tilde{B},\Omega)
|t_1v_1-t_2v_2|
+c C_5(E,x,\tilde{V},\tilde{B},\Omega)\int_0^{t_2} \Delta(\tau,\tau)d\tau,
\end{equation}
for $t_1\in [0,t_{+,x,p_1}[$ and $t_2\in [0,t_{+,x,p_2}[$, $t_1\ge t_2$ and where
$C_4$ and $C_5$ are defined by \eqref{6.47a} and
\eqref{6.47b}.

Let $t_1\in [0,t_{+,x,p_1}[$ and $t_2\in [0,t_{+,x,p_2}[$, $t_1\ge t_2.$
Estimates \eqref{6.46} and $|v_1-v_2|\sigma\le |t_1v_1-t_2v_2|,$ $\sigma \le t_2,$ give in particular
\begin{eqnarray}
\Delta(\sigma,\sigma)&\le&
\label{6.48}C_4
|t_1v_1-t_2v_2|
+c C_5\int_0^{\sigma} \Delta(\tau,\tau)d\tau
\end{eqnarray}
for $\sigma \in[0,t_2].$
Using \eqref{6.48} and using Gronwall's lemma (formulated above) and $\sigma\le{5\delta(\Omega)\over c}$, we obtain
\begin{equation}
\label{6.49}\Delta(\sigma,\sigma)\le C_4
e^{5 \delta(\Omega)C_5}|t_1v_1-t_2v_2|,
\end{equation}
for $\sigma \in[0,t_2].$

Using \eqref{6.49} and \eqref{6.46} and $t_2\le{5\delta(\Omega)\over c}$, we obtain 
$\Delta(t_1,t_2)\le
C_3(E,x,\tilde{V},$ $\tilde{B},\Omega)
|t_1v_1-t_2v_2|,$
for $t_1\in [0,t_{+,x,p_1}[$ and $t_2\in [0,t_{+,x,p_2}[$, $t_1\ge t_2.$

Proposition 6.2 is proved.\hfill$\Box$

\vskip 4mm

\noindent 6.8 {\it Proof of Proposition 6.3.} We shall work in coordinates.
We consider the following infinitely smooth parametrizations of $\S^{n-1},$ $\phi_{i,\pm}:B_{n-1}(0,1)\to \S^{n-1},$ $i=1\ldots n,$ 
defined by

\begin{eqnarray}
\label{56.2d}&&\phi_{i,\pm}(w)=\\
&&\left\lbrace
\begin{array}{l}
\left(w^1,\ldots,w^{i-1},\pm\sqrt{1-\sum_{l=1}^{n-1}{w^l}^2},w^i,\ldots,w^{n-1}\right),\ \textrm{ if }1\le i\le {n-1},\\
\left(w^1,\ldots,w^{n-1},\pm\sqrt{1-\sum_{l=1}^{n-1}{w^l}^2}\right),\ \textrm{ if } i= n
\end{array}
\right.\nonumber
\end{eqnarray}
for $w=(w^1,\ldots,w^{n-1})\in B_{n-1}(0,1)$ and where $B_{n-1}(0,1)$ denotes the unit Euclidean open ball of $\R^{n-1}$ of center $0$.

Let $(t_0,x_0,p_0)\in \Lambda\cap (]0,+\infty[\times {\cal V}_E),$ $p_0=(p_0^1,\ldots,p_0^n).$
Then $(t,x_0,p_0)$ $\in \Lambda$ for all $t\in [0,t_0].$
As $\Lambda$ is an open subset of $\R\times \R^n\times\R^n$, there exists $\ep >0$ such that
$\{(t,x,p)\in \R\times \R^n\times\R^n\ |\  -\ep<t<t_0+\ep,\ \max(|x-x_0|,|p-p_0|)<\ep\}\subseteq \Lambda.$
We denote by $B(x_0,\ep)$ the Euclidean open ball of $\R^n$ of center $x_0$ and radius $\ep.$ 
Let $(U,\phi)$ be an infinitely smooth parametrization of an open neighborhood of ${p_0\over |p_0|}$ in $\S^{n-1},$ and $k=1\ldots n$  
such that 
\begin{eqnarray}
\label{56.2a}&&U \textrm{ is an open subset of }B_{n-1}(0,1),\\
\label{56.2b}&&|p_0^k|\ge n^{-1/2} |p_0|,\\
\label{56.2e}&&\textrm{if }\pm p_0^k>0\textrm{ then }\phi(w)=\phi_{k,\pm}(w) \textrm{ for all }w\in U,
\end{eqnarray}
\begin{equation}
\label{56.2c}(t,x,r_{\tilde V,E}(x)\phi(w))\in \Lambda,
\textrm{ for }(w,t,x)\in U\times]-\ep,t_0+\ep[\times B(x_0,\ep).
\end{equation}

Consider $Q\in C^1(]-\ep,t_0+\ep[\times B(x_0,\ep)\times U,\Omega)$ defined by
\begin{equation}
\label{56.3}Q(t,x,w)=\psi_1(t,x,r_{\tilde V,E}(x)\phi(w)), \ (t,x,w)\in]-\ep,t_0+\ep[\times B(x_0,\ep)\times U,
\end{equation}
where $\psi=(\psi_1,\psi_2)$ is the flow of the differential system \eqref{4}. Let $w_0\in U$ be such that $\phi(w_0)={p_0\over|p_0|}.$
We shall prove that $({\pa Q\over \pa t}(t_0,x_0,w_0),{\pa Q\over \pa w^1}(t_0,x_0,w_0),\ldots,$ ${\pa Q\over \pa w^{n-1}}(t_0,x_0,w_0))$ 
is a basis of $\R^n.$

Note that from \eqref{6.01}, it follows that
\begin{eqnarray}
&&\label{56.4}Q(t,x,w)=x+tg(r_{\tilde V,E}(x)\phi(w))+\int_0^t\left[g\left(r_{\tilde V,E}(x)\phi(w)\right.\right.\\
&&\left.+\int_0^\sigma F(Q(s,x,w),{\pa Q\over \pa s}(s,x,w))ds\right)-\left.g(r_{\tilde V,E}(x)\phi(w))\right]d\sigma\nonumber,
\end{eqnarray}
for $w\in U$ and $(t,x)\in ]-\ep,t_0+\ep[\times B(x_0,\ep)$ (where $g,$ $r_{\tilde V,E}$ and $F$ are defined by \eqref{B},
\eqref{5.02} and \eqref{5.0}).

We shall prove  \eqref{56.9}.

Using \eqref{56.4} we obtain
\begin{eqnarray}
\label{56.5}{\pa Q\over\pa t}(t,x,w)&=&g(r_{\tilde V,E}(x)\phi(w))+\left[g\left(r_{\tilde V,E}(x)\phi(w)\right.\right.\\
&&\left.\left.+\int_0^t F(Q(s,x,w),{\pa Q\over \pa s}(s,x,w))ds
\right)-g(r_{\tilde V,E}(x)\phi(w))\right],\nonumber
\end{eqnarray}
for $w\in U$ and $(t,x)\in ]-\ep,t_0+\ep[\times B(x_0,\ep).$
Combining \eqref{56.5}, \eqref{56.2}, \eqref{5.17}, \eqref{18c} and estimates $|{\pa Q\over \pa t}(t,x,w)|\le c,$ 
$t\le{5\delta(\Omega)\over c}$, 
it follows that
\begin{equation}
\label{56.9}\left|{\pa Q\over \pa t}(t,x,w)-g(r_{\tilde V,E}(x)\phi(w))\right|\le 
{4n^{3/2}c^2\beta(\tilde{V},\tilde{B},\Omega)\over E-\tilde V(x)}t,
\end{equation}
for $w\in U$ and $(t,x)\in [0,t_0+\ep[\times B(x_0,\ep).$

We shall  prove  \eqref{56.24}.
Let $i=1\ldots n-1.$ 
Let $X_i\in C( ]-\ep,t_0+\ep[\times B(x_0,\ep)\times U,\R^n)$ be defined by
\begin{eqnarray}
\label{56.10}
X_i^j(s,x,w)=\sum_{l=1}^n \left({\pa F_j\over \pa x_l'}(x',{\pa Q\over \pa s}(s,x,w))_{|x'=Q(s,x,w)}{\pa Q_l\over \pa w^i}(s,x,w)\right.&&\\
+\left.{\pa F_j\over \pa y_l'}(Q(s,x,w),y')_{|y'={\pa Q\over \pa s}(s,x,w)}{\pa \bar Q_l\over \pa w^i}(s,x,w)\right),\nonumber&&
\end{eqnarray}
for $j=1\ldots n,$ $(s,x,w)\in ]-\ep,t_0+\ep[\times B(x_0,\ep)\times U,$ and where $X_i=(X_i^1,\ldots, X_i^n)$ and  
$\bar Q\in C^1( ]-\ep,t_0+\ep[\times B(x_0,\ep)\times U,\R)$ is
defined by
\begin{equation}
\label{56.11}
\bar Q(s,x,w)=g(\psi_2(s,x,r_{\tilde V,E}(x)\phi(w)))={\pa Q\over \pa s}(s,x,w),
\end{equation}
for $(s,x,w)\in ]-\ep,t_0+\ep[\times B(x_0,\ep)\times U.$

From \eqref{18a}, and \eqref{5.0}, it follows that
\begin{equation}
\label{56.11a}|X_i(\sigma,x,w)|\le \beta(\tilde{V},\tilde{B},\Omega)n\left(2
\sqrt{n}\left|{\pa Q\over \pa w^i}(\sigma,x,w)\right|+c^{-1}\left|{\pa \bar Q\over \pa w^i}(\sigma,x,w)\right|
\right),
\end{equation}
for $(\sigma,x,w)\in ]-\ep,t_0+\ep[\times B(x_0,\ep)\times U.$

We shall estimate $\bar Q.$
Note that from \eqref{6.01}, it follows that
$\bar Q_l(s,x,w)=
g_l(r_{\tilde V,E}(x)\phi(w)+\int_0^s F(Q(\sigma,x, r_{\tilde V,E}(x)\phi(w)),\bar Q(\sigma,x, r_{\tilde V,E}(x)\phi(w)))d\sigma),$
for $(s,$ $x,w)\in ]-\ep,t_0+\ep[\times B(x_0,\ep)\times U$ and $l=1\ldots n.$
From this latter equality and \eqref{56.10}, it follows that
\begin{eqnarray*}
&&{\pa \bar Q_l\over \pa w^i}(s,x,w)=
\nabla g_l\left(r_{\tilde V,E}(x)\phi(w)+\int_0^s F(Q(\sigma,x, r_{\tilde V,E}(x)\phi(w)),\right.\\
&&\left.\bar Q(\sigma,x, r_{\tilde V,E}(x)\phi(w)))d\sigma\right)\circ \left(r_{\tilde V,E}(x){\pa \phi\over \pa w^i}(w)+
\int_0^s X_i(\sigma,x,w)d \sigma\right),
\end{eqnarray*}
for $(s,x,w)\in ]-\ep,t_0+\ep[\times B(x_0,\ep)\times U.$
Hence
\begin{eqnarray}
\label{56.13}&&\left|{\pa \bar Q_j\over \pa w^i}(s,x,w)-r_{\tilde V,E}(x)\nabla g_j(r_{\tilde V,E}(x)\phi(w))\circ
{\pa \phi\over \pa w^i}(w)\right|\le\\
&&r_{\tilde V,E}(x)\left|\left(\nabla g_j\left(r_{\tilde V,E}(x)\phi(w)+\int_0^s F(Q(\sigma,x, r_{\tilde V,E}(x)\phi(w)),\right.\right.
\right.\nonumber\\
&&\left.\left.\bar Q(\sigma,x, r_{\tilde V,E}(x)\phi(w)))d\sigma\right)-\left.\nabla g_j\left(r_{\tilde V,E}(x)\phi(w)\right)\right)\circ 
{\pa \phi \over \pa w^i}(w)\right|\nonumber\\
&&+\left|
\nabla g_j\left(r_{\tilde V,E}(x)\phi(w)+\int_0^s F(Q(\sigma,x, r_{\tilde V,E}(x)\phi(w)),\bar Q(\sigma,x, r_{\tilde V,E}(x)\right.\right.\nonumber\\
&&\left.\left.\phi(w)))d\sigma\right)\circ
\int_0^sX_i(\sigma,x,w)d\sigma\right|,\nonumber
\end{eqnarray}
for $j=1\ldots n$ and $(s,x,w)\in ]-\ep,t_0+\ep[\times B(x_0,\ep)\times U.$
We estimate the second term of the sum on the right-hand side of \eqref{56.13} by using \eqref{5.16}, \eqref{18c}, \eqref{56.2} and $s\le
{5\delta(\Omega)\over c},$  and \eqref{56.11a}.
We estimate the first term of the sum on the right-hand side of \eqref{56.13} by using \eqref{5.18} and \eqref{18c}, \eqref{56.2} and $s\le
{5\delta(\Omega)\over c},$ and the estimate
$\int_0^s F(Q(\sigma,x, r_{\tilde V,E}(x)\phi(w)),\bar Q(\sigma,x, r_{\tilde V,E}(x)\phi(w)))d\sigma\le 2n\beta(\tilde{V},\tilde{B},\Omega)s,$ for 
$(s,x,w)\in ]-\ep,t_0+\ep[\times B(x_0,\ep)\times U.$ We obtain
\begin{eqnarray*}
&&\left|{\pa \bar Q\over \pa w^i}(s,x,w)
-{r_{\tilde V,E}(x)\over \left({E-\tilde V(x)\over c^2}\right)}{\pa \phi\over \pa w^i}(w)\right|\le
2n\sqrt{n}c^3\beta(\tilde{V},\tilde{B},\Omega)\left(12\sqrt{n}+1\right)
\end{eqnarray*}
\begin{eqnarray*}
&&\times{r_{\tilde V,E}(x)\over \left(E-\tilde V(x)\right)^2}
|{\pa \phi \over \pa w^i}(w)|s+{2c^2\beta(\tilde{V},\tilde{B},\Omega)n\sqrt{n}\over E-\tilde{V}(x)}\left(\int_0^s
\left|{\pa Q\over \pa w^i}(\sigma,x,w)\right|d\sigma\right.\nonumber\\
&&\times 2
\sqrt{n}+c^{-1}\left.\int_0^s\left|{\pa \bar Q\over \pa w^i}(\sigma,x,w)-
{r_{\tilde V,E}(x)\over \left({E-\tilde V(x)\over c^2}\right)}{\pa \phi\over \pa w^i}(w)\right|d\sigma
\right),\nonumber
\end{eqnarray*}
for  $(s,x,w)\in [0,t_0+\ep[\times B(x_0,\ep)\times U$ (note that $\left({E-\tilde V(x)\over c^2}\right)^{-1}{\pa \phi\over \pa w^i}(w)
=\left(\nabla g_j(r_{\tilde V,E}(x)\phi(w))\circ{\pa \phi\over \pa w^i}(w)\right)_{j=1\ldots n}$).

From Gronwall's lemma (formulated in Subsection 6.7) and $t_0+\ep \le {5\delta(\Omega)\over c},$ it follows that
\begin{eqnarray}
\label{56.18}\left|{\pa \bar Q\over \pa w^i}(s,x,w)
-{r_{\tilde V,E}(x)\over \left({E-\tilde V(x)\over c^2}\right)}{\pa \phi\over \pa w^i}(w)\right|\le
{2c^2n^{3/2}\beta(\tilde{V},\tilde{B},\Omega)\over E-\tilde V(x)}
e^{{10\delta(\Omega)\beta(\tilde{V},\tilde{B},\Omega)n^{3/2}\over E-\tilde{V}(x)}}&&\\
\times 
\left[c\left(12\sqrt{n}+1\right){r_{\tilde V,E}(x)\over E-\tilde V(x)}
\left|{\pa \phi \over \pa w^i}(w)\right|s
+2\sqrt{n}\int_0^s\left|{\pa Q\over \pa w^i}(\sigma,x,w)\right|d\sigma\right],\nonumber&&
\end{eqnarray}
for  $(s,x,w)\in [0,t_0+\ep[\times B(x_0,\ep)\times U.$

From \eqref{56.11}, it follows that 
$Q(s,x,w)=x+\int_0^s\bar Q(\sigma,x,w)d\sigma,$ for $(s,x,w)$ $\in [0,t_0+\ep[\times B(x_0,\ep)\times U.$
Hence
${\pa Q\over \pa w^i}(s,x,w)=\int_0^s{\pa \bar Q\over \pa w^i}(\sigma,x,w)d\sigma,$
for $(s,x,w)$ $\in [0,t_0+\ep[\times B(x_0,\ep)\times U.$
This latter equality and \eqref{56.18} imply 
\begin{eqnarray}
&&\left|{\pa Q\over \pa w^i}(s,x,w)
-{r_{\tilde V,E}(x)\over \left({E-\tilde V(x)\over c^2}\right)}{\pa \phi\over \pa w^i}(w)s\right|
\label{56.20}\le{2c^2n^{3/2}\beta(\tilde{V},\tilde{B},\Omega)\over E-\tilde V(x)}
\\
&&e^{{10\delta(\Omega)\beta(\tilde{V},\tilde{B},\Omega)n^{3/2}\over E-\tilde{V}(x)}}
\left[c\left(12\sqrt{n}+1\right){r_{\tilde V,E}(x)\over E-\tilde V(x)}
\left|{\pa \phi \over \pa w^i}(w)\right|{s^2\over 2}\right.\nonumber\\
&&\left.+2\sqrt{n}\int_0^s\int_0^\tau\left|{\pa Q\over \pa w^i}(\sigma,x,w)
\right|d\sigma d\tau\right],\nonumber
\end{eqnarray}
for  $(s,x,w)\in [0,t_0+\ep[\times B(x_0,\ep)\times U.$

Note that 
\begin{eqnarray}
\label{56.21}&&\int_0^s\int_0^\tau\left|{\pa Q\over \pa w^i}(\sigma,x,w)\right|d\sigma d\tau
\le s\int_0^s\left|{\pa Q\over \pa w^i}(\sigma,x,w)\right|d\sigma\\
&&\le {5\delta(\Omega)\over c}\int_0^s\left|{\pa Q\over \pa w^i}(\sigma,x,w)
-\sigma{r_{\tilde V,E}(x)\over \left({E-\tilde V(x)\over c^2}\right)}{\pa \phi\over \pa w^i}(w)\right|d\sigma\nonumber\\
&&+{5r_{\tilde V,E}(x)\delta(\Omega)\over c\left({E-\tilde V(x)\over c^2}\right)}\left|{\pa \phi\over \pa w^i}(w)\right|{s^2\over 2},\nonumber
\end{eqnarray}
for  $(s,x,w)\in [0,t_0+\ep[\times B(x_0,\ep)\times U$ (we used that $t_0+\ep \le {5\delta(\Omega)\over c}$).

Let 
\begin{equation}
\label{56.22a}C_3'(E,x,\tilde{V},\tilde{B},\Omega)={20cn^2\beta(\tilde{V},\tilde{B},\Omega)\delta(\Omega)}
e^{{10\delta(\Omega)\beta(\tilde{V},\tilde{B},\Omega)n^{3/2}\over E-\tilde{V}(x)}}
\end{equation}
\begin{eqnarray}
\label{56.22b}C_4'(E,x,\tilde{V},\tilde{B},\Omega)&=&{2c^2n^{3/2}\beta(\tilde{V},\tilde{B},\Omega)}
e^{{10\delta(\Omega)\beta(\tilde{V},\tilde{B},\Omega)n^{3/2}\over E-\tilde{V}(x)}}\\
&&\times 
\left[c\left(12\sqrt{n}+1\right){r_{\tilde V,E}(x)\over E-\tilde V(x)}
+{10r_{\tilde V,E}(x)\sqrt{n}\delta(\Omega)\over c\left({E-\tilde V(x)\over c^2}\right)}\right],\nonumber\\
\label{56.22c}C_5'(E,x,\tilde{V},\tilde{B},\Omega)&=&
C_4'(E,x,\tilde{V},\tilde{B},\Omega)e^{C_3'(E,x,\tilde{V},\tilde{B},\Omega)\over E-\tilde V(x)},
\end{eqnarray}
for $x\in \Omega$.

From \eqref{56.20}-\eqref{56.22c} and Gronwall's lemma (formulated in Subsection 6.7), it follows that
\begin{equation}
\label{56.24}\left|{\pa Q\over \pa w^i}(s,x,w)
-{r_{\tilde V,E}(x)\over \left({E-\tilde V(x)\over c^2}\right)}{\pa \phi\over \pa w^i}(w)s\right|
\le
{C_5'(E,x,\tilde{V},\tilde{B},\Omega)\over E-\tilde V(x)}\left|{\pa \phi \over \pa w^i}(w)\right|{s^2\over 2}
\end{equation}
for  $(s,x,w)\in [0,t_0+\ep[\times B(x_0,\ep)\times U.$

Now we assume without loss of generality that the integer $k$ in \eqref{56.2b} is $n$, and $p_0^n>0.$
We remind that  $w_0\in U$ is defined by $\phi(w_0)={p_0\over |p_0|}.$ We shall prove \eqref{F}.

From \eqref{56.2e}, it follows that
\begin{eqnarray}
\label{56.25a}{\pa \phi\over \pa w^l}(w_0)&=&e_l-{w_0^l\over \sqrt{1-|w_0|^2}}e_n,\\ 
\label{56.25b}|{\pa \phi\over \pa w^l}(w_0)|&\le&\sqrt{1+n},
\end{eqnarray}
for $l=1\ldots n-1$ and where $(e_1,\ldots,e_n)$ is the canonical basis of $\R^n$ and
$w_0=(w_0^1,\ldots,w_0^{n-1})$ (for \eqref{56.25b}, we used the estimate ${p_0^n\over |p_0|}\ge n^{-1/2}$ which implies that 
$1-|w_0|^2\ge {1\over n}$ and we used $|w_0^l|\le |w_0|<1,$ $l=1\ldots n-1$ and we used \eqref{56.25a}).
In addition, using \eqref{56.25a}, we obtain
\begin{equation}
|\sum_{l=1}^{n-1}\mu_l{\pa \phi\over \pa w^l}(w_0)|
\label{56.26}\ge\left|\left(\mu_1,\ldots,\mu_{n-1}\right)\right|,
\end{equation}
for all $\left(\mu_1,\ldots,\mu_{n-1}\right)\in \R^{n-1}.$

Using the fact that $\phi(w_0)\in \S ^{n-1}$ is orthogonal to ${\pa \phi\over \pa w^l}(w_0),$ $l=1\ldots n-1,$ and using \eqref{56.26},
we obtain
\begin{equation}
|\mu_1\phi(w_0)+\sum_{l=1}^{n-1}\mu_{l+1}{\pa \phi\over \pa w^l}(w_0)|=
\sqrt{\mu_1^2+|\sum_{l=1}^{n-1}\mu_{l+1}{\pa \phi\over \pa w^l}(w_0)|^2}
\label{56.27}
\ge n^{-1/2}\sum_{l=1}^n |\mu_l|,
\end{equation}
for all $\left(\mu_1,\ldots,\mu_n\right)\in \R^n.$     

Note that 
\begin{eqnarray}
\label{56.28}
&&\left|\lambda_1{\pa Q\over \pa t}(t_0,x_0,w_0)+\sum_{l=1}^{n-1}\lambda_{l+1}{\pa Q\over \pa w^l}(t_0,x_0,w_0)\right|
\ge\\
&&\left|\lambda_1g(r_{\tilde V,E}(x_0)\phi(w_0))+\sum_{l=1}^{n-1}\lambda_{l+1}{c^2r_{\tilde V,E}(x_0)t_0\over E-\tilde V(x_0)}{\pa \phi\over \pa
w^l}(t_0,x_0,w_0)\right|\nonumber\\
&&-|\lambda_1|\left|{\pa Q\over \pa t}(t_0,x_0,w_0)-g(r_{\tilde V,E}(x_0)\phi(w_0))\right|\nonumber\\
&&-\sum_{l=1}^{n-1}|\lambda_{l+1}|\left|{\pa Q\over \pa w^l}(t_0,x_0,w_0)-
{c^2r_{\tilde V,E}(x_0)t_0\over E-\tilde V(x_0)}{\pa \phi\over \pa w^l}(t_0,x_0,w_0)\right|,\nonumber
\end{eqnarray}
for $(\lambda_1,\ldots,\lambda_n)\in \R^n.$

We estimate the first term on the right-hand side of \eqref{56.28} by using \eqref{56.27} (note that $g(r_{\tilde V,E}(x_0)\phi(w_0))={c^2r_{\tilde V,E}(x_0)\over
E-\tilde V(x_0)}\phi(w_0)$). We estimate the second term  and third term on the right-hand side of \eqref{56.28} 
by using \eqref{56.9} and \eqref{56.24} and \eqref{56.25b}. Using also the estimate $t_0\le{5\delta(\Omega)\over c}$, we finally obtain
\begin{equation}
\label{F}\left|\lambda_1{\pa Q\over \pa t}(t_0,x_0,w_0)+\sum_{l=1}^{n-1}\lambda_{l+1}{\pa Q\over \pa w^l}(t_0,x_0,w_0)\right|
\ge|\lambda_1|{cC_6\over \sqrt{n}}+
t_0\sum_{l=1}^{n-1}|\lambda_{l+1}|{C_7\over \sqrt{n}},
\end{equation}
for $(\lambda_1,\ldots,\lambda_n)\in \R^n.$ 
This latter inequality and \eqref{56.2} and $t_0>0$ imply that the family $\left({\pa Q\over \pa t}(t_0,x_0,w_0),
{\pa Q\over \pa w^1}(t_0,x_0,w_0)\right.,\ldots,$$\left.{\pa Q\over \pa w^{n-1}}(t_0,x_0,w_0)\right)$ is free.
Then using inverse function theorem, it follows that $\varphi_E$ is a local $C^1$ diffeomorphism at $(t_0,x_0,p_0).$

Proposition 6.3 is proved.\hfill$\Box$

\vskip 4mm

\noindent 6.9 {\it Proof of Proposition 6.4.}
Before proving Proposition 6.4, we shall first prove the following Lemma 6.1.
\vskip 2mm

{\bf Lemma 6.1.} {\it
Assume that 
\begin{equation}
\label{56.32}\begin{array}{l}
E\ge C_1(\tilde V,\tilde B,\Omega),\\
C_8(E,\tilde V,\tilde B,\Omega)>0,
\end{array}
\end{equation}
where $C_1$ and $C_8$ are defined by \eqref{6.6ia} and \eqref{56.41c}.

Let $x\in \Omega$ and $p\in \S^{n-1}_{x,E}.$
Then
\begin{equation}
\label{56.33}\psi_2(t_{+,x,p},x,p)\circ N(\psi_1(t_{+,x,p},x,p))>0,
\end{equation}
where $t_{+,x,p}$ is defined by \eqref{6.t+} and $\psi=(\psi_1,\psi_2)$ is the flow of the differential system \eqref{4},
and where $N(y)$ denotes the unit outward normal vector of $\pa D$ at $y\in \pa D$ ($\circ$ denotes the usual scalar product on $\R^n$).
}

\begin{proof}[Proof of Lemma 6.1]
Consider  the function $m\in C^2(\R,\R)$ defined by
\begin{equation}
\label{56.34}m(t)=\chi_\Omega(\psi_1(t,x,p)),\ t\in \R,
\end{equation}
where $\chi_\Omega$ is a $C^2$ defining function for $\Omega$ (see definition of $C_8$, \eqref{56.41c}).
Derivating twice \eqref{56.34} and using \eqref{4}, we obtain
\begin{eqnarray}
&&\label{56.36}\ddot m(t)=Hess\chi_\Omega(\psi_1(t,x,p))(g(\psi_2(t,x,p)),g(\psi_2(t,x,p)))\\
&&+\left(1+{|\psi_2(t,x,p)|^2\over c^2}\right)^{-1/2}\nabla \chi_\Omega(\psi_1(t,x,p))\circ
F\left(\psi_1(t,x,p),\right.\nonumber\\
&&\left.g(\psi_2(t,x,p))\right)-{\psi_2(t,x,p)\circ F\left(\psi_1(t,x,p),g(\psi_2(t,x,p))\right)\over c^2\left(1+{|\psi_2(t,x,p)|^2\over c^2}\right)^{3/2}}
\nonumber\\
&&\times\nabla \chi_\Omega(\psi_1(t,x,p))\circ \psi_2(t,x,p),\nonumber
\end{eqnarray}
for $t\in \R$ and where $g$ is the function defined by \eqref{B}. From \eqref{56.36}, conservation of energy and \eqref{18c} and $|g(\psi_2(t,x,p))|<c,$ 
it follows that
$\ddot m(t_{+,x,p})\ge c^2C_8(E,\tilde V,\tilde B,\Omega)>0$
(we used \eqref{56.32}).

For all $t\in[0,t_{+,x,p}[,$ $\psi_1(t,x,p)\in \Omega.$ Hence 
$m(t)<0,$  $t\in[0,t_{+,x,p}[.$ This estimate and the estimate $\ddot m(t_{+,x,p})>0$ and 
Taylor expansion of $m$ at $t_{+,x,p}$ ($
m(t)=\dot m(t_{+,x,p})(t-t_{+,x,p})+{1\over 2}\ddot m(t_{+,x,p})(t-t_{+,x,p})^2+o((t-t_{+,x,p})^2),\ t\in \R$)
imply that $\dot m(t_{+,x,p})>0$.
 
Lemma 6.1 is proved.
\end{proof}

Now we are ready to prove Proposition 6.4.

Let $x\in \Omega$.
As $n\ge 2,$ the set $\Omega\backslash\{x\}$ is connected and we shall prove that the set 
\begin{eqnarray*}
A_x&=&\{y\in \Omega\backslash\{x\}\ |\ \textrm{there exists }p\in \S^{n-1}_{x,E} \textrm{ and }t>0 \textrm{ such that }(t,x,p)\in \Lambda \\
&&\textrm{ and } \psi_1(t,x,p)=y\}
\end{eqnarray*}
is a closed and open nonempty subset of $\Omega\backslash\{x\}$ (where $\psi=(\psi_1,\psi_2)$ is the differential flow of \eqref{4}).
Then we will have $A_x=\Omega\backslash\{x\},$ which will prove Proposition 6.4.

Note that $A_x$ is nonempty since for $p\in \S^{n-1}_{x,E}$, $(0,x,p)\in \Lambda$ and ${\pa \psi_1\over \pa t}(t,x,$ $p)_{|t=0} =g(p)\not=0.$ Hence there exists $\ep>0$ such that $(\ep,x,p)\in
\Lambda$ and $\psi_1(\ep,x,p)\not=\psi_1(0,x,p)=x.$

Note also that $A_x$ is an open subset of $\Omega\backslash\{x\}.$ Let $y\in A_x$. Then there exists $p\in \S^{n-1}_{x,E}$ and $t>0$ such
that $(t,x,p)\in \Lambda$ and $\psi_1(t,x,p)=y.$  
From \eqref{56.41} and Proposition 6.3, it follows, in particular, that there exists an open neighborhood $U\subseteq \Omega\backslash\{x\}$ 
of $y$ such that $U\subseteq A_x.$

It remains to prove that $A_x$ is a closed subset of $\Omega\backslash\{x\}$.
Consider a sequence $(y_k)$ of points of $\Omega\backslash\{x\}$ which converges to some $y\in \Omega\backslash\{x\}$ as $k\to +\infty.$
For each $k$, there exists $p_k\in\S^{n-1}_{x,E}$ and $t_k>0$ such that $(t_k,x,p_k)\in \Lambda$ and
\begin{equation}
\label{56.42}\psi_1(t_k,x,p_k)=y_k.
\end{equation}
From Proposition 6.1, it follows that $t_k\in[0,{5\delta(\Omega)\over c}]$ for all $k$.
Using compactness of $[0,{5\delta(\Omega)\over c}]$ and compactness of $\S^{n-1}_{x,E}$, we can assume that
$(t_k)$ converges to some $t\in[0,{5\delta(\Omega)\over c}]$ and that $(p_k)$ converges to some $p\in \S^{n-1}_{x,E}.$
Using \eqref{56.42} and continuity of $\psi_1$, we obtain
\begin{equation}
\label{56.43}y=\lim_{k\to+\infty}y_k=\lim_{k\to+\infty}\psi_1(t_k,x,p_k)=\psi_1(t,x,p).
\end{equation}
Note that $t>0$ since $y\not=x.$
Let $s\in[0,t[.$ Then using that $t_k\to t$ as $k\to +\infty,$ we obtain that there exists a rank $N_s$ such that $s<t_k$ for $k\ge
N_s.$ Hence $(s,x,p_k)\in \Lambda$ for $k \ge N_s$ and, in particular, $\psi_1(s,x,p_k)\in \Omega$ for $k\ge N_s.$
Hence we obtain that
\begin{equation}
\label{56.44}\psi_1(s,x,p)=\lim_{k\to +\infty}\psi_1(s,x,p_k)\in \bar \Omega,\textrm{ for }s\in [0,t[.
\end{equation}
 
Using Lemma 6.1 with \eqref{56.43} ($y\in \Omega$) and \eqref{56.44}, we obtain that $(t,x,p)$ $\in \Lambda\cap
\left(]0,+\infty[\times\{x\}\times\S^{n-1}_{x,E}\right)$ and $\psi_1(t,x,p)=y.$ Hence $y\in A_x.$

Proposition 6.4 is proved. \hfill$\Box$

\section{The nonrelativistic case} 

\noindent 7.1 {\it Nonrelativistic Newton equation in electromagnetic field.}
Consider the classical nonrelativistic Newton equation in a static electromagnetic field in an open subset $\Omega$ of $\R^n,$
$n\ge 2,$
\begin{equation}
\label{7.1}\begin{array}{l}
\ddot x=-\nabla V(x)+ B(x)\dot x,\\
\end{array}
\end{equation}
where $x=x(t)$ is a $C^1$
function with values in $\Omega,$ $\dot x={dx\over dt},$ and 
$V\in C^2(\bar \Omega,\R)$,
$B\in {\cal F}_{mag}(\bar \Omega)$.

The equation (7.1) is an equation for $x=x(t)$ and is the equation of motion in $\R^n$ of a nonrelativistic particle of mass
$m=1$ and charge $e=1$ in an external electromagnetic field described by  $V$ and $B$. 
In this equation $x$ is the position of the particle, $\dot x$ is its velocity, $t$ is
the time.

For the equation (7.1) the energy
\begin{equation}
\label{7.2}E={1\over 2}|\dot x(t)|^2+V(x(t))
\end{equation}
is an integral of motion.

\vskip 4mm
\noindent 7.2 {\it Inverse boundary problem.}
Consider equation (7.1) under condition \eqref{1.3a}.

One can prove that at
sufficiently large energy $E$ (i.e. $E>E^{\it nr}(\|V\|_{C^2,D},$ $\|B\|_{C^1,D},D)\ge \sup_{x\in D} V(x)$ where
real constant $E^{\it nr}(\|V\|_{C^2,D},\|B\|_{C^1,D},D)$ also has properties \eqref{1.6*} and \eqref{1.5*}), the solutions $x$ of energy
$E$ have properties \eqref{2.1} and \eqref{2.2} (the proof is obtained by slight modifications of proofs of Section 6).
Then at fixed energy $E>E^{\it nr}(\|V\|_{C^2,D},\|B\|_{C^1,D},D),$ one can define $s^{\it nr}_{V,B}(E,q_0,q),$
$k_{0,V,B}^{\it nr}(E,q_0,q),$ $ k_{V,B}^{\it nr}(E,q_0,q),$  as were defined $s_{V,B}(E,q_0,q),$
$k_{0,V,B}(E,q_0,q),$ $ k_{V,B}(E,q_0,q),$ in Section 2 for any
$q_0,q\in\bar D,$ $q_0\not=q.$
Further one can consider the following nonrelativistic version of Problem 1 formulated in Introduction
\begin{equation*}
\begin{array}{l}
{\bf Problem\ }1': 
\textrm{ given }k^{\it nr}_{V,B}(E,q_0,q),\ k^{\it nr}_{0,V,B}(E,q_0,q) \textrm{ for all } q_0,q\in\pa D,\\
q_0\not=q,\textrm{ at fixed
 sufficiently large energy }E,
\textrm{ find } V \textrm{ and }B.
\end{array}
\end{equation*}
The following uniqueness theorem  holds
\vskip 2mm
{\bf Theorem 7.1.} {\it At fixed} $E>E^{\it nr}(\|V\|_{C^2,D},\|B\|_{C^1,D},D)$, {\it the boundary data} $k^{\it nr}_{V,B}(E,q_0,q)$, 
$(q_0,q)\in \pa D\times \pa D,$ $q_0\not=q$,
{\it uniquely determine $V,B$. 

At fixed} $E>E^{\it nr}(\|V\|_{C^2,D},\|B\|_{C^1,D},D)$, {\it the boundary data} 

\noindent $k_{0,V,B}^{\it nr}(E,q_0,q)$, $(q_0,q)\in \pa D\times \pa D,$ $q_0\not=q$,
{\it uniquely determine $V,B$. 
}

\vskip 2mm
Theorem 7.1 is proved in Subsection 7.6.

\vskip 4mm

\noindent 7.3 {\it Inverse scattering problem.}
We consider equation (7.1) under condition \eqref{1.3b}.

The following is valid (see, for example, [LT] where classical scattering of particles in a long-range magnetic field is studied, and see [S] where
classical scattering of particles in a short-range electric field is studied): for any 
$(v_-,x_-)\in \R^n\times\R^n,\ v_-\neq 0,$
the equation (7.1)  has a unique solution $x\in C^2(\R,\R^n)$ such that
\begin{equation}
{x(t)=v_-t+x_-+y_-(t),}\label{7.6}
\end{equation}
where $\dot y_-(t)\to 0,\ y_-(t)\to 0,\ {\rm as}\ t\to -\infty;$  in addition for almost any 
$(v_-,x_-)\in \R^n\times \R^n,\ v_-\neq 0,$
\begin{equation}
{x(t)=v_+t+x_++y_+(t),}\label{7.7}
\end{equation}
 where $v_+\neq 0,\ v_+=a^{\it nr} (v_-,x_-),\ x_+=b^{\it nr}(v_-,x_-),\ \dot y_+(t)\to 0,\ \ y_+(t)\to 0,{\rm\ as\ }t \to +\infty$.

For an energy $E>0,$ the map 
$S_E^{\it nr}: \S_E^{\it nr}\times\R^n \to \S_E^{\it nr}\times\R^n $ (where 
$\S_E^{\it nr}=\{v\in \R^n\ |\ |v|=\sqrt{2E} \}$)
given by the formulas
\begin{equation}
{v_+=a^{\it nr}(v_-,x_-),\ x_+=b^{\it nr}(v_-,x_-)},\label{7.8}
\end{equation}
is called the scattering map at fixed energy $E$ for the equation (7.1) under condition \eqref{1.3b}. 
By ${\cal D}(S_E^{\it nr})$ we denote the domain of definition of $S_E^{\it nr}.$  The data
$a^{\it nr}(v_-,x_-),$ $b^{\it nr}(v_-,x_-)$ for $(v_-,x_-)\in {\cal D}(S_E^{\it nr})$ are called the scattering data at fixed energy $E$ for the equation 
(7.1) under condition \eqref{1.3b}.
We consider the following inverse scattering problem at fixed energy for the equation (7.1) under condition 
\eqref{1.3b}:
$$
{\bf Problem\ }2': {\rm\ given\ }S_E^{\it nr} {\rm\ at\ fixed\ energy\ }E, {\rm\ find\ }V{\rm\ and\ }B.
$$
From Theorem 7.1 and property \eqref{1.5*}, we obtain

\vskip 2mm

{\bf Theorem 7.2.}
{\it Let $\lambda\in \R^+$ and let $D$ be a bounded strictly convex (in the strong sense) open domain of $\R^n$ with $C^2$ boundary.  Let $V_1, V_2\in C^2_0(\R^n,\R),$
$B_1, B_2\in C^1_0(\R^n,A_n(\R))\cap {\cal F}_{mag}(\R^n)$
$\max(\|V_1\|_{C^2,D},\|V_2\|_{C^2,D},\|B_1\|_{C^1,D} ,\|B_2$ $\|_{C^1,D})\le \lambda,$ and ${\rm supp}(V_1)\cup {\rm supp}
(V_2)$ $\cup{\rm supp}(B_1)\cup{\rm supp}(B_2)\subseteq D.$ Let $S^\mu_E$ be the (nonrelativistic) scattering map at fixed energy $E$ subordinate to 
$(V_\mu,B_\mu)$ for $\mu=1,2.$ 
Then there exists a nonnegative real
constant $E^{\it nr}(\lambda,D)$ such that for any
$E>E^{\it nr}(\lambda,D),$ $(V_1,B_1)\equiv (V_2,B_2)$ if and only if $S^1_E\equiv S^2_E.$
}

\vskip 4mm 

\noindent 7.4 {\it Classical Hamiltonian mechanics.}
For $x\in \bar D$ and for $E>V(x),$ we define 
$$r_{V,E}^{\it nr}(x)=\sqrt{2(E-V(x))}.$$
Let $\A\in {\cal F}_{pot}(D, B)$. 
The equation (7.1) in $D$ is the Euler-Lagrange equation for the Lagrangian $L^{\it nr}$ defined by
$L^{\it nr}(\dot x,x)={1\over 2}|\dot x|^2
+\A(x)\circ\dot x-V(x),$ $\dot x\in \R^n$ and $x\in D,$ where $\circ$ denotes the usual scalar product on $\R^n.$
The Hamiltonian $H^{\it nr}$ associated to the Lagrangian $L^{\it nr}$ by Legendre's transform (with
respect to $\dot x$) is $H^{\it nr}(P,x)={1\over 2}|P-\A(x)|^2+V(x)$ where $P\in \R^n$ and $x\in D.$
Then equation (1.1) in $D$ is equivalent to the Hamilton's equation
\begin{equation}
\begin{array}{l}
\dot x={\pa H^{\it nr}\over \pa P}(P,x),\\
\dot P=-{\pa H^{\it nr}\over \pa x}(P,x),
\end{array}
\label{7.01}
\end{equation} 
for $P\in \R^n,$ $x\in D.$

For a solution $x(t)$ of equation (7.1) in $D,$ we define the impulse vector 
$$
P^{\it nr}(t)=\dot x(t)+\A(x(t)).
$$
Further for $q_0,q\in \bar D,$ $q_0\not= q,$  and $t\in[0,s^{\it nr}(E,q_0,q)],$ we consider
\begin{equation}
P^{\it nr}(t,E,q_0,q)=\dot x^{\it nr}(t,E,q_0,q)
+\A(x^{\it nr}(t,E,q_0,q)),
\end{equation}
where $x^{\it nr}(.,E,q_0,q)$ is the solution given by \eqref{2.2} in the nonrelativistic case.
From Maupertuis's principle (see [A]), it follows that if $x(t),$ $t\in[t_1,t_2],$ is a solution of $\eqref{7.1}$ in $D$ with energy
$E$, then $x(t)$ is a critical point of the functional ${\cal A}(y)=\int_{t_1}^{t_2}\left[r_{V,E}^{\it nr}(y(t))|\dot y(t)|+\right.$ $
\left.\A(y(t))\circ\dot y(t)
\right] dt$ defined on the set of the functions $y\in C^1([t_1,t_2], D),$ with boundary conditions $y(t_1)=x(t_1)$ and $y(t_2)=x(t_2).$
Note that for $q_0,q\in D,$ $q_0\not=q,$ functional ${\cal A}$ taken along the trajectory of the solution $x^{\it nr}(.,E,q_0,q)$ 
given by \eqref{2.2} is equal  to the reduced action ${\cal S}_{0_{V,\A,E}}^{\it nr}(q_0,q)$ from
$q_0$ to $q$ at fixed energy $E$  for \eqref{7.01}, where
\begin{equation}
{\cal S}_{0_{V,\A,E}}^{\it nr}(q_0,q)=\left\lbrace
\begin{array}{ll}
0,& \textrm{ if } q_0=q,\\
\int_0^{s^{\it nr}(E,q_0,q)}P^{\it nr}(s,E,q_0,q)\circ\dot x^{\it nr}(s,E,q_0,q)ds,&\textrm{ if } q_0\not=q,
\end{array}
\right.
\end{equation}
for $q_0,q\in \bar D$.

\vskip 4mm
\noindent 7.5 {\it Properties of the reduced action at a fixed and sufficiently large energy.} 
The reduced action at fixed and sufficiently large energy  for \eqref{7.01} has the same properties 
that those given in Proposition 3.1, 3.2 for the reduced action at fixed and sufficiently large energy
for the relativistic case.
 
Let 
$E>E^{\it nr}(\|V\|_{C^2,D}, \|B\|_{C^1,D},D).$ The reduced action ${\cal S}_{0_{V,\A,E}}^{\it nr}$ at fixed energy $E$ has the following properties: 
\begin{eqnarray}
&&{\cal S}_{0_{V,\A,E}}^{\it nr}\in C(\bar D\times \bar D,\R),\label{7.05}\\
&&{\cal S}_{0_{V,\A,E}}^{\it nr}\in C^2((\bar D\times \bar D)\backslash \bar G,\R),\label{7.06}\\
&&{\pa {\cal S}_{0_{V,\A,E}}^{\it nr}\over \pa x_i}(\zeta,x)=k^{{\it nr},i}_{V,B}(E,\zeta,x)+\A_i(x),\label{7.07}\\
&&{\pa {\cal S}_{0_{V,\A,E}}^{\it nr}\over \pa \zeta_i}(\zeta,x)=-k^{{\it nr},i}_{0,V,B}(E,\zeta,x)-\A_i(\zeta),\label{7.08}\\
&&{\pa^2 {\cal S}_{0_{V,\A,E}}^{\it nr}\over \pa \zeta_i\pa x_j}(\zeta,x)=-{\pa k^{{\it nr},i}_{0,V,B}\over \pa x_j}(E,\zeta,x)=
{\pa k^{{\it nr},j}_{V,B}\over \pa \zeta_i}(E,\zeta,x)
,\label{7.09}
\end{eqnarray}
for $(\zeta,x)\in (\bar D\times \bar D)\backslash \bar G,$ $\zeta=(\zeta_1,..,\zeta_n),$ $x=(x_1,..,x_n),$ and $i,j=1\ldots n.$
In addition, 
\begin{eqnarray}
&&\max(|{\pa {\cal S}_{0_{V,\A,E}}^{\it nr}\over \pa x_i}(\zeta,x)|,|{\pa {\cal S}_{0_{V,\A,E}}^{\it nr}\over \pa \zeta_i}(\zeta,x)|)\le M_1,\label{7.010}\\
&&|{\pa^2 {\cal S}_{0_{V,\A,E}}^{\it nr}\over \pa \zeta_i\pa x_j}(\zeta,x)|\le {M_2\over |\zeta -x|},\label{7.011}
\end{eqnarray}
for $(\zeta,x)\in (\bar D\times \bar D)\backslash \bar G,$ $\zeta=(\zeta_1,..,\zeta_n),$ $x=(x_1,..,x_n),$ and $i,j=1\ldots n,$ 
and where $M_1$ and $M_2$ depend on $V,$ $B$ and $D.$

In addition,
the map $\nu_{V,B,E}:\pa D\times D\to \S^{n-1},$  defined by
\begin{equation}
\nu_{V,B,E}^{\it nr}(\zeta,x)=-{k_{V,B}^{\it nr}(E,\zeta,x)\over|k_{V,B}^{\it nr}(E,\zeta,x)|}, \ {\rm for }\ (\zeta,x)\in \pa D\times D,
\label{7.012}
\end{equation}
has the following properties:
\begin{equation}
\begin{array}{l}
\nu_{V,B,E}^{\it nr}\in C^1(\pa D\times D,\S^{n-1}),\\
{\rm the\ map\  }\nu_{V,B,E,x}^{\it nr}:\pa D\to \S^{n-1},\ \zeta\mapsto \nu_{V,B,E}^{\it nr}(\zeta,x),{\rm \ is\ a}\\
C^1\ {\rm orientation\ preserving\ diffeomorphism\ from\ }\pa D \ {\rm onto}\ \S^{n-1}
\end{array}
\label{7.013}
\end{equation}
for $x\in D$ (where we choose the canonical orientation of $\S^{n-1}$ and the orientation of $\pa D$ given by the canonical orientation of $\R^n$ and 
the unit outward normal vector).

\vskip 2mm
{\bf Remark 7.1.} Equalities \eqref{7.07} and \eqref{7.08} 
are known formulas of classical Hamiltonian mechanics (see Section 46 and further Sections
of [A]). 
\vskip 2mm

{\bf Remark 7.2.} Taking account of \eqref{7.07} and \eqref{7.08}, we obtain the following formulas: 
at $E>E^{\it nr}(\|V\|_{C^2,D}, \|B\|_{C^1,D},D),$ for any $x,\ \zeta \in \bar D,$ $x\not=\zeta,$
\begin{eqnarray*}
B_{i,j}(x)&=&-{\pa k^{{\it nr },j}_{V,B}\over \pa x_i}(E,\zeta,x)+{\pa k^{{\it nr },i}_{V,B}\over \pa
x_j}(E,\zeta,x),\\
B_{i,j}(x)&=&-{\pa k^{{\it nr },j}_{0,V,B}\over \pa x_i}(E,x,\zeta)+{\pa  k^{{\it nr },i}_{0,V,B}\over \pa
x_j}(E,x,\zeta).
\end{eqnarray*}
\vskip 4mm

\noindent 7.6  {\it Proof of Theorem 7.1.}
We define  the $n-1$ differential form $\omega_{0,V,B}^{\it nr}$ on $\pa D\times D$ as was defined the $n-1$ differential form 
$\omega_{0,V,B}$ on $\pa D\times D$ in Subsection 3.3.

Now let $\lambda\in \R^+$ and $V_1, V_2\in C^2(\bar D,\R),$ $B_1, B_2\in {\cal F}_{mag}(\bar D)$ such that $\max(\|V_1\|_{C^2,D},\|V_2\|_{C^2,D},
\|B_1\|_{C^1,D},\|B_2\|_{C^1,D})\le \lambda.$  For $\mu=1,2,$ let $\A_\mu\in {\cal
F}_{pot}(D, B_\mu).$ 

Let $E>E^{\it nr}(\lambda,\lambda,D)$ where $E^{\it nr}(\lambda,\lambda,D)$ is defined by the nonrelativistic formulation of
\eqref{1.5*}.
Consider $\beta^{{\it nr},1},$ $\beta^{{\it nr},2}$ the differential one forms defined on $(\pa D\times \bar D)\backslash{\bar G}$ by
\begin{equation}
\label{7.016}\beta^{{\it nr},\mu}(\zeta,x)=\sum_{j=1}^n k^{{\it nr},j}_{V_\mu,B_\mu}(E,\zeta,x)dx_j,
\end{equation}
for $(\zeta,x)\in (\pa D\times \bar D)\backslash{\bar G},$ $x=(x_1,\ldots,x_n)$ and $\mu=1,2.$

Define the differential forms $\Phi_0^{\it nr}$ and $\Phi^{\it nr}$ as were defined $\Phi_0$ and $\Phi$ in Subsection 3.3 (replace 
$\beta^\mu,$ ${\cal S}_{0_{V_\mu,\A_\mu,E}}$ by $\beta^{{\it nr},\mu},$ ${\cal S}_{0_{V_\mu,\A_\mu,E}}^{\it nr}$, $\mu=1,2)$.

Then Lemma 3.1 and Theorem 3.1 remain valid by replacing $\Phi_0,$ $\Phi,$  $r_{V_\mu,E}$, $\omega_{0,V_\mu,B_\mu}$ and 
$\bar k_{V_\mu,B_\mu}$ 
by $\Phi_0^{\it nr},$ $\Phi^{\it nr},$ $r_{V_\mu,E}^{\it nr},$ $\omega_{0,V_\mu,B_\mu}^{\it nr}$ 
and $k_{V_\mu,B_\mu}^{\it nr}$, and the proof of these results are
obtained by slight modifications of proof of Lemma 3.1 and Theorem 3.1.

Hence Theorem 7.1 follows from the nonrelativistic formulation of Theorem 3.1.\hfill $\Box$

\section*{References}

\parindent=-1.7cm 
\leftskip=-\parindent 

\leavevmode \hbox to 1.7cm {}

\leavevmode \hbox to 1.7cm {[A]\hfill}V. I. Arnold, {\it Mathematical Methods of Classical Mechanics},

\noindent Springer Verlag New York
Heidelberg Berlin, 1978.

\leavevmode \hbox to 1.7cm {[B]\hfill}G. Beylkin, Stability and uniqueness of the solution of the inverse kinematic problem of seismology in higher
dimensions, {\it Zap. Nauchn. Sem. Leningrad. Otdel. Mat. Inst. Steklov. (LOMI)} {\bf 84}, 3-6 (1979) (Russian). English transl.: {\it J. Soviet
Math.} {\bf 21}, 251-254 (1983).  

\leavevmode \hbox to 1.7cm {[BG]\hfill}I. N. Bernstein, M. L. Gerver, A condition for distinguishing metrics from hodographs, {\it Comput.
Seismology} {\bf 13}, 50-73 (1980) (Russian). 

\leavevmode \hbox to 1.7cm {[E]\hfill}A. Einstein, \" Uber das Relativit\" atsprinzip und die aus demselben gezogenen Folgerungen, 
{\it Jahrbuch der Radioaktivit\" at und Elektronik} {\bf 4},  411-462 (1907).

\leavevmode \hbox to 1.7cm {[ER]\hfill}G. Eskin, J. Ralston,
Inverse scattering problem for the Schr\"odinger equation with magnetic potential at a fixed energy,
{\it Comm. Math. Phys.} {\bf 173},  199-224 (1995).

\leavevmode \hbox to 1.7cm {[GN]\hfill}M.L. Gerver, N. S. Nadirashvili, Inverse problem of mechanics at high energies, 
{\it Comput. Seismology} {\bf 15}, 118-125 (1983) (Russian).

\leavevmode \hbox to 1.7cm {[I]\hfill}H. Isozaki, Inverse scattering theory for Dirac operators, {\it Ann. Inst. H. Poincar\'e
Phys. Th\'eor.}  {\bf 66}:2, 237-270  (1997). 

\leavevmode \hbox to 1.7cm {[J1]\hfill}A. Jollivet, On inverse scattering for the multidimensional relativistic Newton equation at 
high energies, {\it J. Math. Phys. }  {\bf 47}(6), 062902 (2006).

\leavevmode \hbox to 1.7cm {[J2]\hfill}A. Jollivet, On inverse scattering in electromagnetic field in classical relativistic mechanics at high
energies, 2005 preprint, 

\noindent /math-ph/0506008.

\leavevmode \hbox to 1.7cm {[J3]\hfill}A. Jollivet, On inverse problems for the multidimensional relativistic Newton equation at fixed 
energy, 2006 preprint, /math-ph/0607003 (to appear in {\it Inverse Problems}, 2007).

\leavevmode \hbox to 1.7cm {[LL]\hfill}L.D. Landau, E.M. Lifschitz, {\it The Classical Theory of Fields}, Pergamon Press New York, 
1971.

\leavevmode \hbox to 1.7cm {[LT]\hfill}M. Loss, B. Thaller, Scattering of particles by long-range magnetic fields, {\it Ann. Physics} {\bf 176}, 159-180 (1987).

\leavevmode \hbox to 1.7cm {[MR]\hfill}R. G. Muhometov, V.G. Romanov, On the problem of determining an isotropic Riemannian metric in 
$n$-dimensional space, {\it Dokl. Akad. Nauk SSSR} {\bf 243}:1, 41-44 (1978) (Russian). English transl.: {\it Soviet math. Dokl.} {\bf 19}, 1330-1333
(1978). 

\leavevmode \hbox to 1.7cm {[NSU]\hfill}A. Nachman, J. Sylvester, G. Uhlmann, An $n$-dimensional 

\noindent Borg-Levinson theorem, {\it Comm. Math. 
Phys.}  {\bf 115}:4, 595-605 (1988).

\leavevmode \hbox to 1.7cm {[NaSuU]\hfill}G. Nakamura, Z. Sun, G. Uhlmann, Global identifiability for an inverse problem for the
Schr\"odinger equation in a magnetic field, {\it Math. Ann.} {\bf 303}, 377-388 (1995). 

\leavevmode \hbox to 1.7cm {[N1]\hfill}R.G. Novikov, A multidimensional inverse spectral problem for the equation
$-\Delta\psi+(V(x)-Eu(x))\psi=0,$ {\it Funktsional. Anal. i} 

\noindent {\it Prilozhen} {\bf 22}:4, 11-22, 96 (1988). English transl.: {\it Funct. Anal.
Appl.} {\bf 22}:4, 263-272 (1988).  

\leavevmode \hbox to 1.7cm {[N2]\hfill}R.G. Novikov, Small angle scattering and X-ray transform
in classical mechanics, {\it Ark. Mat.} {\bf 37},  141-169 (1999).

\leavevmode \hbox to 1.7cm {[N3]\hfill}R.G. Novikov, The $\overline\partial$-approach to approximate inverse scattering at fixed energy 
in three dimensions, {\it IMRP Int. Math. Res. Pap.}  {\bf 6}, 287-349 (2005). 

\leavevmode \hbox to 1.7cm {[S]\hfill}B. Simon, Wave operators for classical particle scattering.
{\it Comm. Math. Phys.} {\bf 23}, 37-48 (1971).

\leavevmode \hbox to 1.7cm {[Y]\hfill}K. Yajima, Classical scattering for relativistic particles, {\it J. Fac. Sci., Univ. Tokyo, 
Sect. I A} {\bf 29}, 599-611 (1982).
\vskip 8mm
\noindent A. Jollivet

\noindent Laboratoire de Math\'ematiques Jean Leray (UMR 6629)

\noindent Universit\'e de Nantes 

\noindent F-44322, Nantes cedex 03, BP 92208,  France

\noindent e-mail: jollivet@math.univ-nantes.fr
\end{document}